\documentclass[letterpaper,11pt]{article}
\usepackage{jheppub}
\usepackage{epstopdf}
\usepackage{slashed}
\usepackage{graphicx} 
\usepackage{soul}
\usepackage{amsmath}
\usepackage{multirow}
\usepackage{tablefootnote}
\usepackage{comment}
\usepackage{color}
\usepackage[utf8]{inputenc}
\usepackage{caption,subcaption}
\usepackage[normalem]{ulem}
\usepackage[T1]{fontenc}

\usepackage{graphicx}

\usepackage{amsmath}
\usepackage{amsfonts}
\usepackage{amssymb}
\usepackage{cleveref}
\usepackage{color}
\usepackage{xcolor}

\newcommand{\beq}{\begin{equation}}
\newcommand{\eeq}{\end{equation}}
\newcommand{\bea}{\begin{eqnarray}}
\newcommand{\eea}{\end{eqnarray}}
\newcommand{\ba}{\begin{array}}
\newcommand{\ea}{\end{array}}

\def\m1{M_1}
\def\m2{M_2}
\def\m3{M_3}

\def\ch10{\tilde \chi^0_1}

\def\mm{\mu^+\mu^-}

\def\mev{\,{\rm MeV}}

\def\to{\rightarrow}

\newcommand{\lsim}{\mathrel{\mathop{\kern 0pt \rlap
  {\raise.2ex\hbox{$<$}}}
  \lower.9ex\hbox{\kern-.190em $\sim$}}}
\newcommand{\gsim}{\mathrel{\mathop{\kern 0pt \rlap
  {\raise.2ex\hbox{$>$}}}
  \lower.9ex\hbox{\kern-.190em $\sim$}}}

\def\abi{\,{\rm ab}^{-1}}

\title{\Large{\bf WIMPs at High Energy Muon Colliders}
}

\author[a]{Tao Han,}
\author[b]{Zhen Liu,}
\author[c]{Lian-Tao Wang}
\author[d]{and Xing Wang}

\affiliation[a]{PITT PACC, Department of Physics and Astronomy, University of Pittsburgh, Pittsburgh, PA 15217, USA}
\affiliation[b]{Maryland Center for Fundamental Physics, Department of Physics,
University of Maryland, College Park, MD 20742, USA}
\affiliation[c]{Department of Physics, Enrico Fermi Institute, and Kavli Institute for Cosmological Physics, University of Chicago, Chicago, IL 60637, USA}
\affiliation[d]{Department of Physics, University of California at San Diego, La Jolla, CA 92093, USA}

\preprint{PITT-PACC-2009}

\abstract{
The Weakly Interacting Massive Particle (WIMP) paradigm is one of the most compelling scenarios for particle dark matter (DM).
We show in this paper that a high energy muon collider can make decisive statements about the WIMP DM, and this should serve as one of its main physics driver cases. We demonstrate this by employing the DM as the lightest member of an electroweak (EW) multiplet, which is a simple, yet one of the most challenging WIMP scenarios given its minimal collider 
 signature and high thermal target mass scale of 1 TeV--23 TeV.
We perform a first study of the reach of high energy muon colliders, focusing on the simple, inclusive and conservative signals with large missing mass, through the mono-photon, VBF di-muon and a novel mono-muon channel. 
Using these inclusive signals, it is possible to cover the thermal targets of doublet and triplet with a 10 TeV muon collider. Higher energies, 14 TeV--75 TeV, would ensure a  $5 \sigma$ reach above the thermal targets for the higher EW multiplets. 
We also estimate the reach of a search for disappearing tracks, demonstrating the potential significant enhancement of the sensitivity.
}

\makeatletter
\gdef\@fpheader{}
\makeatother

\begin{document}

\maketitle

\noindent
\section{Introduction}

There is mounting evidence for the existence of dark matter (DM) from the astronomical and cosmological observations \cite{Bertone:2004pz,Drees:2012ji}. Yet, the nature of dark matter remains to be one of the most outstanding puzzles in contemporary particle physics.
Weakly Interacting Mass Particles (WIMPs), present in many theories beyond the Standard Model (BSM), are natural cold DM candidates \cite{Jungman:1995df,Arcadi:2017kky,Roszkowski:2017nbc}. 
Among the WIMP candidates, one particularly simple case is the dark matter particle being the lightest member of an electroweak (EW) multiplet. 
The mass scale set by the requirement of saturating the thermal relic abundance is in the range of 1~TeV--23~TeV, while lower masses can still be viable with non-thermal production mechanisms. The mass splitting among the members of the same multiplet is controlled by the electroweak symmetry breaking, which is small in comparison with the overall mass scale.  Moreover, in the minimal case, the splitting is generated only at one-loop level. Hence, it is generic to expect that the masses of the members of the same multiplet are highly degenerate.  Both the high mass scale and near degeneracy render the DM searches at colliders extremely challenging. The model-independent mono-$X$ signals ($X=g,\gamma,W/Z, h...$) are not expected to reach beyond two to three hundred GeV at the high luminosity upgrade of the LHC (HL-LHC)~\cite{Low:2014cba,Han:2018wus}, while disappearing track based searches can extend the coverage up to 900 GeV for a triplet (Wino)~\cite{CidVidal:2018eel}. In the non-minimal cases where there are more than one multiplets reachable at the LHC and there is a sizable mass difference between them, HL-LHC will be able to reach up to 800 GeV for doublet (Higgsino) and 1200 GeV for wino \cite{Canepa:2020ntc}. 
At a future 100 TeV hadron collider, SppC or FCC$_{hh}$, one may hope to extend the coverage to 
a 1.5 (6) TeV for the Higgsino 
(Wino)~\cite{Benedikt:2018csr,CEPCStudyGroup:2018ghi,Strategy:2019vxc}. 

Recently, high energy muon colliders have been gathering steam and new benchmark running scnearios have been proposed, especially after the endorsement of its R$\&$D by the European strategy and the subsequent formation of the Muon Collider Collaboration~\cite{muCcoll:2020}. 
Being 200 times heavier than the electron, the muons could be readily accelerated to much higher energies before reaching the radiation loss barrier. In the past decade, there were studies on the physics potential of a muon collider~\cite{Ankenbrandt:1999cta}, mostly focused on a low energy version running on the Higgs pole as a Higgs factory \cite{Barger:1995hr,Barger:1996jm,Han:2012rb}. 
A muon collider operating at much higher energies, however, offers great potential in reaching new physics up to its energy thresholds, and thus provides tremendous opportunities to produce and discover new heavy EW particles. 
In this paper, we consider the following benchmark choices of the collider energies and the corresponding integrated luminosities,
\beq
\sqrt{s} = 3,\ 6,\ 10,\ 14, \  30 \ {\rm and}\ 100\ {\rm TeV},\quad {\mathcal L} = 1,\ 4,\ 10,\ 20,\ 90, \ {\rm and}\ 1000\ {\rm ab}^{-1} .
\label{eq:para}
\eeq
Here we follow the basic assumption that the integrated luminosity of high energy colliders scales linearly with $s$.
Motivated by this exciting perspective, we perform a first study on the reach of muon colliders for WIMP DM. Particles in an EW multiplet, especially those charged,  can be copiously produced. Depending on its lifetime, it would either decay into slow-moving (soft) particles or leaving (disappearing) charged tracks in the detector. These signals will certainly be an integral part of the DM searches at a muon collider. However, due to the large beam induced background (BIB) expected at a muon collider, it would be difficult to precisely assess the reach by relying on these signals. 
In this paper, we first focus on the universal and inclusive signals, where the particles in an EW multiplet are  produced in association with at least one energetic SM particle.  The soft particles or disappearing tracks are treated as invisible.  The most obvious channel is the pair production of the EW multiplet in association with a photon. 
In addition to the standard mono-photon channel, which dominates the sensitivity to higher-dimensional EW multiplet, we also consider a few other vector boson fusion (VBF) channels unique to a high-energy muon collider \cite{Costantini:2020stv}. 
 In particular, the mono-muon channel shows the most promise.  Special care is needed for computing the signal production and estimating the background in this channel. 
After considering the inclusive signatures, we perform a phenomenological estimate of the size of the disappearing track signal. Without detailed knowledge of the beam induced background, we give the reach if 20$-$50 signal events can be isolated with background on the order of 
100. This should be thought of as a performance target rather than a precise projection. 
We find that the future high energy muon collider can draw firm conclusions on the EW multiplet WIMP DM by a combination of the multiple search channels considered in this paper. Furthermore, there are many potential improvements to enhance the sensitivities. Given the promising results for the most challenging set of signals considered, we conclude that the high energy muon collider could make a  huge impact in our search for the thermal dark matter, and it should serve as one of the main physics drivers for a high energy muon collider program. 

The rest of this paper is organized as follows. In \autoref{sec:WIMP}, we set up our theoretical framework for the EW multiplets and lay out some general considerations for the WIMP DM issues. 
In \autoref{sec:WIMPmuC}, we present our analyses at a muon collider with a variety of energy and luminosity benchmarks. 
We summarize our results and discuss directions for further exploration in \autoref{sec:Sum}.
We also provide the details of our results for the $2\sigma$ and $5\sigma$ sensitivities for the EW multiplets with various machine parameter choices in an appendix.

\section{General Considerations of WIMP Benchmarks}
\label{sec:WIMP}

Many scenarios of WIMP DM have been put forward, among which the most frequently considered are the color-neutral fermionic EW gauge multiplets. The best-known examples include the SU(2) doublet and triplet, also known as the Higgsino and Wino in supersymmetric theories. In addition to these, we also consider a broader class of DM candidates, including higher SU(2) representations \cite{Cirelli:2005uq,Cirelli:2009uv,DiLuzio:2018jwd}, the so-called ``minimal dark matter'' scenario.

More specifically, we will consider multiplets $(1, n, Y)$ under the Standard Model (SM) gauge group SU(3)$_{\rm C} \otimes$SU(2)$_{\rm L} \otimes$U(1)$_{\rm Y}$. 
The $i^{th}$ member of this multiplet has electric charge $Q_i=t^3_i+Y$, where $t^3_i$ is the corresponding SU(2)$_{\rm L}$ isospin component. First, we consider fermionic multiplets. In this case, they only have gauge interactions at the renormalizable level. The mass scale of the EW multiplet is set by the vector-like mass parameter $M$. After electroweak symmetry breaking, the mass spectrum of the multiplet is not exactly degenerate. Minimally, the degeneracy will be lifted by EW loop corrections \cite{Thomas:1998wy,Buckley:2009kv,Cirelli:2005uq,Cirelli:2009uv,Ibe:2012sx}. If there is more than one EW multiplet or with additional singlets, the mixing among them can also shift the mass spectrum. Note that there is also an upper limit on the dimension of a multiplet. A large representation will lead to a breakdown of perturbative expansion, which sets a limit $n<16$ \cite{Di_Luzio_2019}. At the same time, for $n > 7$, the Landau pole will be about one order of magnitude above the mass of EW multiplet \cite{DiLuzio:2015oha}, which makes the model quite contrived. For this reason, we limit ourselves to $n \leq 7$. 
We begin with odd-dimensional multiplets, $(1, n=2T+1, Y)$, with a positive integer $T$. To focus on the minimal scenario, we will only consider the mass splitting generated by EW loop corrections. As we will discuss carefully in \autoref{sec:DT}, if we choose the hypercharge $Y=0$, the electrically neutral member is always the lightest mass eigenstate in the multiplet. In this case, fermions in these multiplets can be either Majorana or Dirac, as we listed in the left column of \autoref{tab:WIMP}. Beyond the renormalizable level, there can be additional contributions to the mass splitting. We will assume these effects to be sub-leading in comparison with the EW loop corrections.  Moreover, there could be operators that will allow the dark matter particle to decay. Here, we assume that additional symmetries can be imposed so that the neutral particle remains a good dark matter candidate. It has been proposed \cite{DelNobile:2015bqo} that the stability can be also guaranteed by introducing a small hyper-charge $Y=\epsilon$. Obviously, the dark matter candidate in the resulting Dirac EW multiplet will have a small electric charge $Q=\epsilon$ in this case.  Here, we will not insist on a particular mechanism for the stability of the dark matter, since such a mechanism will not have an impact on the collider signals to be investigated in this paper. We will, however, adopt the notation $(1, n=2T+1, \epsilon)$ to label a Dirac multiplet, and correspondingly $(1, n=2T+1, 0)$ for a Majorana multiplet. 

\begin{table}[t]
  \centering
    \begin{tabular}{c|c|r|c|c|c|c}
    \hline
    \multicolumn{2}{c|}{\multirow{2}[-3]{*}{Model}} & \multicolumn{1}{c|}{\multirow{2}[-3]{*}{Therm.}} & \multicolumn{4}{c}{5$\sigma$ discovery coverage (TeV)}  \\ \cline{4-7}
    \multicolumn{2}{c|}{$({\rm color}, n, Y)$} & target   & \multicolumn{1}{c|}{mono-$\gamma$} & \multicolumn{1}{c|}{mono-$\mu$} & \multicolumn{1}{c|}{di-$\mu$'s} & \multicolumn{1}{c}{disp. tracks} \\  \hline
    (1,2,1/2)$$ & Dirac & 1.1 TeV & ---  & 2.8  & ---  & $3.2-8.5$  \\ \hline  \hline
    (1,3,0)$$ & Majorana & 2.8 TeV & ---   & 3.7  & ---  & $13-14$  \\ \hline
    (1,3,$\epsilon$)$$ & Dirac & 2.0 TeV & 0.9  & 4.6  & ---  & $13-14$  \\ \hline  \hline
    (1,5,0)$$ & Majorana & 14 TeV & 3.1  & 7.0  & 3.1  & $10-14$ \\ \hline
    (1,5,$\epsilon$)$$ & Dirac & 6.6 TeV &  6.9 & 7.8 & 4.2  & $11-14$   \\ \hline  \hline
    (1,7,0)$$ & Majorana & 23 TeV  & 11   & 8.6   & 6.1  & $8.1-12$ \\ \hline
    (1,7,$\epsilon$)$$ & Dirac & 16 TeV & 13  & 9.2 & 7.4 & $8.6-13$  \\ \hline  \hline

    \end{tabular}%
    \caption{
    Generic minimal dark matter considered in this paper and a brief summary of their $5\sigma$ discovery coverage at a 30 TeV high energy muon collider with the three individual channels. Further details of individual and combined channels, the $2\sigma$ and 5$\sigma$ reaches, and different collider parameter choices,  including $\sqrt s=$3, 6, 10, 14, 30, 100~TeV are provided in the summary plots in \autoref{fig:channels}, \autoref{fig:summary}, and in the appendix.
    \label{tab:WIMP}
    }
\end{table}%
For an even-dimensional $n$-plet, setting $Y=(n-1)/2$ ensures the lightest eigenstate of the EW multiplet to be neutral.\footnote{For smaller values of $Y$ for the even $n$-plet, one might need to rely on some additional splitting generating mechanisms to change the lightest state being charged to neutral for $n\geq 4$. For a more detailed discussion on the splittings and hyper-charges, see \autoref{sec:DT}.} In the minimal case, the limits from direct detection rule out all cases with $Y\neq 0$.\footnote{The only exception is the case with tiny hyper-charge discussed above.} Hence, to make the even-dimensional multiplet a viable scenario, we could go beyond the minimality and introduce another state which mixes with the multiplet after EW symmetry breaking and generates a small Majorana mass splitting between the neutral Dirac fermion pair \cite{Cirelli:2009uv}.  It is also possible to have such a splitting, while the EW loop corrections still dominate the mass splitting between the neutral and the charged members of the multiplet. For example, if a dimension-5 operator generates a mass splitting after integrating out the new physics with a mass scale $M$, we have $\Delta m \propto v^2/M$. Requiring this to be smaller than the loop contributions and yet large enough to protect against the direct detection bounds puts $M \sim (10$--$1000)$ TeV. 
Whether such additional new physics can also be probed at a high-energy muon collider is a model-dependent question that we will not pursue further.    For the rest of our analyses, we will present the EW doublet (Higgsino) results while implicitly making the assumptions above. It is the smallest even-dimensional multiplet and also present in SUSY. The results for higher even-$n$ multiplets are included in the appendix. The main features of the collider signals in these cases are similar to those odd-dimensional multiplets discussed in detail in this paper. 

In principle, both real and complex scalar EW multiplets can contain viable dark matter candidates. The discussion of EW loop corrections to the mass splitting parallels to that of the fermions. The stability constraint due to non-renormalizable operators tends to be stronger. However, it can also be circumvented either by introducing more symmetries or assuming a tiny hyper-charge. One main difference is that the scalar can have more couplings in addition to gauge interactions at the renormalizable level, of the form $\chi \chi^\dagger H H^{\dagger}$ with different ways of contracting SU(2)$_{\rm L}$ indices. Such couplings can induce sizable splittings in the EW multiplet after the EW symmetry breaking. Hence, there are more parameters and model dependences in comparison with the case of fermionic EW multiplets. While there is certainly rich physics to be studied here, we will leave a full exploration to a future study. With the simplifying assumption that the scalars only have gauge interactions, the feature of its signal and the reach are similar to those of the fermions. We note here the scalar pair production is dominated by the $p$-wave process and has fewer degrees of freedom than the fermion cases, and hence the reach near the kinematic threshold of $m_\chi\sim \sqrt{s}/2$ is reduced. On the other hand, new contact interactions of $VV\chi\chi$ type, where $VV$ are standard model gauge boson and $\chi\chi$ are scalar EW multiplet pairs, lead to additional vector-boson-fusion production contributions.

The interactions of the EW multiplet with the SM particles set the thermal relic abundance of the cold DM.  Requiring thermal relic abundance matches today's observation \cite{Aghanim:2018eyx} can determine the mass of the dark matter, which we refer to as the {\it thermal target}. 
We list the multiplets according to the SM gauge quantum numbers under SU(3)$_{\rm C}\otimes$SU(2)$_{\rm L} \otimes$U(1)$_{\rm Y}$ and the predicted thermal targets in \autoref{tab:WIMP}, which set the benchmark for searches at future colliders. We note here that perturbative calculation of the thermal targets for many of the EW multiplets receives large corrections from the Sommerfeld enhancement~\cite{Belotsky:2005dk,Hisano:2006nn,Cirelli:2007xd} as well as bound state effects~\cite{An:2016gad,Mitridate:2017izz}. 
In detail, we follow Ref.~\cite{DiLuzio:2018jwd} in thermal target calculation for most of the benchmark points, which are mainly derived from Ref.~\cite{DelNobile:2015bqo} with Sommerfeld corrections taken into account. One exception is the 5-plet Majorana fermion, where the bound state effects are also included, lifting the thermal target from 9 TeV to 14 TeV~\cite{Mitridate:2017izz}. The bound state effect for the triplet Majorana fermion does not shift the thermal target compared to the Sommerfeld calculation~\cite{Mitridate:2017izz}. For the 7-plet Majorana fermion, we use the relation that in the SU(2)$_{\rm L}$ invariant calculation, the degrees of freedom decrease by a factor of two, pushing the thermal target $\sqrt{2}$ higher compared to the Dirac case available from Ref.~\cite{DelNobile:2015bqo}.
We note that these thermal targets have some theoretical uncertainties due to the non-perturbative effects mentioned above. Nevertheless, they can serve as useful targets.  Below the target mass, the thermal relic of EW multiplets does not over-close the universe, while some non-thermal productions or other DM species are assumed to produce the correct relic abundance of the cold dark matter. 
Reaching these targets marks a great triumph for future colliders in probing WIMP dark matter, with the potential of the next milestone discovery.

\section{WIMP Phenomenology at High Energy Muon Colliders}
\label{sec:WIMPmuC}

With only gauge interactions,
the production and decay for the EW multiplets are highly predictable. Since the mass splittings  between the charged and neutral states are expected to be small, typically of the order of a few hundred MeV, the decay products will be very soft, most likely escaping the detection. In this case, the main signal at high energy muon colliders is large missing energy-momenta. We note that, unlike in the high energy hadronic collisions where only the missing transverse momenta can be reconstructed by the momentum conservation, the four-momentum of the missing particle system can be fully determined in leptonic collisions because of the well-constrained kinematics. Importantly, a large missing invariant mass can be inferred. We thus introduce the ``missing mass'' defined as
\begin{equation}
m^2_{\rm missing} \equiv (p_{\mu^+} + p_{\mu^-} - \sum_i p_i^{\rm obs})^2,
\label{eq:recoilM}
\end{equation}
where $p_{\mu^+}, p_{\mu^-}$ are the momenta for the initial colliding beams, and $p_i^{\rm obs}$ is the momentum for the $i^{th}$ final state particle observed. If the EW multiplet particles are not detected, $m_{\rm missing}$ for the signal will have a threshold at twice the dark matter mass. We thus call this characteristic signature the ``missing-mass'' signal. In the first part of our analyses, we focus on the missing-mass signature in three leading channels, namely the mono-photon plus missing mass in~\autoref{sec:monophoton},  a novel channel of mono-muon plus missing mass in~\autoref{sec:monomuon}, and VBF di-muon plus missing mass in~\autoref{sec:VBF}.
We shall see that these three channels are complimentary to each other. In particular, the mono-muon channel provides very competitive sensitivities for EW multiplets for the doublet and the triplets, enabling coverage for the thermal targets with relatively lower center of mass energies of the muon collider. There are also dedicated studies in the search for the exotic signatures at hadron colliders, such as the disappearing track signal, which could help to significantly enhance the reach. 
While being susceptible to the beam induced background~(BIB),
we do expect these set of signals, being quite unique to this class of models, will play an important role in the searches at muon colliders, and we discuss them in detail in \autoref{sec:DT}. 


The signal and background have been generated using the Monte Carlo generator {\tt MadGraph}~\cite{Alwall:2014hca}. 
The EW multiplet model files are generated using {\tt FeynRules}~\cite{Alloul:2013bka} with many properties cross-checked with our own calculation. 
While there is no concrete design for a detector at the muon collider yet, due to the need of shielding, we conservatively assume that the detector would have good coverage in the range of $10^\circ - 170^\circ$. 
As we will describe in detail later, we are focusing on energetic photons and muons as our main final states. For these objects, we assume the beam induced and other detector generated background will not significantly affect the particle ID and reconstruction quality after the reconstructed object surpassing 10s of GeV of energy threshold. 

In the rest of this section, we describe the main channels for the EW multiplet production and their SM backgrounds included in our study and report the reach. 

\subsection{Mono-Photon}
\label{sec:monophoton}

\begin{figure}[tb]
\centering
\begin{subfigure}[t]{0.22\textwidth}\centering
\includegraphics[width=\textwidth]{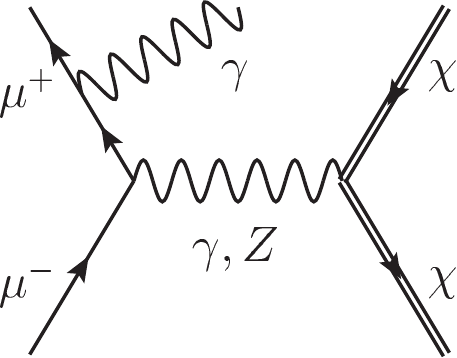}
\caption{}
\end{subfigure} \ \ 
\begin{subfigure}[t]{0.22\textwidth}\centering
\includegraphics[width=\textwidth]{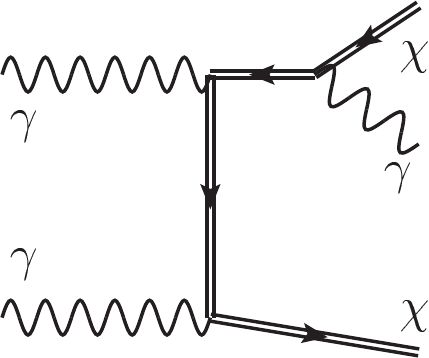}
\caption{}
\end{subfigure} \ \ 
\begin{subfigure}[t]{0.22\textwidth}\centering
\includegraphics[width=\textwidth]{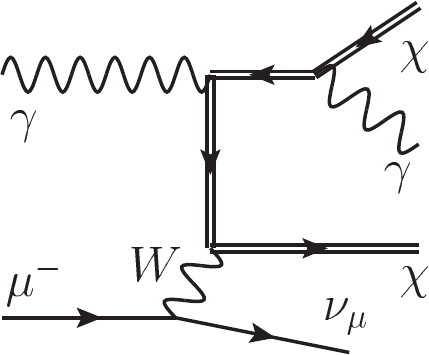}
\caption{}
\end{subfigure} \ \ 
\begin{subfigure}[t]{0.22\textwidth}\centering
\includegraphics[width=\textwidth]{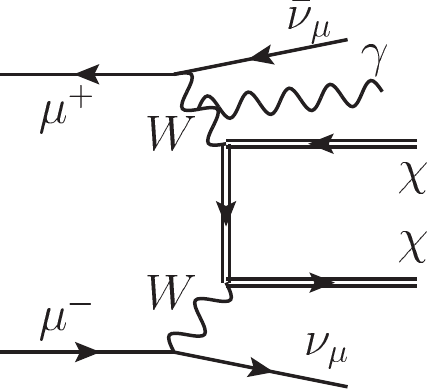}
\caption{}
\end{subfigure}
\caption{Representative Feynman diagrams for the mono-photon signal from a variety of $\chi\chi$ production channels (a) $\mm$ annihilation, (b) $\gamma\gamma$ fusion, (c) $\gamma W$ fusion, and (d) $WW$ fusion.}
\label{fig:Feynmonogm}
\end{figure}

We first consider the mono-photon signal. The  members of the electroweak multiplet, both charged and neutral,  can be produced either via $s$-channel $\gamma$ and $Z$ or via the vector boson fusion processes. The charged states will in turn decay into the lightest state and some soft particles, or leave a charge track if the charged states are long lived. As we stated above, we will consider these soft particles to be unobservable for now. Hence, the most obvious signal would be to have an additional photon recoiling against the EW multiplet in the production process. In the following, we will study this mono-photon channel in detail. 

\begin{figure}[tb]
\centering
\begin{subfigure}[t]{0.22\textwidth}\centering
\includegraphics[width=\textwidth]{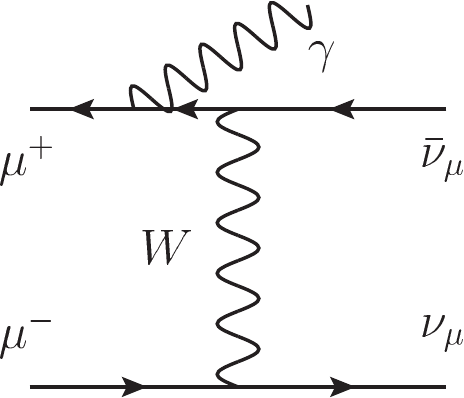}
\caption{}
\end{subfigure} \ \
\begin{subfigure}[t]{0.22\textwidth}\centering
\includegraphics[width=\textwidth]{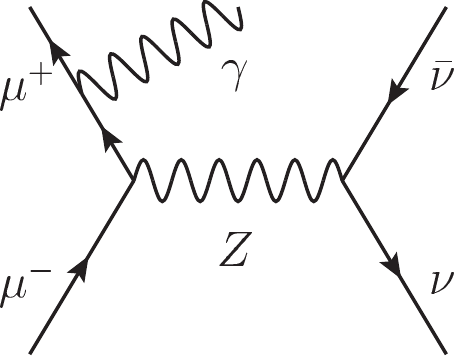}
\caption{}
\end{subfigure}
\caption{Representative Feynman diagrams for the SM mono-photon background (a) from 
$W$-exchange, and (b) from $Z\to \nu\bar\nu$.} 
\label{fig:FeynmonogmSM}
\end{figure}

We consider the following signal processes
\begin{eqnarray}
 \mu^+ \mu^-  &\rightarrow& \gamma \chi \chi \quad {\rm via\ annihilation}\ \mu^+\mu^- \to \chi\chi , \\
\gamma \gamma & \rightarrow& \gamma \chi \chi \quad {\rm via}\ \gamma\gamma\to \chi\chi , \\
 \gamma \mu^\pm & \rightarrow &\gamma \nu \chi \chi \quad {\rm via}\ \gamma W\to \chi \chi, \\
 \mu^+ \mu^- & \rightarrow& \gamma \nu \nu \chi \chi \quad {\rm via}\ WW\to \chi\chi\ {\rm and}\ 
 \mu^+\mu^- \to \chi\chi Z.
\end{eqnarray}
where $\chi$ represents any state within the $n$-plet and $\chi\chi$ represents any combination of a pair of the $\chi$ states allowed by the gauge symmetries. 
We show the representative Feynman diagrams for the mono-photon signal corresponding to the above processes in \autoref{fig:Feynmonogm}.
Apart from the initial state radiation (ISR) or final state radiation (FSR) photon, the signal rate and kinematics are mainly determined by the underlying two-to-two processes. 
For a heavy $\chi$, the direct $\mu^+\mu^-$ annihilation remains to be the dominant production source via $\gamma^*,Z^* \to \chi\chi$ (dubbed as a Drell-Yan process due to its similarity to $pp\to \gamma^*/Z^* \to \ell^+\ell^-$ at hadron colliders). 
For the next two processes in $\gamma\gamma$ and $\gamma W$ fusion, photons are treated as initial state partons. This is appropriate since there are large fluxes of photons coming from collinear radiation of the high-energy muon beams. 
We modify {\tt MadGraph} to include photons from muons using its encoded improved effective photon approximation~\cite{Frixione:1993yw} with a dynamical scale $Q=\sqrt{\hat s}/2$, where $\sqrt{\hat s}$ is the partonic center-of-mass (c.m.) energy. 
The process (d) in \autoref{fig:Feynmonogm} inherits both the $WW$ VBF and $ \chi\chi Z$ with $Z\to \nu\bar \nu$. 
For simplicity, we will not invoke the EW parton distribution functions (PDFs) for the massive vector bosons \cite{Han:2020uid} in this study and will perform the tree-level fixed order calculations. 

As for the signal identification, we first require a photon in the final state to be in the detector acceptance 
\begin{equation}
10^\circ < \theta_\gamma < 170^\circ .
\label{eq:angle}
\end{equation}
Taking into account the invariant mass of the dark matter pair system being greater than $2 m_\chi$, we impose further selective cuts on the energy of the photon and on the missing mass 
\begin{equation}
 E_\gamma > 50~{\rm GeV} ,\ \ \ 
m^2_{\rm missing} \equiv (p_{\mu^+} + p_{\mu^-} - p_\gamma)^2 > 4m_\chi^2 .
\label{eq:ecut}
\end{equation}
The missing-mass cut is equivalent to an upper limit on the energy of the photon $E_\gamma < (s - 4 m_\chi^2)/2\sqrt{s}$, 
where $\sqrt{s}$ is the collider c.m.~energy.

We consider multiple sources of the SM background, with some representative Feynman diagrams shown in \autoref{fig:FeynmonogmSM}. The most significant SM background, after the selection cuts, is
\begin{equation}
\mu^+ \mu^- \rightarrow \gamma \nu \bar{\nu}, 
\end{equation}
dominantly from contributions via the $t$-channel $W$-exchange. 

\begin{figure}[tb]
\centering
\begin{subfigure}{0.48\textwidth}
\includegraphics[width = \textwidth]{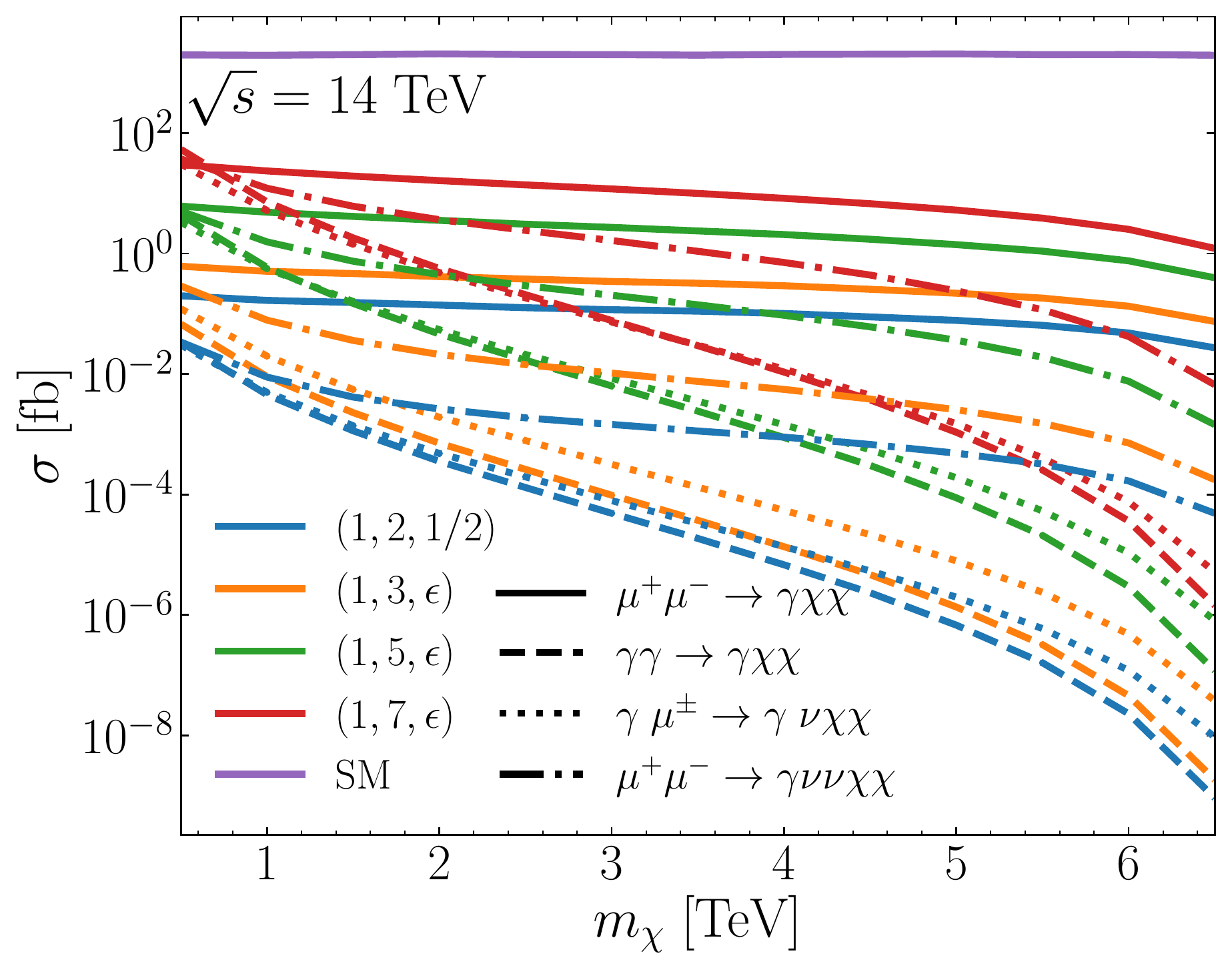}
\caption{}
\end{subfigure}
\begin{subfigure}{0.48\textwidth}
\includegraphics[width = \textwidth]{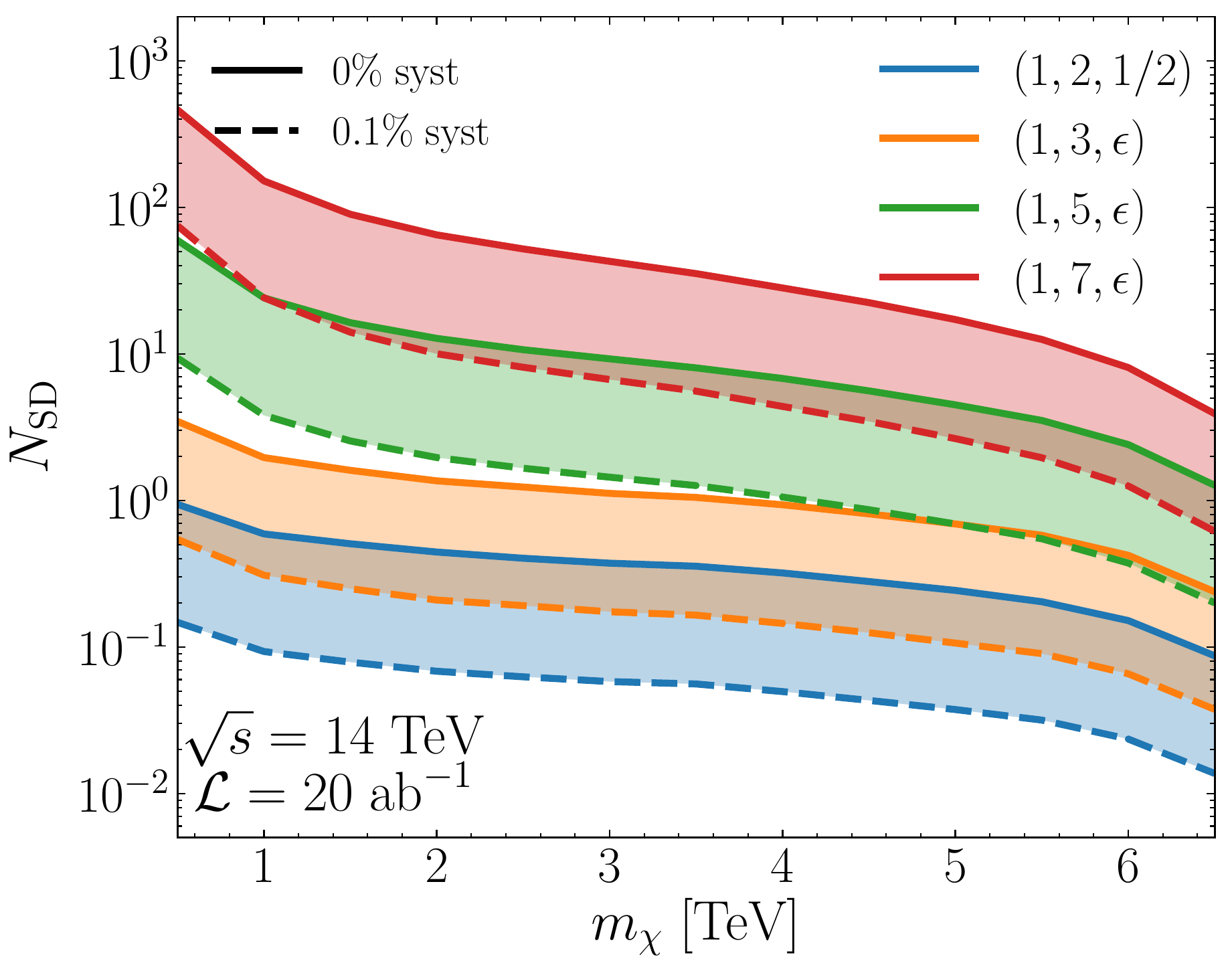}
\caption{}
\end{subfigure}
\caption{(a) Total cross section and (b) the significance defined in \autoref{eq:NSD} for a pair of EW multiplets plus a mono-photon at a muon collider with $\sqrt s=14$ TeV. In (b) the solid and dashed lines correspond to the systematic uncertainties of 0\% and 0.1\%, respectively. 
}
\label{fig:monogm}
\end{figure}

In \autoref{fig:monogm}, we show the cross sections for the signal processes with a variety of the EW multiplets as labeled on the figures.
For simplicity, we only plot the Dirac EW multiplets. The pair-production cross section for Majorana fermions will be a factor of two smaller than the Dirac fermions with same quantum numbers.
The dominant process is the Drell-Yan pair production with an additional ISR or FSR photon. 
For a fixed muon collider center of mass energy, the Drell-Yan (DY) process cross section is rather insensitive to the DM mass except for the near threshold regime with $\beta_\chi=\sqrt{1-{4m_\chi^2}/s}$ suppression, as an $s$-wave process. 
The VBF production for the EW multiplet, on the other hand, are characterized by the infrared behavior of the initial state gauge bosons and the cross section is thus scaled with the heavy DM mass as approximately $1/m_\chi^4$ for $m_\chi \ll \sqrt{s}/2$. This scaling can be understood as the following. From the vector boson PDF point of view, the parton luminosity scales as $1/\tau$. Near the threshold, 
$\tau \propto m_\chi^2$.
In addition, the underlying $VV\rightarrow \chi\chi$ cross section of the hard partonic cross section is suppressed by the flux factor of $1/(4m_\chi^2)$. 
The SM backgrounds are dominated by low energy ISR photons, and hence is insensitive to the DM mass-dependent missing-mass cut. The post-cut background rate shown in \autoref{fig:monogm} in purple line is hence flat. 
As the energy of the muon collider increases, the Drell-Yan process rate will decrease. At the same time, the rate of the VBF process will increase and asymptote to increasing  logarithmically with energy as $\sqrt{s} \gg 2 m_\chi $.
This renders the relative importance between different signal production channels to change.

\begin{figure}[tb]
\centering
\begin{subfigure}{0.48\textwidth}
\includegraphics[width = \textwidth]{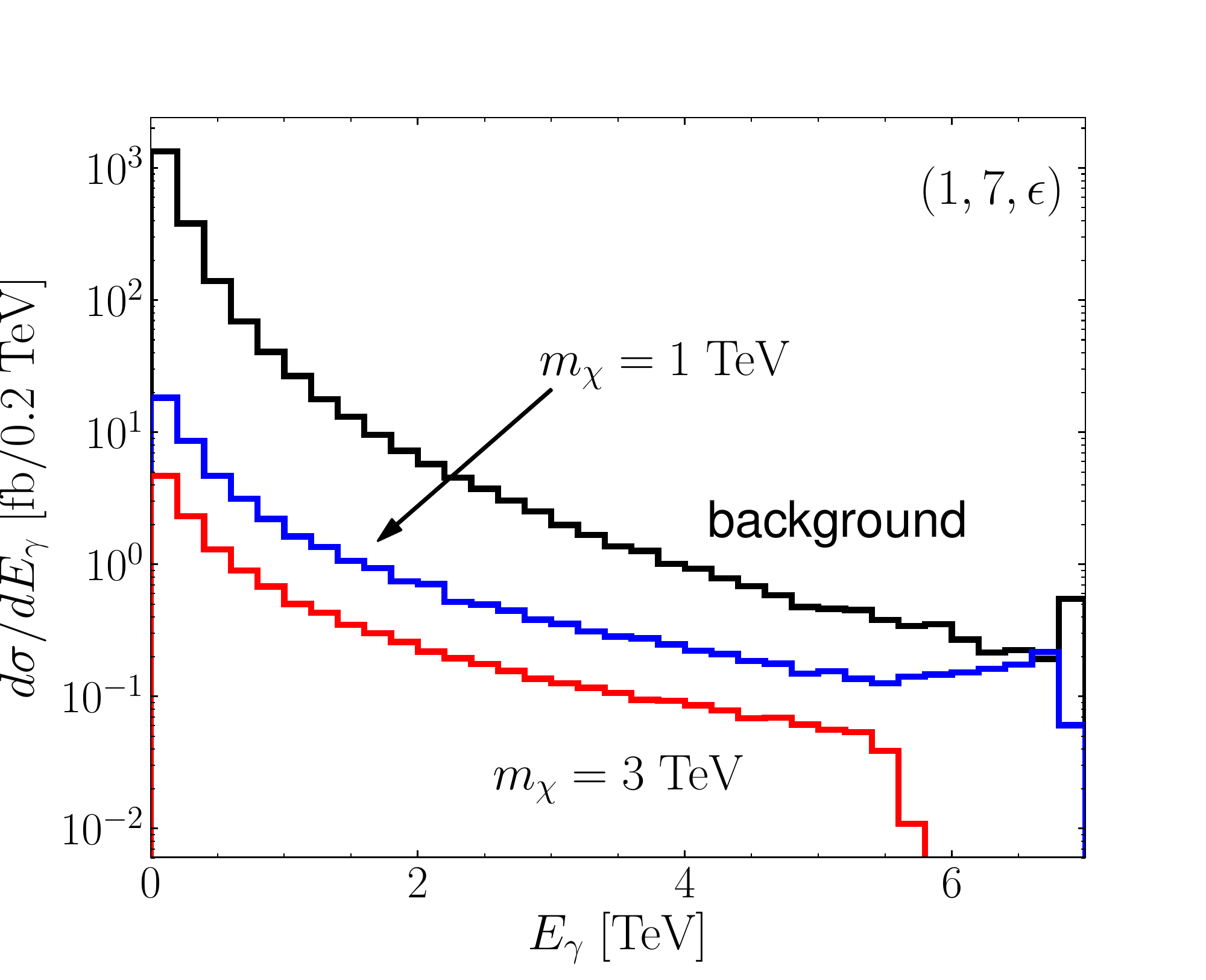}
\caption{}
\end{subfigure}
\begin{subfigure}{0.48\textwidth}
\includegraphics[width = \textwidth]{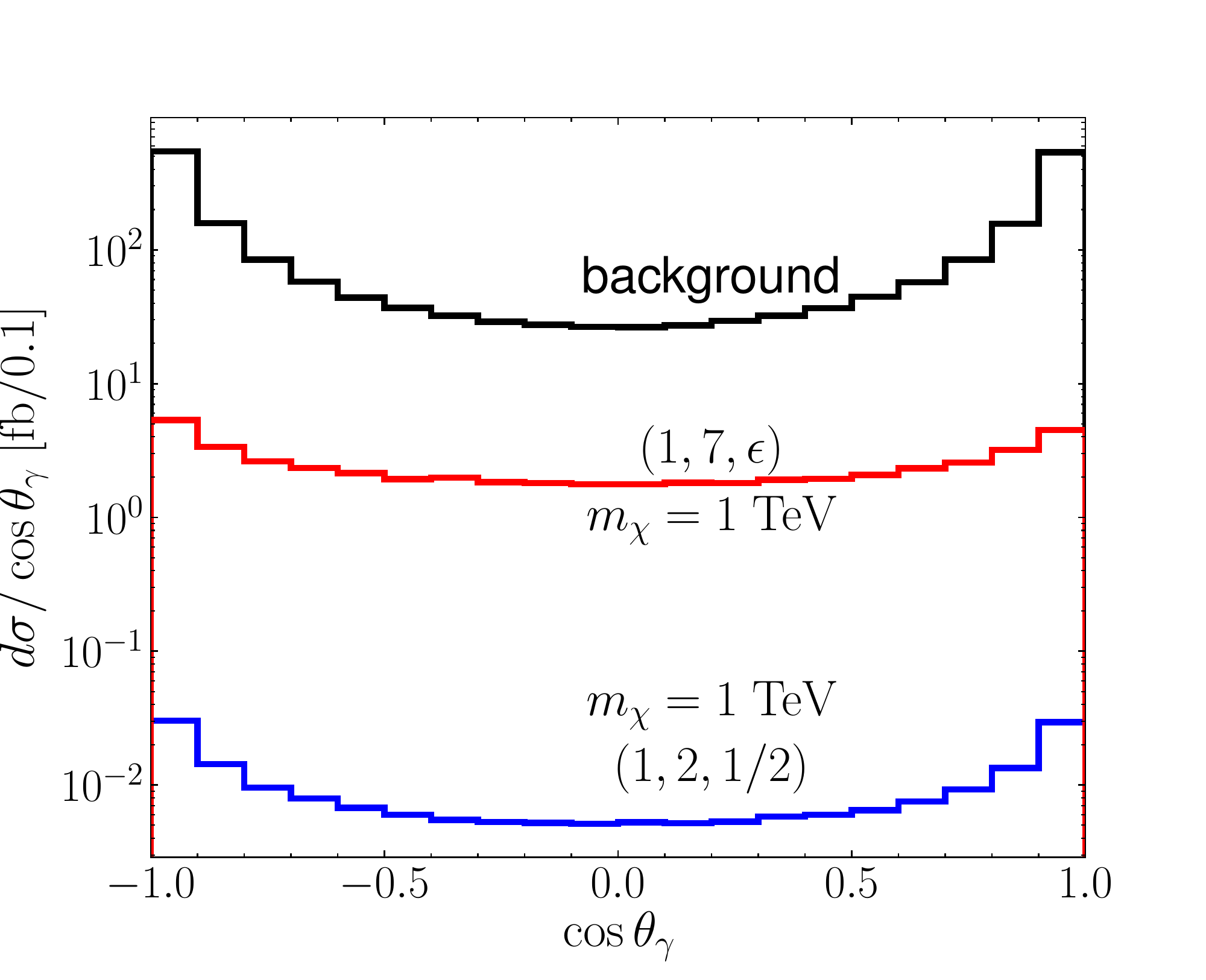}
\caption{}
\end{subfigure}
\begin{subfigure}{0.48\textwidth}
\includegraphics[width = \textwidth]{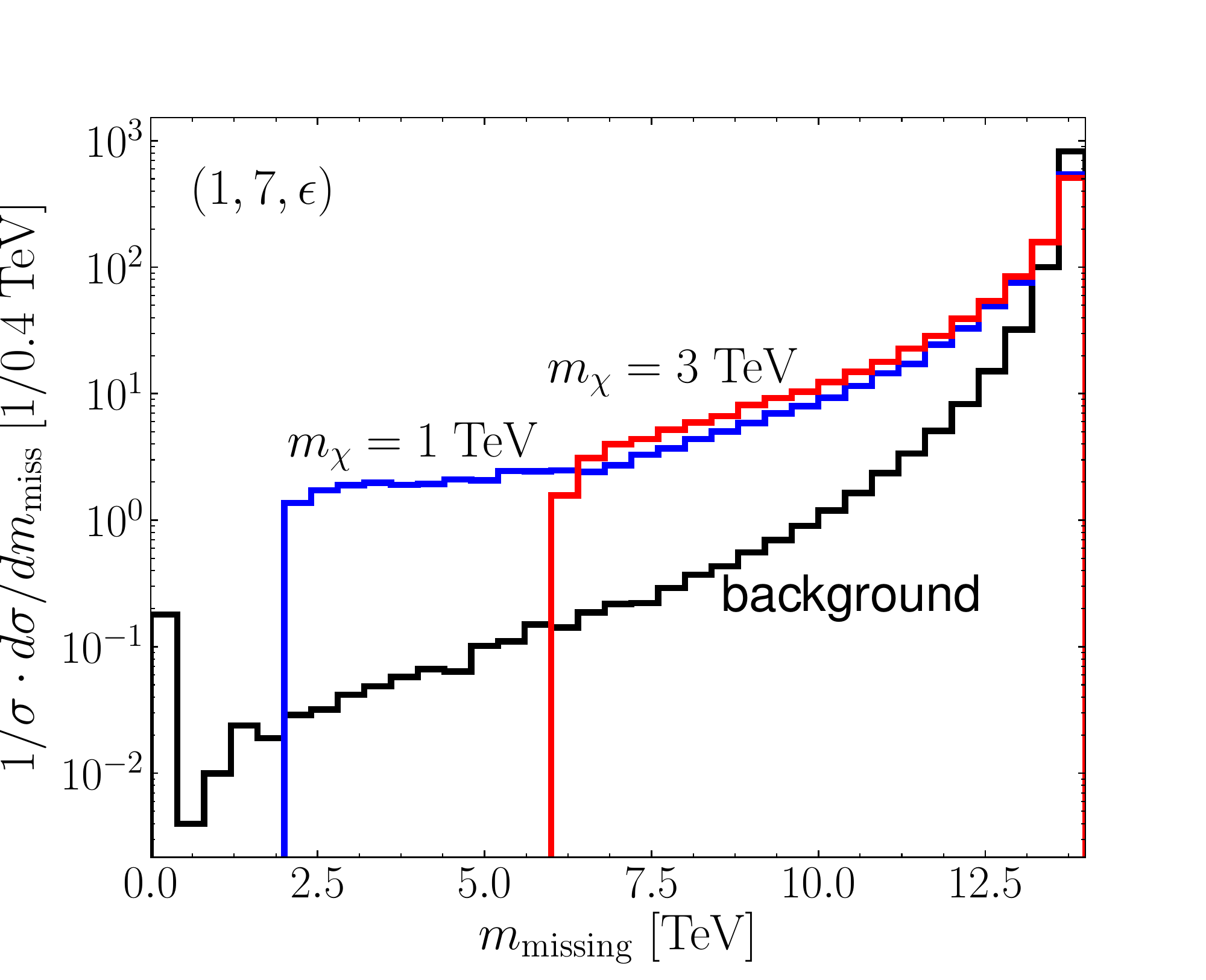}
\caption{}
\end{subfigure}
\caption{Kinematic distributions for the mono-photon process at $\sqrt{s}=14$~TeV with cuts in \autoref{eq:angle}, for (a) the energy distributions of the photon for the background and two representative benchmarks for 7-plet $(1, 7,\epsilon)$ with $m_\chi = 1$ TeV (blue) and 3 TeV (red), respectively; (b) the angular distributions of the photon for the background and two representative benchmarks for doublet $(1, 2,1/2)$ (blue) and 7-plet $(1, 7,\epsilon)$ (red) with $m_\chi = 1$ TeV; (c) normalized missing-mass distributions for the signals and backgrounds.}
\label{fig:monogm_dist}
\end{figure}

In \autoref{fig:monogm_dist}(a), we show the energy distributions of the photon at $\sqrt{s}=14$~TeV, for the background and two representative benchmarks of the 7-plet $(1, 7,\epsilon)$ with $m_\chi = 1$ TeV and 3 TeV, respectively. We see a mono-chromatic peak for the background process near $E_\gamma\approx {\sqrt s}/2$, that is due to the two-body kinematics from the contribution of $\mu^+\mu^- \to \gamma Z$. The sharp energy endpoint in the signal distribution is determined by the masses of heavy missing particles as discussed at \autoref{eq:ecut}. \autoref{fig:monogm_dist}(b) shows the angular distribution of the photon at $\sqrt{s}=14$~TeV. For the background, the photon is mostly along the forward or backward direction, due to the nature of the ISR. For the signal, the photon can be emitted from both ISR and FSR. Although suppressed by the multiplet mass $m_\chi$, the FSR could be enhanced by the large electric charge in a higher dimensional multiplet, and the photon becomes more central, as shown by the red line in the middle in \autoref{fig:monogm_dist}(b).
Importantly, we show the normalized missing-mass distributions for signal and background in \autoref{fig:monogm_dist}(c). We see the threshold near twice of the EW multiplet mass, and this sharp rise could serve as the characteristic signature for the signal parameter identification.

\begin{figure}[tb]
\centering
\begin{subfigure}{0.48\textwidth}
\includegraphics[width = \textwidth]{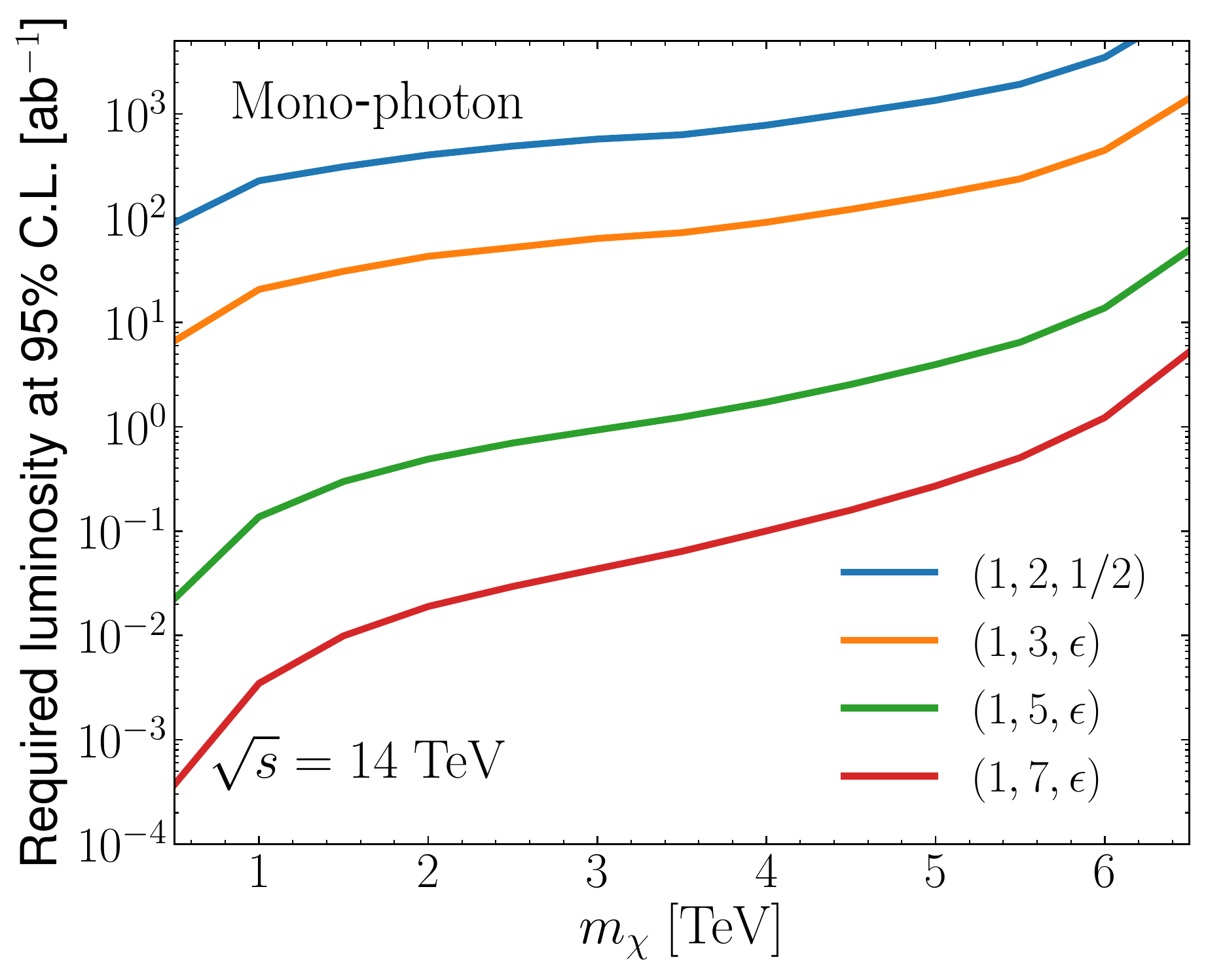}
\caption{}
\end{subfigure}
\begin{subfigure}{0.48\textwidth}
\includegraphics[width = \textwidth]{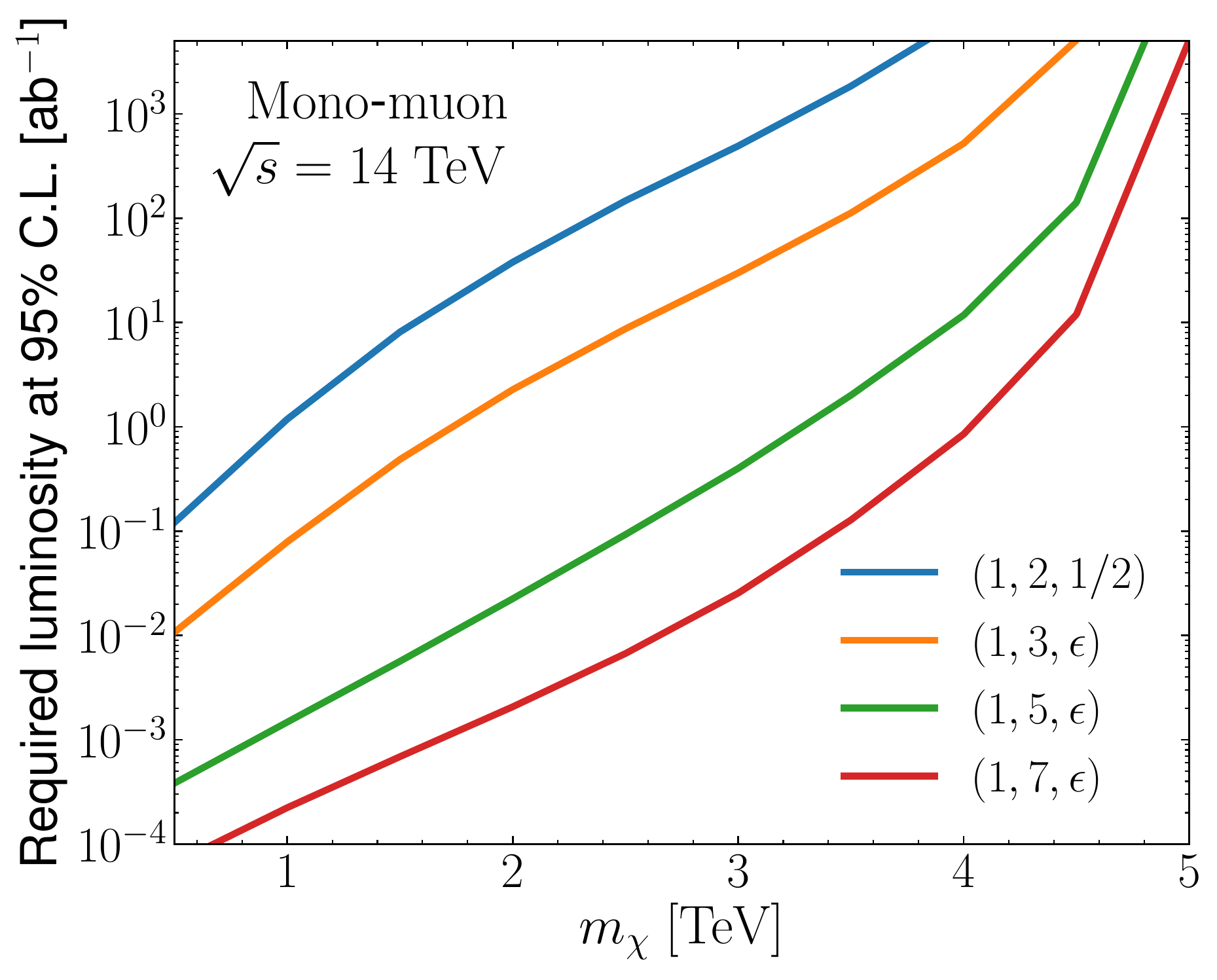}
\caption{}
\end{subfigure}
\caption{Integrated luminosities needed for (a) mono-photon and (b) mono-muon channels, to reach $2\sigma$ statistical significance at $\sqrt s=14$ TeV.
}
\label{fig:monogmL}
\end{figure}

For the expected reach, we take a conservative approach to  estimate the significance as 
\bea
{\rm N_{\rm SD}} = {S\over \sqrt{S+B + (\epsilon_S S)^2 + (\epsilon_B B)^2}},
\label{eq:NSD}
\eea
where $S$ and $B$ are the numbers of events for the signal and background, and $\epsilon_S$ and $\epsilon_B$ are the corresponding coefficients for systematic uncertainties, respectively. It is clear from this equation that, in a statistical uncertainty-dominated scenario ($\epsilon_S=\epsilon_B=0$), the significance scales as $S/\sqrt{S+B}$, and in a systemic uncertainty-dominated scenario, the significance scales as $S/(\epsilon_B B)$. In processes where the $S/\sqrt{B}$ is high, but $S/B$ is tiny, one needs to pay special attention to the uncertainty arising from the systematics.
Our results for the reach of mono-photon channel are shown in \autoref{fig:monogm}(b), with and without the systematic uncertainties $\epsilon_{S}=\epsilon_{B} = 0.1\%$. In \autoref{fig:monogmL}(a), we also show the integrated luminosities needed to reach $2\sigma$ statistical significance for the mono-photon channel at $\sqrt s=14$ TeV, in absence of systematic uncertainties. 
We see, for instance, that we could reach the $2\sigma$ sensitivity with this channel alone for the 5-plet EW DM to its thermal target mass of 6.6 TeV with about 50 ab$^{-1}$. 
The coverage for the higher representation of the 7-plet would be better.

We note that the signal-to-background ratio is low in this channel, $S/B < 10^{-2}$, which demands a very good control of the systematic error. The theoretical uncertainties are anticipated to be small with higher order calculations of the electroweak process. With a large event sample, typically  
about $10^7$ background events, it is hopeful that the systematics can be modeled by a sideband with similar rate to control the error to be less than $10^{-3}$. 

\subsection{Mono-Muon}
\label{sec:monomuon}
\begin{figure}[tb]
\centering
\begin{subfigure}[t]{0.25\textwidth}\centering
\includegraphics[width=\textwidth]{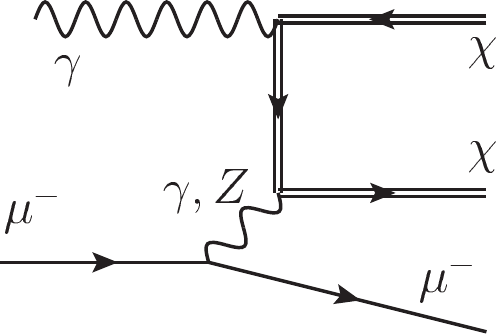}
\caption{}
\end{subfigure} \ \
\begin{subfigure}[t]{0.20\textwidth}\centering
\includegraphics[width=\textwidth]{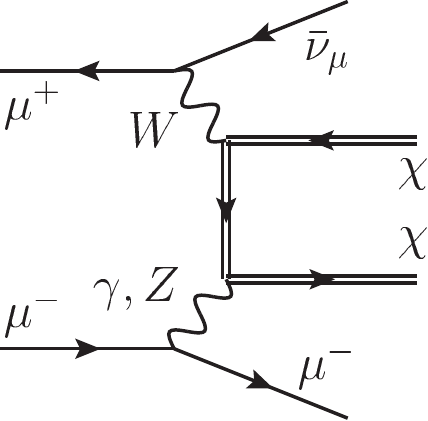}
\caption{}
\end{subfigure}
\caption{Representative Feynman diagrams for the mono-muon signal (a) from 
$\gamma Z$ fusion and (b) from $WZ/W\gamma$ fusion. 
}
\label{fig:FeynmonoMu}
\end{figure}

While the mono-photon is a generic dark matter signal for all high energy colliders, mono-muon signal to be studied in this section is unique to muon colliders. 
The leading signal processes are 
\begin{equation}
\aligned
\gamma \;\mu^\pm & \rightarrow \mu^\pm \chi \chi\quad {\rm via}\ \ \gamma Z \to \chi \chi ,\\ 
\mu^+ \mu^- &\rightarrow \mu^\pm \nu \chi \chi \quad {\rm via}\ \ \gamma W, ZW \to \chi \chi,
\endaligned
\end{equation}
where $\chi$'s represent any states within the $n$-plet, and $\chi\chi$ represents any combination of a pair of the $\chi$ states allowed by gauge symmetries. The $\mu^\pm$ is required to be in the detector coverage as in \autoref{eq:angle}. Some representative Feynman diagrams of such a signal, from $\gamma Z$ fusion and $WZ/W\gamma$ fusion,  are shown in \autoref{fig:FeynmonoMu}. 

\begin{figure}[tb]
\centering
\begin{subfigure}[t]{0.22\textwidth}\centering
\includegraphics[width=\textwidth]{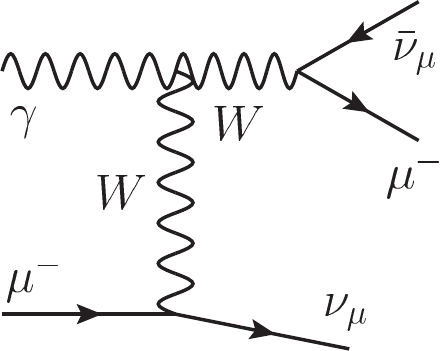}
\caption{}
\end{subfigure} \ \
\begin{subfigure}[t]{0.22\textwidth}\centering
\includegraphics[width=\textwidth]{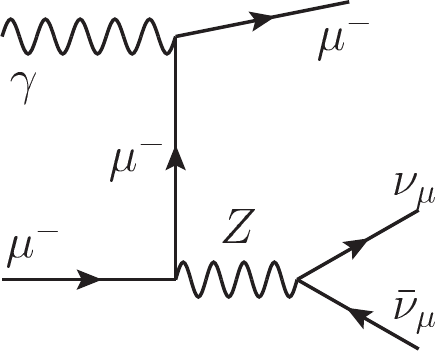}
\caption{}
\end{subfigure}
\caption{Representative Feynman diagrams for the SM mono-muon backgrounds (a) from $W\to \mu\bar\nu$ and (b) from $Z\to \nu \bar\nu$.
}
\label{fig:FeynmonoMuB}
\end{figure}

The main background comes from processes in which a charged particle (mostly muon) escapes detection in the forward direction, due to the finite angular acceptance of the detector. The dominant process is 
\begin{equation}
\gamma \;\mu^\pm \rightarrow \mu^\pm \nu \bar{\nu},
\label{eq:gmutomununu}
\end{equation}
resulting from both $Z\to \nu \bar\nu$ and $W\to \mu\bar\nu$, where the muon from which the photon radiates missed the detection, as shown in \autoref{fig:FeynmonoMuB}.

\begin{figure}[tb]
\centering
\begin{subfigure}{0.48\textwidth}
\includegraphics[width = \textwidth]{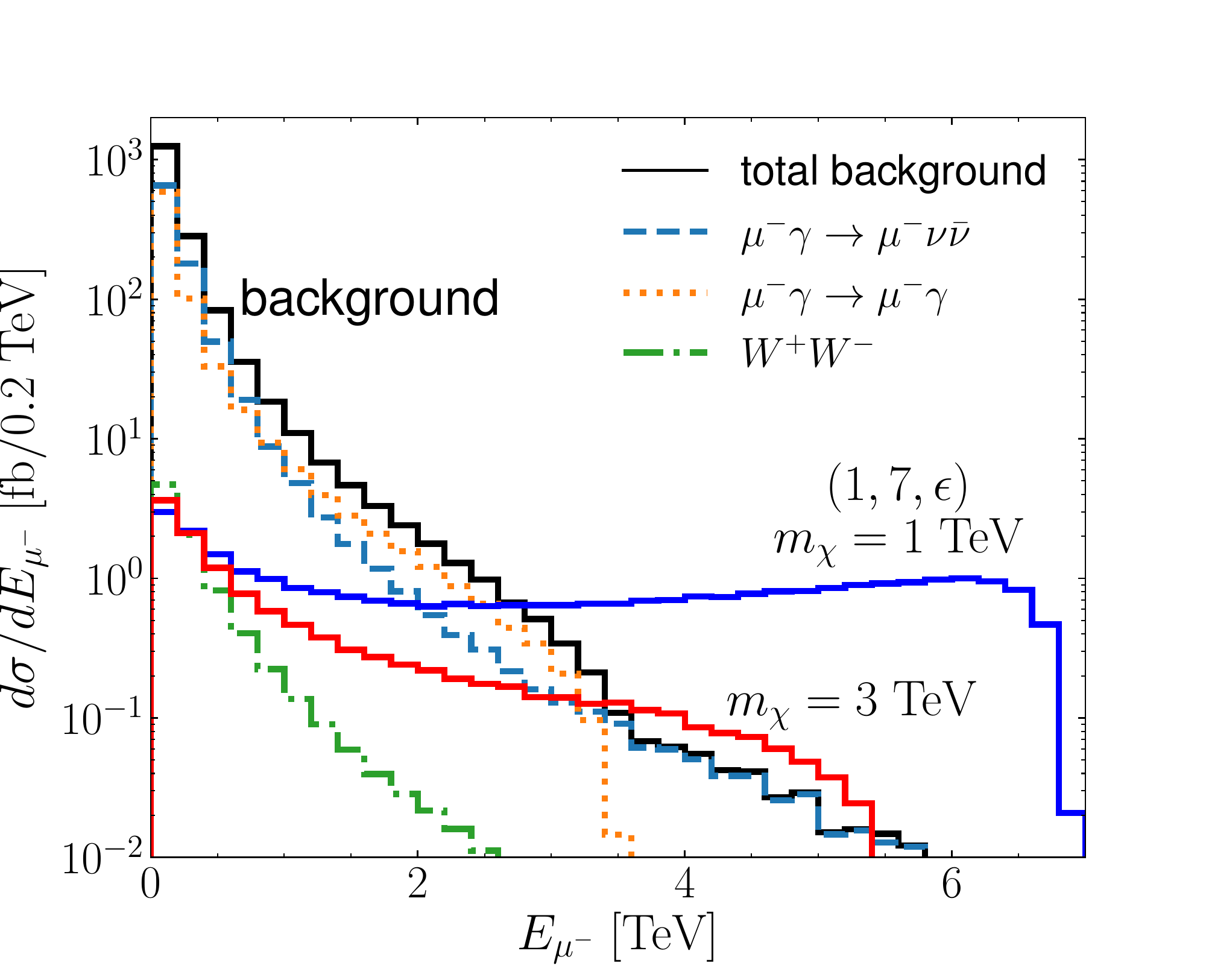}
\caption{}
\end{subfigure}
\begin{subfigure}{0.48\textwidth}
\includegraphics[width = \textwidth]{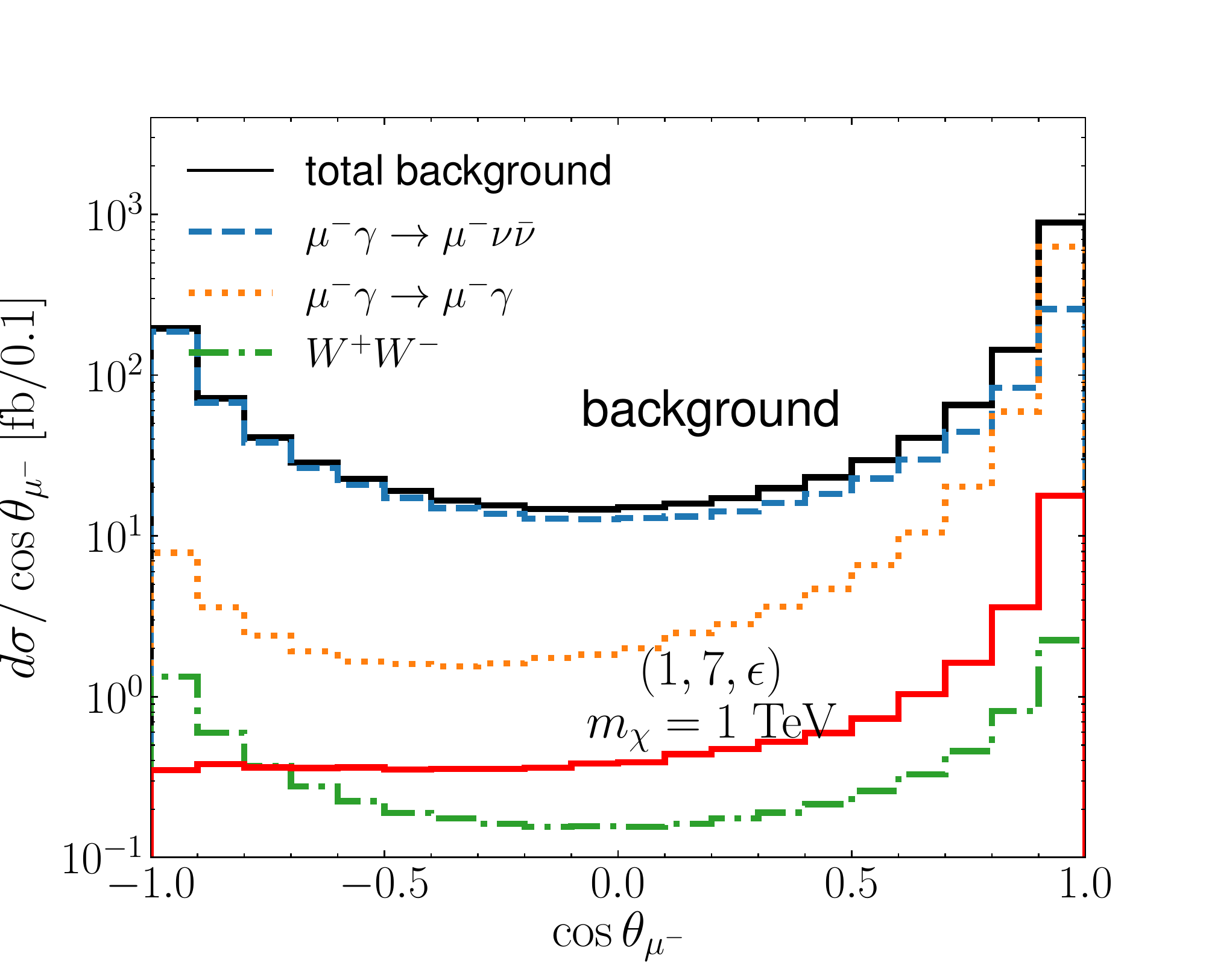}
\caption{}
\end{subfigure}
\begin{subfigure}{0.48\textwidth}
\includegraphics[width = \textwidth]{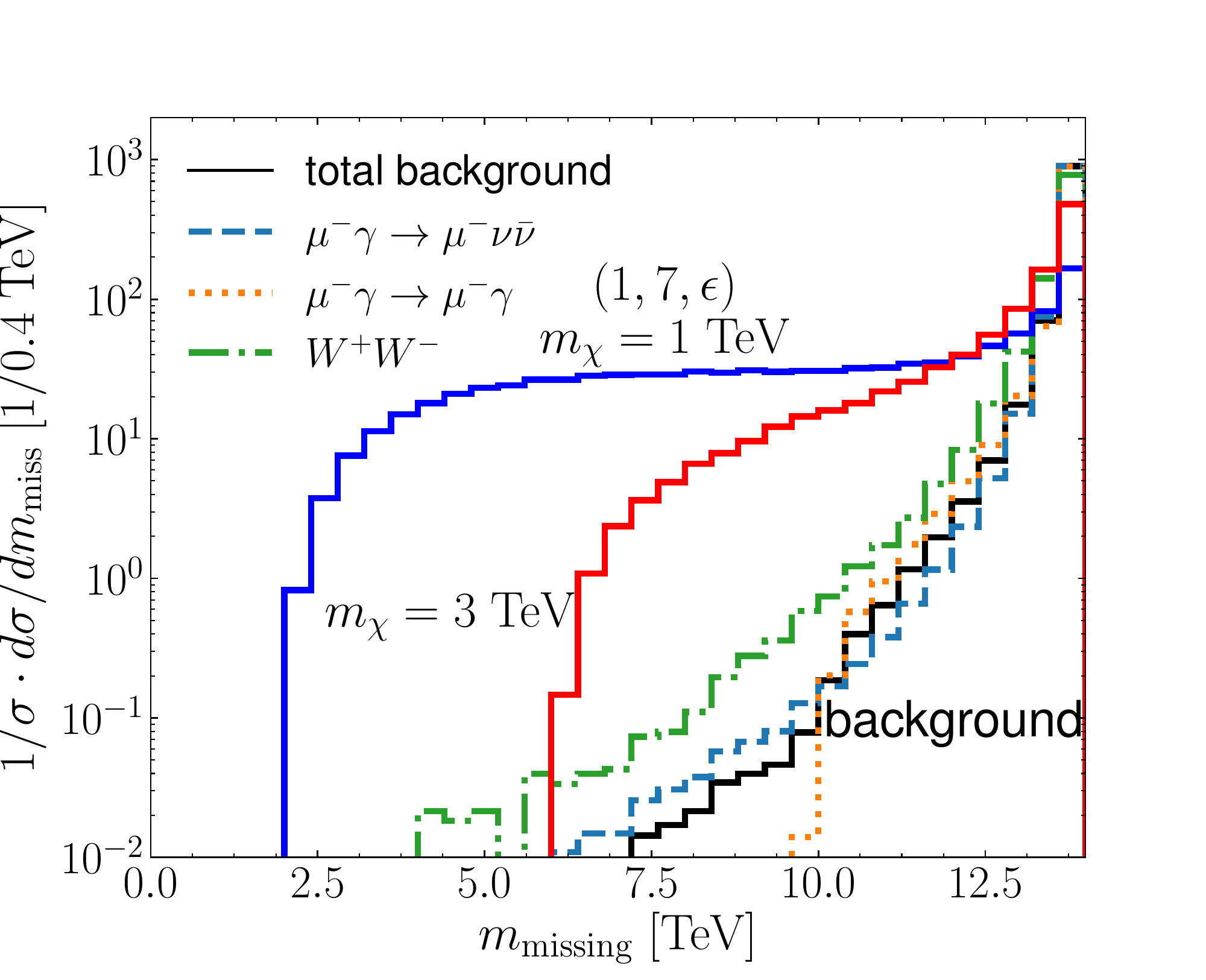}\\
\caption{}
\end{subfigure}
\caption{(a) The energy distributions of the $\mu^-$ at $\sqrt{s}=14$~TeV, for the backgrounds and two representative benchmarks for 7-plet $(1, 7,\epsilon)$ with $m_\chi = 1$ TeV (blue) and 3 TeV (red), respectively; (b) the angular distributions of the $\mu^-$ at $\sqrt{s}=14$~TeV, for the backgrounds and 7-plet $(1, 7,\epsilon)$ (red) with $m_\chi = 1$ TeV; (c) normalized missing-mass distributions  for the signals and backgrounds.}
\label{fig:monomu_dist}
\end{figure}
Many other processes also have the property that some final state particles prefer to go forward, and they can potentially contribute to the background. This leads us to consider high-rate processes with muons and missing energy in the final state, such as 
\beq
\gamma\ \mu^\pm \rightarrow \gamma\ \mu^\pm 
\label{eq:gmutogmu}
\eeq 
where the photon is missed. There are also various di-boson production processes with subsequent leptonic decays to contribute to the backgrounds. The $W^+W^-$ background is clearly orders of magnitude smaller than other processes discussed above.
The process shown in \autoref{fig:FeynmonoMuB} for the dominant background $\gamma \;\mu^- \rightarrow \mu^- \nu \bar{\nu}$ yields very different kinematic behavior.
The initial state photon  is radiated off an incoming muon and tends to be soft. 
The process in the left panel is dominated by the soft $W$-exchange, and hence the final state $W$ decay into muons is more symmetric. The process in the right panel is also dominated by the soft $\mu$ exchange, and hence the final state muon is soft as well. With a hard muon energy cut, the backgrounds in both \autoref{eq:gmutomununu} and \autoref{eq:gmutogmu} can be effectively suppressed, as shown in 
panel (a) of \autoref{fig:monomu_dist}. Hence, we require 
\begin{equation}
E_{\mu^\pm} > 0.71,\; 1.4,\; 2.3,\; 3.2,\; 6.9,\; 22.6~{\rm TeV},\quad {\rm for }~\sqrt{s}=3,\; 6,\; 10,\; 14,\; 30,\; 100~{\rm TeV},
\end{equation}
With respect to the  dominant background $\gamma \;\mu^- \rightarrow \mu^- \nu \bar{\nu}$ and the sub-dominant background $\gamma\ \mu^\pm \rightarrow \gamma\ \mu^\pm$, the signal significance can also be enhanced somewhat by requiring the $\mu^-$ to be in the forward direction (with respect to the initial $\mu^-$), as shown in \autoref{fig:monomu_dist}(b). 
Therefore, the following selection cuts are applied:
\begin{equation}
10^\circ < \theta_{\mu^-} < 90^\circ,\quad 90^\circ < \theta_{\mu^+} < 170^\circ,
\end{equation}
where the polar angle is defined with respect to the incoming $\mu^-$.

The missing mass is also very useful for the mono-muon channel as shown in panel (c) of \autoref{fig:monomu_dist}, where we see the threshold effect near twice of the EW multiplet mass. This sharp rise could serve as the characteristic signature for the signal identification.  We will impose the missing mass cut
\begin{equation}
m^2_{\rm missing} = (p^{\rm in}_{\mu^+} + p^{\rm in}_{\mu^-} - p^{\rm out}_{\mu^\pm})^2 > 4m_\chi^2.
\end{equation}

\begin{figure}[tb]
\centering
\begin{subfigure}{0.48\textwidth}
\includegraphics[width = \textwidth]{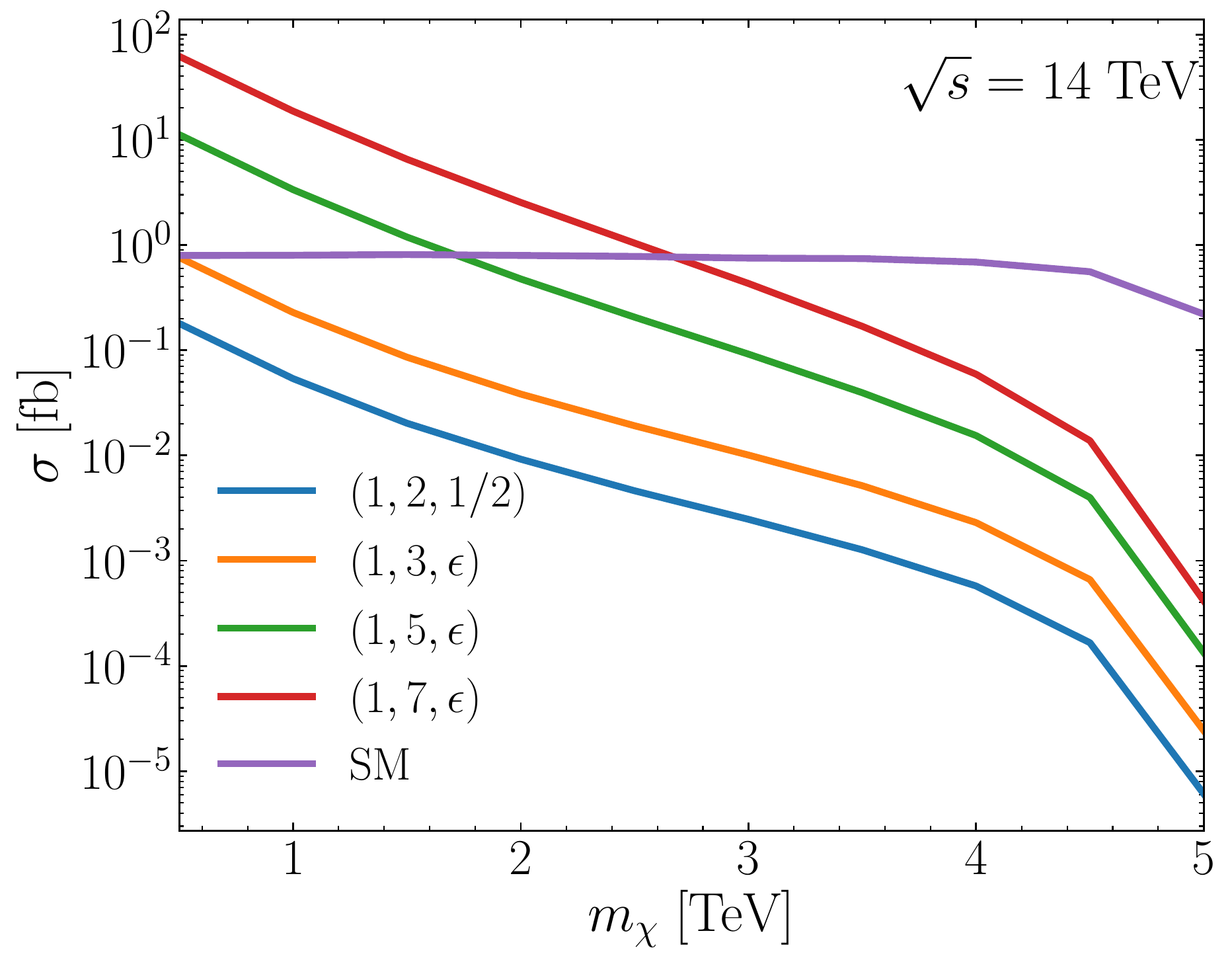}
\caption{}
\end{subfigure}
\begin{subfigure}{0.48\textwidth}
\includegraphics[width = \textwidth]{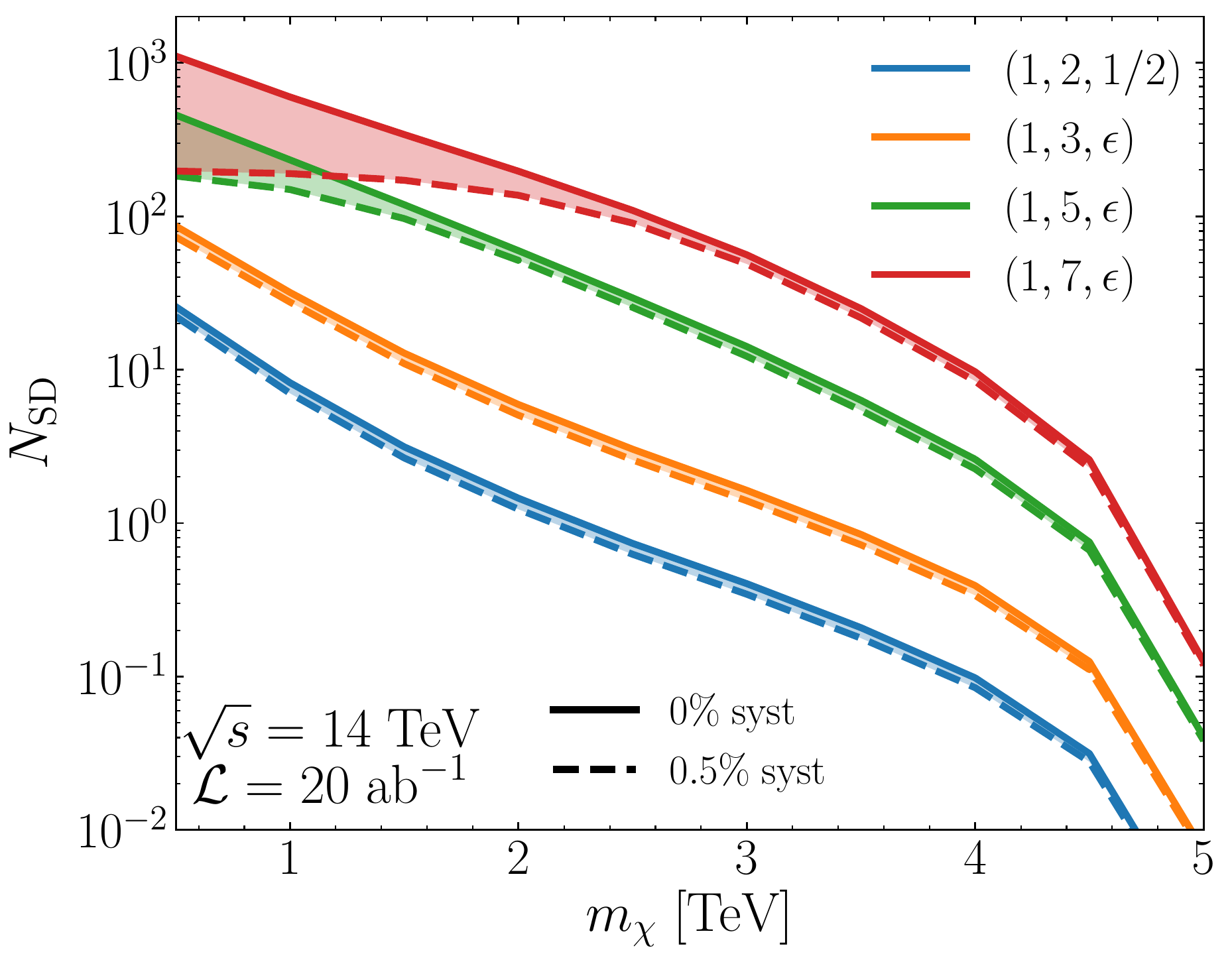}
\caption{}
\end{subfigure}
\caption{(a) Total cross section and (b) the significance defined in \autoref{eq:NSD} for a pair of EW multiplets plus a mono-muon at a muon collider with $\sqrt s=14$ TeV. In (b), the solid and dashed lines correspond to the systematic uncertainties of 0\% and 0.5\%, respectively.}
\label{fig:monomu}
\end{figure}

In \autoref{fig:monomu}(a), we show the cross sections for the signal processes with a variety of the EW multiplets after the selection cuts above. In comparison with the Drell-Yan production in the case of mono-photon signal, the cross section of mono-muon decrease significantly for larger dark matter masses as $\sim{1 / m_\chi^4}$, for the same reason as discussed in the previous section about the VBF component of the mono-photon signal. 
In comparison with the mono-photon, a notable feature of the mono-muon channel is that the background is much lower, resulting a much larger $S/B$, making it more robust against systematic uncertainties. 

In \autoref{fig:monomu}(b), we show the significance of the mono-muon processes. 
For relatively light dark matter mass, the reach in the mono-muon channel is better than the mono-photon. This is particularly interesting for the lower-dimensional multiplets, such as the Higgsino, where the target thermal mass is relatively low. \autoref{fig:monogmL}(b) shows the integrated luminosities needed to reach $2\sigma$ 
statistical significance for the mono-muon channel at $\sqrt s=14$ TeV, in the absence of systematic uncertainties.
We see that with a luminosity of about 2 ab$^{-1}$, we could reach the $2\sigma$ sensitivity with this channel alone for the doublet (Dirac triplet) EW DM to its thermal target mass of 1.1 (2.0) TeV. 
The coverage for the higher representations would be better.

\subsection{VBF Di-Muon}
\label{sec:VBF}
\begin{figure}[tb]
\centering
\begin{subfigure}[t]{0.22\textwidth}\centering
\includegraphics[width=\textwidth]{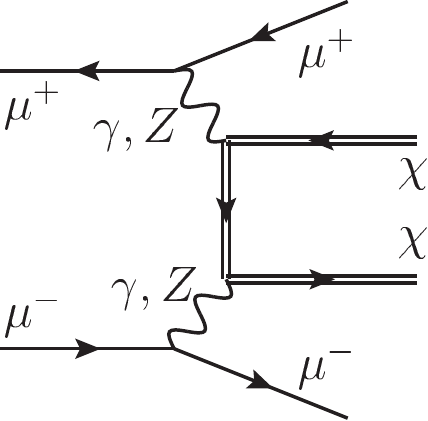}
\caption{}
\end{subfigure} \ \
\begin{subfigure}[t]{0.22\textwidth}\centering
\includegraphics[width=\textwidth]{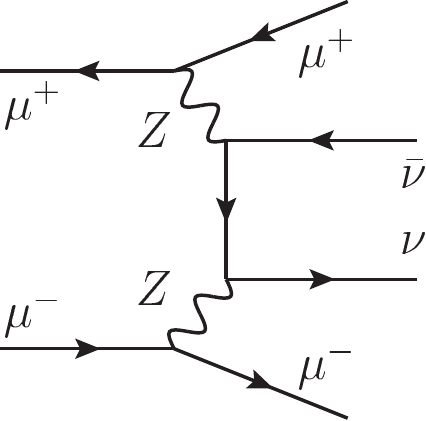}
\caption{}
\end{subfigure}
\caption{Representative Feynman diagrams for (a) the di-muon signal and (b) the SM background.}
\label{fig:FeyndiMu}
\end{figure}

Beyond the single muon signature, one could also consider to tag both muons in the final state to account for other additional contributing channels via the VBF
\begin{equation}
\mu^+ \mu^- \rightarrow \mu^+ \mu^- \chi \chi \quad {\rm via\ fusion}\ \ \gamma^*\gamma^*, \gamma^* Z, ZZ \to \chi \chi 
\end{equation}
where $\chi$ represents any state within the $n$-plet, as depicted in \autoref{fig:FeyndiMu}(a).
We require both final state muons to be in the detector coverage as in \autoref{eq:angle}. This effectively suppresses the backgrounds that are dominated by low momentum transfer.  For a $\gamma^*$ initiated process, the cross section with finite angle scattering falls at higher energies of the final state muons as $1/(p_T^\mu)^2$ for each tagged muon.
Although $p_T^\mu\sim M_Z$ for a $Z$-initiated process, the muons can still be highly boosted due to the large beam energy, with a scattering angle of the order 
$\theta_\mu \sim M_Z/E_{\mu f}$~\cite{Han:2020pif}, likely outside the detector coverage. The energy of the muons in the low energy-transfer processes is almost the beam energy. For a final state muon to be observable, our requirement in \autoref{eq:angle} can be translated into $\sin\theta>0.17$ and the corresponding $p_T$ is then $O(1)$~TeV for a 14~TeV muon collider, providing the huge suppression of background.\footnote{One can go beyond our conservative assumption of effective experimental detector coverage of $|\eta|<2.5$ (correspond to $10^\circ<\theta< 170^\circ$), and try to tag muons in the more forward regions if advanced detector designs allows for it. Covering the forward region will improve the signal efficiency and help further separating different types of background for the searches discussed in this section.} 
Similarly, this angular requirement also cause some signal efficiency loss. For heavier EW multiplets, due to the large momentum transfer, and lower final state muon energy, the efficiency loss due to the angular cut is much less severe.\footnote{This can also be seen in \autoref{fig:monomu_dist} for the signal energy and angular distribution, although it is strictly for the process where only one final state muon is detectable.}

The leading irreducible background is
\begin{equation}
\mu^+ \mu^- \rightarrow \mu^+ \mu^- \nu \bar{\nu}.
\end{equation}
The dominant contributions are from both $\gamma^*\gamma^*, \gamma^* Z, ZZ$ fusion processes as well as $ZZ\to \mu^+ \mu^-\ \nu\bar \nu$, as shown in \autoref{fig:FeyndiMu}(b).
 To suppress the large non-fusion background primarily from a $Z$ decay to leptons,  the muons are required to have
\begin{equation}
m_{\mu^+\mu^-} > 300~{\rm GeV} ,\quad 
m_{\rm missing} = (p^{\rm in}_{\mu^+} + p^{\rm in}_{\mu^-} - p^{\rm out}_{\mu^+} - p^{\rm out}_{\mu^-})^2 > 4m_\chi^2 .
\end{equation}

\begin{figure}[tb]
\centering
\begin{subfigure}{0.48\textwidth}
\includegraphics[width = \textwidth]{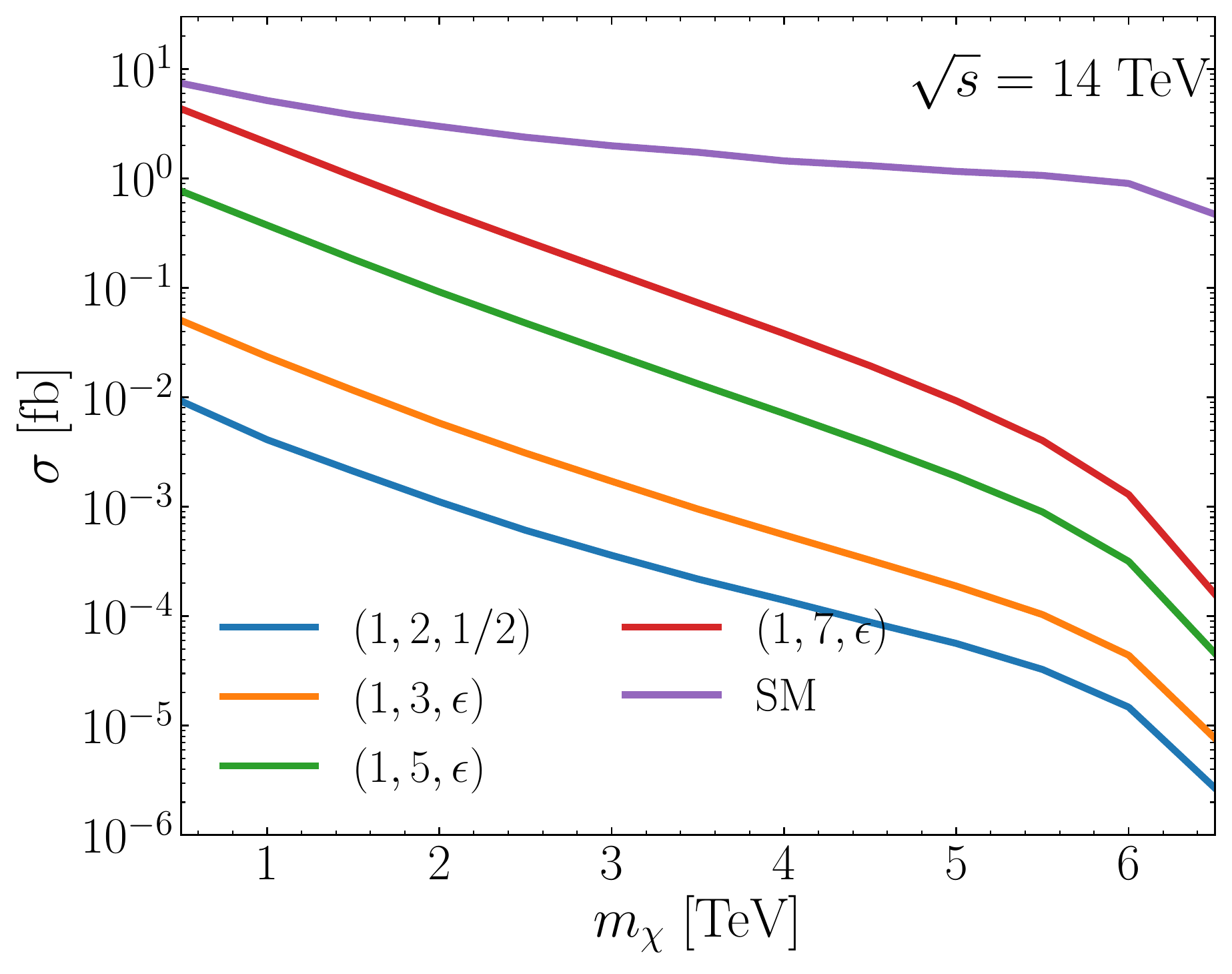}
\caption{}
\end{subfigure}
\begin{subfigure}{0.48\textwidth}
\includegraphics[width = \textwidth]{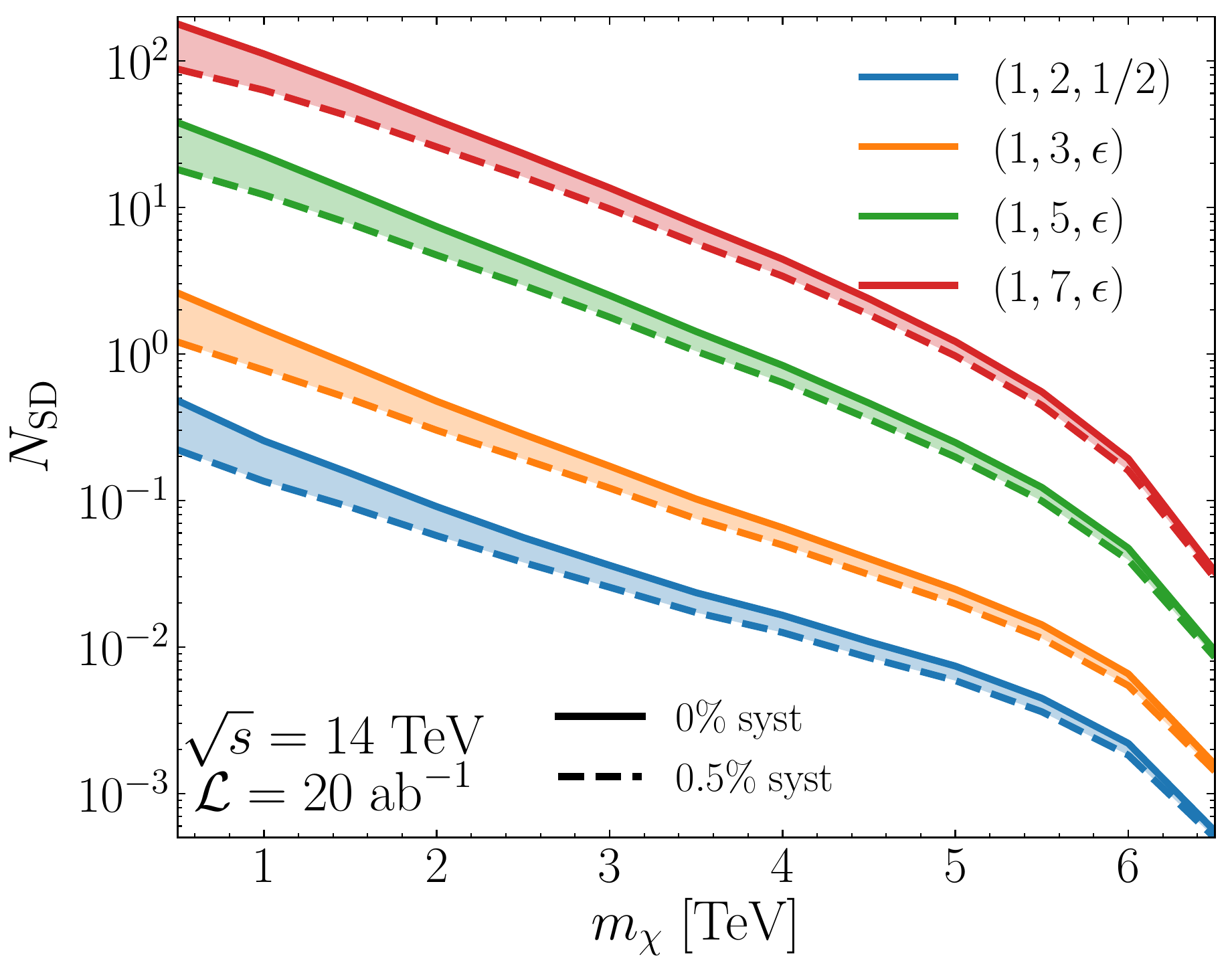}
\caption{}
\end{subfigure}
\caption{(a) Total cross section and (b) the significance defined in \autoref{eq:NSD}  for a pair of EW multiplets plus di-muon at a muon collider with $\sqrt s=14$ TeV. In (b) the solid and dashed lines correspond to the systematic uncertainties of 0\% and 0.5\%, respectively. }
\label{fig:monomumu}
\end{figure}
In \autoref{fig:monomumu}(a), we show the cross sections for the signal processes with a variety of the EW multiplets. We see that the cross sections span over a large range and at a fixed muon collider energy, fall as high power of the EW multiplet mass. For comparison, the background cross section is also shown as the curve on top. In \autoref{fig:monomumu}(b), we show the statistical significance of the di-muon signals.

\subsection{Disappearing tracks and other signatures}
\label{sec:DT}
In our phenomenological analyses thus far, we have considered the most pessimistic scenario that all the members of an EW multiplet cannot be detected after being produced, yielding a large missing mass. With a pair of missing particles in the final state and an unknown mass, the signal events cannot be fully kinematically reconstructed. In this sense, the ``minimal dark matter'' under consideration serves as a ``nightmare'' scenario for weakly interacting dark matter. In this section, we take a step beyond the inclusive signals. 

First, we briefly summarize the  mass splittings and transition rates between different states, validating the assumption that the EW multiplets cannot be reconstructed as SM particle objects.  Furthermore, the charged $\pm1$ states have a macroscopic lifetime in collider detectors, resulting in a ``disappearing track'' upon its decay which is an additional unusual signature to enhance the reach. Afterwards, we proceed with some basic considerations of properties of the ``disappearing track'' signatures at a high energy muon collider, and provide an estimate of the sensitivity reach from the mono-photon plus ``disappearing track''. In the last part of this section, we comment on the backgrounds and possible new signatures to explore in the future. 
Ultimately, the collider design and the detector performance will dictate the reach for this kind of searches. Being at this very early stage on planning for a high energy muon collider, we will focus on the performance target and highlight the challenges to achieve those goals. We show the enhancement of the reach if such goals are achieved with the main purpose for motivating the design effort. 

For the EW multiplet minimal dark matter, EW-loop effects induce a universal mass splitting among the component states~\cite{Thomas:1998wy,Buckley:2009kv,Cirelli:2005uq,Cirelli:2009uv,Ibe:2012sx}.  At the 1-loop order and in the limit of $m_Q \gg m_W, m_Z$, we have 
\begin{equation}
  \Delta m_{Q,Q^\prime}\equiv m_Q-m_{Q^\prime}\simeq(Q-Q^\prime) \left(Q+Q^{\prime}+ \frac {2Y} {\cos{\theta_W}}\right) \delta m,
  \label{eq:splitting}
\end{equation}
where $Q$ and $Q^\prime$ are the electric charges of the given components in an EW multiplet, $Y$ is the hyper-charge of the EW multiplet, and 
\beq
\delta m =\frac {g^2} {4\pi} m_W \sin^2\frac {\theta_W}{2} \approx 160 \textrm{--}170~{\rm MeV}. 
\eeq
Using the electroweak input parameters from Ref.~\cite{Zyla:2020zbs} defined at the $Z$-pole, we take $\delta m = 165$~MeV in this study. In principle, there is quite a bit of freedom in choosing the hyper-charge $Y$. We would like to have a neutral state $Q=0$ in the multiplet as the dark matter candidate, which can be achieved by requiring the hyper-charge to take on (half-)integer values for (even) odd-dimensional multiplets, and $Y \leq T$ for EW-multiplet of $(2T+1,Y)$.\footnote{Without loss of generality, we assume $Y\geq0$.} At the same time, at least in the minimal scenarios in which the mass splitting is dominated by the EW loop contributions, there are additional constraints from requiring the neutral state to be the lightest one. As an example, we can consider the EW multiplets with a zero hyper-charge,  $(1,2T+1,0)$, where $T$ is an integer.  This choice also satisfies the direct detection limits. 
For even-dimensional  EW multiplets with a non-zero hyper-charge $Y$, $(1, 2n, Y)$, the highest charged state has an electric charge of $n/2+Y$. The electrically neutral state is automatically the lightest eigenstate when $Y= n-1/2$.\footnote{If a charged state is the lightest state, it cannot be the dark matter. In this case, a new class of signature of heavy stable charge particles can be the most powerful probe where the lepton collider can reach the EW $n$-plet for discovery up to the kinematic limit. Furthermore, it is possible for the neutral state to pick up a milli-charge through loop effects and still being a viable dark matter  candidate~\cite{DelNobile:2015bqo}. We will not pursue these possibilities further in this paper.} 

\begin{figure}[t]
\centering
\includegraphics[width = 0.48\textwidth]{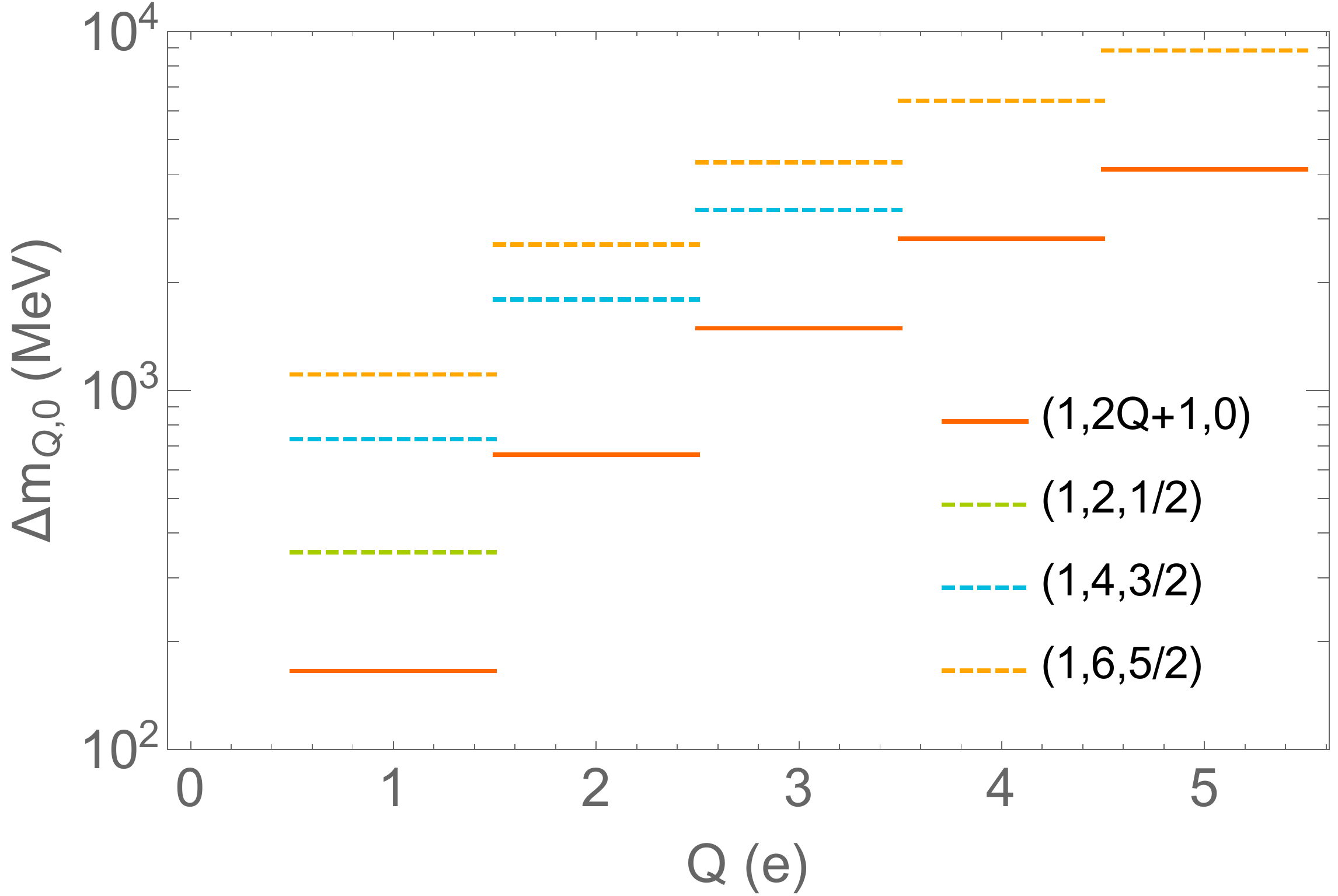}
\includegraphics[width = 0.485\textwidth]{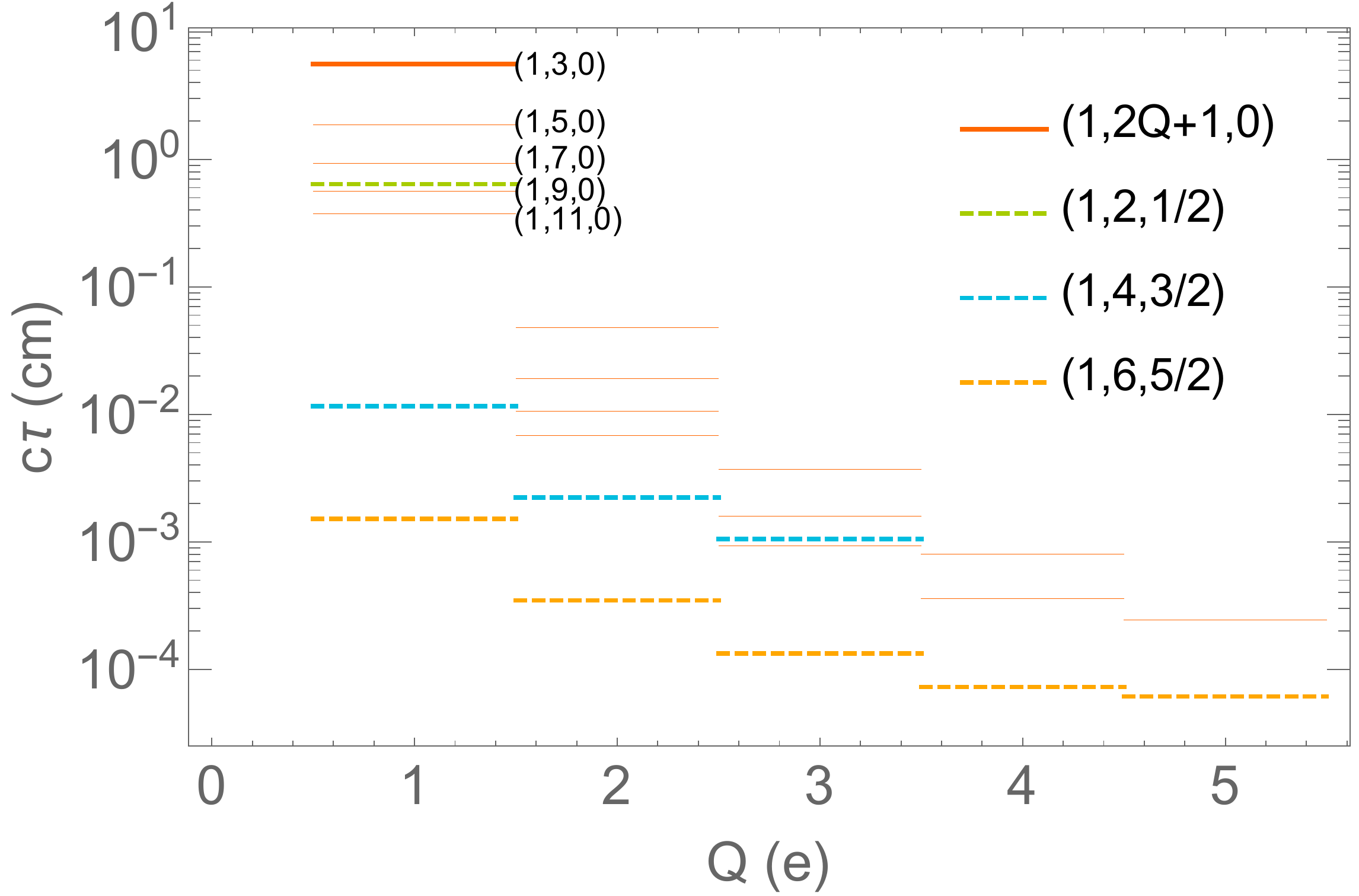}
\caption{Mass splittings (left panel) and proper decay lengths (right panel) of the multiplet components as a function of their electric charge. The odd(even) representations of the EW multiplets are shown in solid (dashed) lines. 
}
\label{fig:lifetime}
\end{figure}

In the left panel of \autoref{fig:lifetime}, we show the mass splittings between the charged and neutral states $\Delta m_{Q,0}$ calculated with \autoref{eq:splitting} for the generic EW multiplets discussed above. 
The particles with the same charge $Q$ in all odd-dimensional representations share the same mass splitting, so long as the state is present (a charge $Q$ requires at least $2Q+1$ representation for $Y=0$).
Importantly, we can see that the smallest splitting in an $n$-plet is between the charge $|Q|=1$ and the neutral states, around $165$~MeV for $(1,2T+1,0)$. For $Y\neq 0$ states, the smallest splitting is for the Higgsino-like case of $(1, 1/2, 1/2)$, around $354$~MeV. The splitting between the charged states are a factor of few larger. 

Such a small EW-loop induced splitting between states implies the transitions between these states to be slow, via two-body and three-body processes to leptons and mesons through an off-shell $W$-boson. Hence, the leading decays are between $Q$ and $Q-1$ states. For the two-body decay process, for $Q>1$, we have\footnote{For $Q=1$ case, the two-body and three-body decay formulae are consistent with Refs.~\cite{Hisano:2006nn,Chen:1996ap,Chen:1999yf}}
\bea
c\tau(\chi^{Q}\rightarrow \chi^{Q-1} \pi^+)&\simeq& c\tau(\pi^\pm) \frac {\kappa_W m_\pi m_\mu^2} {16\Delta m_{Q, Q-1}^3} \left(1-\frac{m_\mu^2}{m_\pi^2}\right)^2 \left(1-\frac{m^2_\pi} {\Delta m^2_{Q,Q-1}}\right)^{-1/2}
\label{eq:ctau}
\\
&=& 5.7\kappa_W \left(\frac{(165~\mev)^2-(134~\mev)^2} {\Delta m^2_{Q,Q-1}-m_\pi^2}\right)^{1/2}\left(\frac {165~\rm MeV} {\Delta m_{Q,Q-1}}\right)^2 {\rm cm},\nonumber
\eea
where $k_W$ is the normalized coupling involved in the process. For a state of EW-multiplet $(1,2T+1,Y)$, it is 
\beq
\kappa_W = \frac {2} {(T-Q+Y+1)(T+Q-Y)},
\label{eq:kw}
\eeq
for $-T+Y< Q \leq T+Y$.
In the second line of \autoref{eq:ctau}, we normalize it in terms of the Wino-like splitting between the charged state and the neutral one, yielding a lifetime around 5.7~cm (in length units). For Higgsino-splitting of 354~MeV, the partial decay width is around 0.68~cm. For the transition between higher charged states, a new channel of $\chi^{Q}\rightarrow \chi^{Q-1} K^+$ also opens up, and the rate estimation can be done with a replacement of the $\pi^{\pm}$ mass by the $K^{\pm}$ mass between the last parentheses in \autoref{eq:ctau}. 

For the three-body decay process with $Q\geq 1$, we have 
\beq
c\tau(\chi^{Q}\rightarrow \chi^{Q-1} e^+ \nu_e) \simeq \frac {15\kappa_W\pi^3} {2 G_F^2 \Delta m_{Q, Q-1}^5}.
\eeq
This process provides an additional transition mode that is a factor of 18 (48) slower than the two-body decay mode discussed above for the Higgsino-like (Wino-like) splittings.  Again, a new channel of $\chi^{Q}\rightarrow \chi^{Q-1} \mu^+ \nu_\mu$ also opens up when $\Delta m_{Q, Q-1} >m_\mu$. Including both the 2-body and 3-body decays, the doublet (pure Higgsino) lifetime is 0.64~cm, which we will use for the rest of this study.

In the right panel of \autoref{fig:lifetime}, we show the proper decay lengths for the states within different EW multiplets including the two-body and three-body channels. 
Due to the factor $\kappa_W$ in \autoref{eq:kw}, the lifetime of a charged particle in higher odd-dimensional representations is shorter. 
The mass splitting and anticipated lifetimes allow us to develop the following very {\it simple} strategy for a phenomenological estimation for the signal rate.
First, the charge $\pm1$ states will have macroscopic lifetime from the collider perspectives, generating the signature of ``disappearing tracks'' typically associated with long-lived particles. Second, although the doubly charged state in the $Y=0$ multiplets has a lifetime as large as $0.5$~mm, it would be difficult to reach the tracker due to the typical low boost of $\gamma=E_\chi/m_\chi$ for a heavy $\chi$ at a muon collider.\footnote{We discuss the potential double displacement signature in the last part of this section.}As a result,  the decay of  states with a charge $\pm 2$ or more into the lower charged states can be treated as  {\it prompt}, and only the charge $\pm 1$ states have a relevant long lifetime. Hence, all the EW pair productions considered in the previous sections, including the production of the states with charge $\geq 2$,  gives rise to  long-lived charged $\pm 1$ particles in the final state. 

\begin{figure}[t]
\centering
\includegraphics[width = 0.495\textwidth]{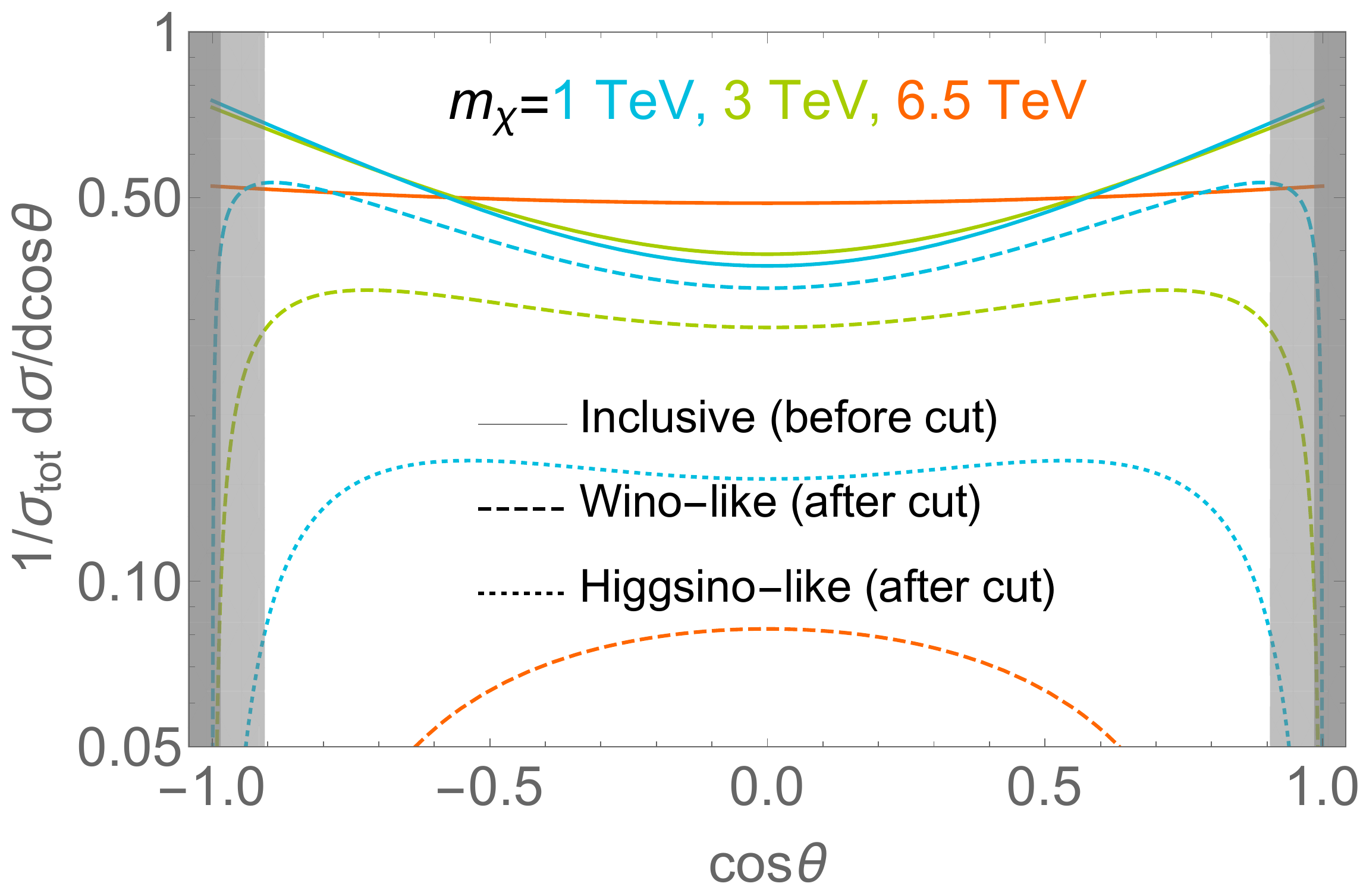}
\includegraphics[width = 0.48\textwidth]{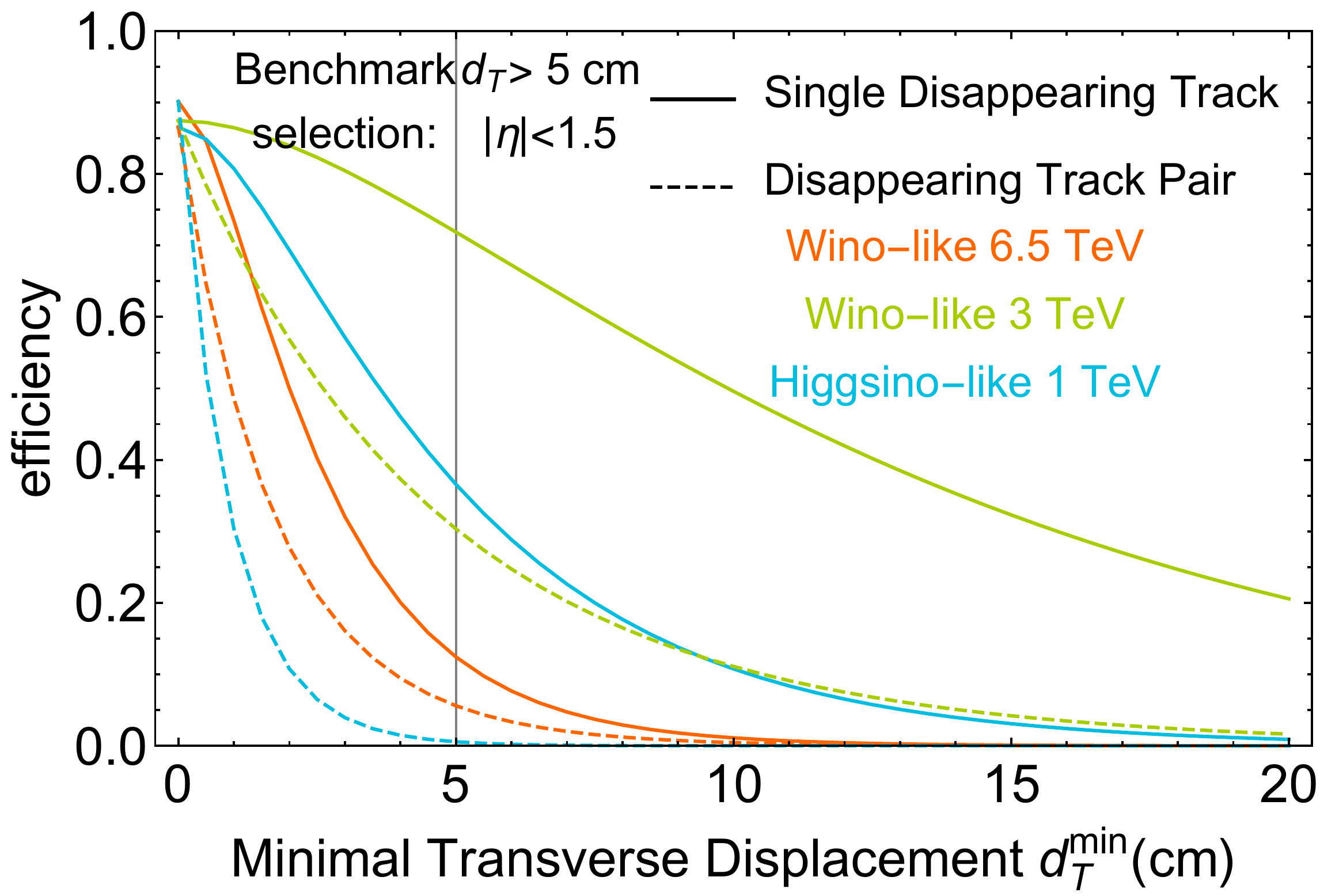}
\caption{
(a) Angular distributions (left panel) for $\mm \to \gamma^* \to \chi^+\chi^-$ before cuts (solid lines) and after cuts in \autoref{eq:dtcut} for the Wino-like scenario (dashed lines) and Higgsino-like scenario (dotted lines), normalized with respect to the inclusive cross section before cuts. The vertical gray region represents the hard-to-detect region due to shielding and detector geometry. 
(b) Disappearing track reconstruction efficiency (right panel) as a function of minimal transverse displacement cut $d_T^{\rm min}$ for single disappearing track reconstruction (solid)  and double disappearing track reconstruction (dashed).
For illustrative purpose, we take $m_\chi=1,3,6.5$ TeV at a 14 TeV muon collider. 
}
\label{fig:eff_diff}
\end{figure}

We proceed to understand the kinematics of these disappearing track signals. 
 For those heavy states, the DY production mechanism dominates, as shown in \autoref{fig:monogm}. In the left panel of \autoref{fig:eff_diff}, we show the differential distribution of the signal (including $s$-channel off-shell photon exchange only for simplicity) as a function of scattering angle $\cos\theta$ (solid curves) at a 14 TeV muon collider. For a DM mass of $m_{\rm dm} =1$~TeV (cyan) and $m_{\rm dm} =3$~TeV (lime), the distribution has the typical vector-like behavior of $(1+\cos^2\theta)$, with a small correction from the finite mass effect. 
In contrast, the large chirality flipping effect for heavy dark matter makes the angular distribution flattened, as shown in the red curve for a 6.5 TeV EW multiplet mass. In the dark and light shaded area, we show the region where it is hard to reconstruct the signal due to the detector acceptance ($\theta<10^\circ$ or $\theta>170^\circ$) and long distance in $z$ direction for the first layer ($|\eta|>1.5$, a typical end-cap region), these regions cut away $O(10\%)$ of the signal rate.

The disappearing track signature can be reconstructed in collider experiments via a series of inner tracker hits (usually pixel detector hits, or ``stubs'' for some double-layer structures) that are not followed by hits in the outer layers with a consistent curvature. Hence, for disappearing tracks, it is critical to have a few inner tracker layers close to the interaction point to suppress backgrounds while maintaining a high signal reconstruction efficiency.
The reconstruction probability of a signal event with one disappearing track is
\beq
\epsilon_\chi (\cos\theta,\gamma, d_T^{\rm min})=\exp\left(\frac {-d_T^{\rm min}} {\beta_T \gamma c\tau}\right),
\eeq
where $\gamma=E_\chi/m_{\chi}$\footnote{For the dominant signal from the DY process, the charged particle energy in the lab frame is $E_\chi \approx \sqrt{s}/2$.} 
and 
$\beta_T=\sqrt{1-1/\gamma^2} \sin\theta$, which is the transverse velocity in the lab frame.
Clearly, the reconstruction efficiency favors central signal events. 
In the left panel of \autoref{fig:eff_diff}, we show the differential signal efficiency for the cases with Wino-like (dashed) and Higgsino-like (dotted) splittings, with a minimal transverse displacement $d_T$ requirement on the signal of 5~cm.
Especially for heavy states due to a low boost, only central events can pass the selection, as shown in the red dashed curve in \autoref{fig:eff_diff} for the case with Wino-like splitting. For the case with Higgsino-like splitting, due to its shorter lifetime, only the boosted signals with 1 TeV mass can efficiently pass the selection cut. In the right panel of \autoref{fig:eff_diff}, we show the integrated reconstruction efficiency, considering only the central regions $|\eta|<1.5$ as a function of the minimal transverse displace requirement $d_T^{\rm min}$, which determines the selection efficiency and will be used to estimate the reach. 
 While requiring the reconstruction of a pair of disappearing tracks (dashed lines) lower the signal reconstruction efficiency compared to only requiring single reconstruction (solid lines), it would help to suppress backgrounds further. We can see that for Wino-like splittings, one can still have $O(20\%)$ signal reconstruction efficiency with $d_T^{\rm min}$ as large as 20~cm. 

\begin{figure}[h!]
\centering
\includegraphics[width = 0.48\textwidth]{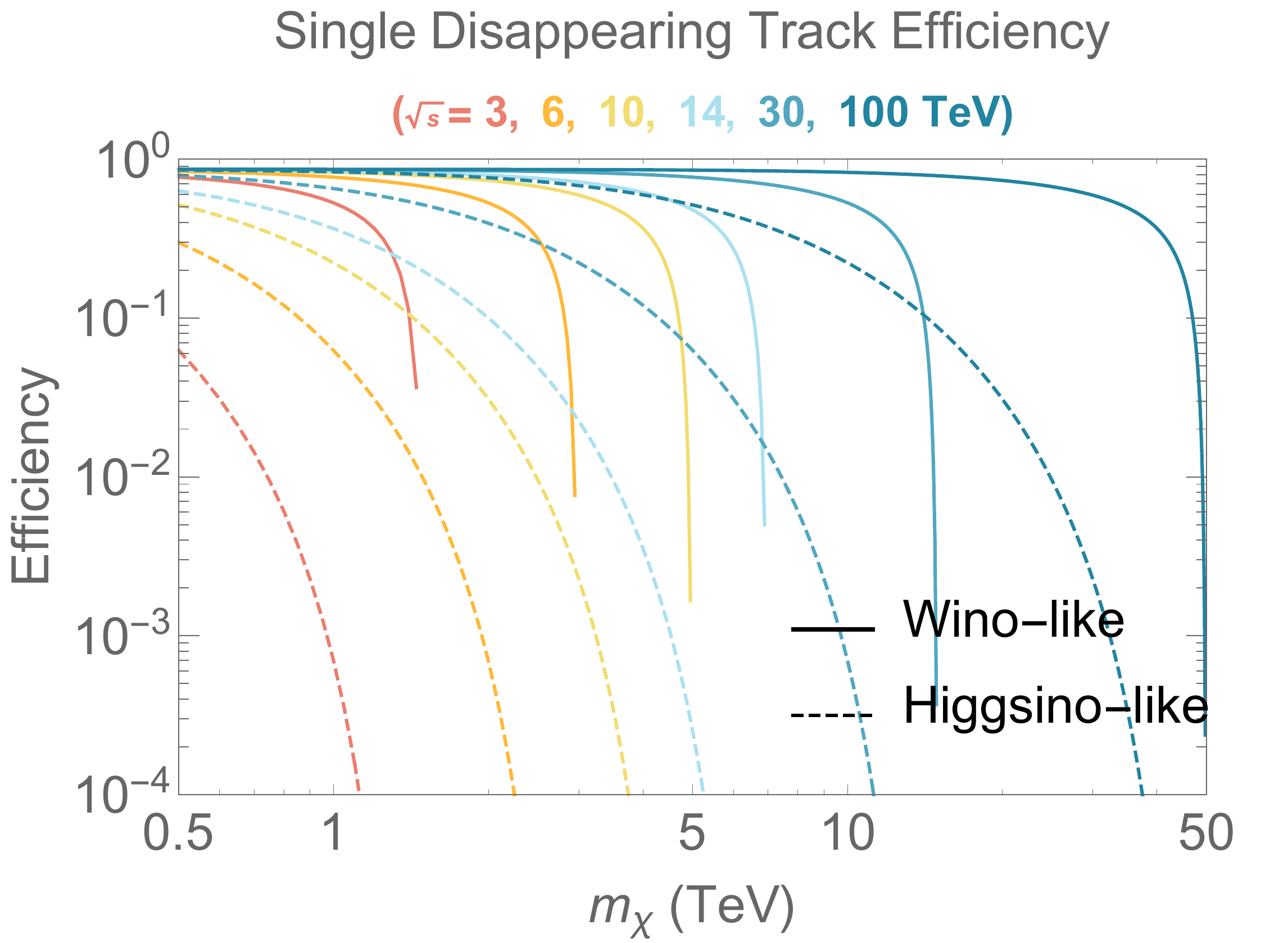}
\includegraphics[width = 0.48\textwidth]{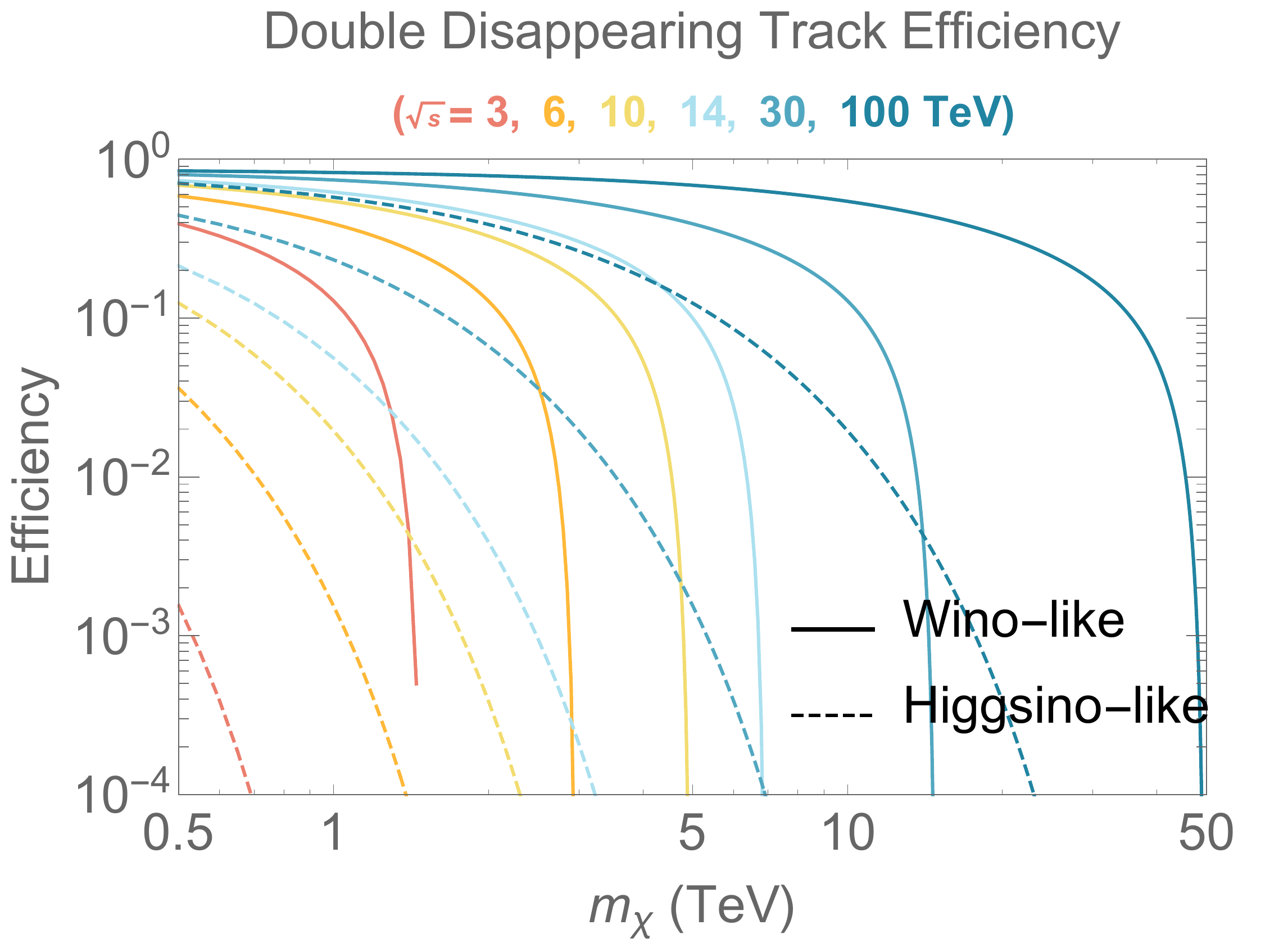}
\caption{
Reconstruction efficiencies for at least one disappearing track (left panel) and two disappearing tracks (right panel) with a  reconstruction cut $d_T^{\rm min}=5$ cm for various muon collider center of mass energies as a function of the minimal dark matter mass. The solid and dashed line represent the Wino-like and Higgsino-likescenarios, respectively.
}
\label{fig:eff}
\end{figure}

In \autoref{fig:eff} we show the overall signal reconstruction efficiencies as a function of the EW multiplet mass for the benchmark muon collider center of mass energies, labelled by the different color codes. In the left and right panels, we show the efficiency of reconstructing (at least) one disappearing track and two disappearing tracks, respectively. The Wino-like and Higgsino-like results are shown in solid and dashed curves, respectively. Requiring the reconstruction of a pair of the disappearing tracks lowers the efficiencies, especially for the Higgsino-like case and in the low-boost regime. 
At a higher energy muon collider, due to the large boost, we can obtain higher signal reconstruction efficiencies. The efficiency degrades quickly when EW multiplet mass approaches the kinematical threshold $m_{\chi}\sim\sqrt{s}/2$. In comparison with the wino-like scenario, the efficiencies are lower for the Higgsino-like signals.  
Higher center of mass energy helps to increase the efficiency for a fixed mass by orders of magnitude as we are catching more events in the exponential decay tails from the boost. 

In the following, we will discuss the experimental identification of the disappearing track signals. 
The disappearing track signals of the minimal dark matter are particularly challenging due to the short lifetimes, especially for Higgsino-like signals, and the moderate-to-low boost for heavy minimal dark matter particles when their masses are close to  the kinematic boundary $m_{\rm dm} \sim \sqrt{s}/2$. Hence, it is critical to push the limit in the detector design to enhance such signals. In particular, the number of tracker layers close to the interaction point would be crucial.
Current disappearing track searches at the LHC requires 3 to 4 hits~\cite{Aaboud:2017mpt,ATL-PHYS-PUB-2019-011} in order to effectively suppress the backgrounds. New proposals for high-luminosity LHC and FCC-hh are envisioning two-hit signals while  the background is still under control~\cite{Saito:2019rtg}. Hence, we would anticipate the needs for hitting the track twice or three times for a disappearing track signal to be identified. 
Some of the current studies of the detector performance for muon colliders ~\cite{DiBenedetto:2018cpy,Bartosik:2019dzq,Bartosik:2020xwr} have used a setup in which there  are 5 tracker layers from 3 cm to 12.9 cm. To set a performance target needed for the search of the minimal dark matter, and in the absence of a concrete design,  we will adopt 
\beq
d^{\rm min}_T=5\ {\rm cm}~{\rm with~} |\eta_\chi|<1.5
\label{eq:dtcut}
\eeq 
as the minimal transverse distance for a charged partner of the dark matter to travel and then to be identified as a disappearing track (with a minimal of 2--3 hits, depending on the detector design). The dependence of the signal efficiency on the $d_T$ is shown in the right panel of \autoref{fig:eff_diff}. 

A unique challenge for a muon collider in identifying the disappearing track signal is the high level of the beam-induced background.
The disappearing tracks would be identified with hits on the first few layers of the pixel detector. However, the soft BIB resulting from SM particles like electrons, charged hadrons, etc., could fake the signal if they also appear to be ``prompt''-originating from the beam interaction point (within the beam spread). Moreover, these soft particles may induce many hits (or ``stubs'') in the tracker, providing a non-negligible chance for these hits to be connected and reconstructed as tracks. To this end, preliminary studies~\cite{DiBenedetto:2018cpy,Bartosik:2019dzq,Bartosik:2020xwr} (based on a 1.5 TeV muon collider) have demonstrated that more than ${\mathcal O} (10^2)$ hits per cm$^2$ are expected at the first layer. The number of hits can fall to a lower level of 
for the subsequent layers. A muon collider running at higher energies, such as the ones we study in this paper, is expected to have lower level of BIB. More aggressive designs with layers closer to the beam spot will probably make the situation worse. At the same time, such a high occupancy level may not be prohibitive for track reconstruction. The upgrade of the ALICE detector~\cite{Rossegger:2012yd} at the LHC will be able to reconstruct tracks with a similar occupancy level. We also note that most of the reconstructed tracks from the BIB tend to be very soft, while the signal from a heavy charge particle would yield a higher momentum of $m_\chi \beta_T \geq $ O(100~GeV). Of course, reconstructing such a track with only a few hits with high efficiency while keeping a low fake rate is very different from reconstructing the tracks from the ordinary charged particles of leptons and hadrons. The actual performance will have to come from detailed studies from BIB with a concrete design of the machine and the detector.
Here, we use {\it 20 (50)} identified signal events for $2 \sigma~(5 \sigma)$ reach, which would be consistent with around 100 background events, as performance benchmarks for exclusion and discovery for the mono-photon plus disappearing track searches. We require these amount of signal events after imposing the mono-photon selection cuts (\autoref{eq:angle} and \autoref{eq:ecut}) and the disappearing track selection cuts (\autoref{eq:dtcut}). For reference, we also show in the appendix the reaches for disappearing track searches if backgrounds are negligible.

Beyond the disappearing track signals, one may seek other characteristic signatures to improve the coverage of electroweak multiplets. More delicate studies may include,
(a) doubly displaced tracks, e.g., from a charge-two state decaying to a charge-one state in the Wino-like scenario;
(b) appearing tracks from the soft and displaced pions as decay products of the charged EW-multiplet~\cite{Fukuda:2019kbp} or electrons (mind the BIB!);
(c) delayed appearing track for signals produced with low boost~\cite{Liu:2018wte} (mind the BIB!);
(d) soft-prompt or displaced particles for generic compressed spectra, instead of the
hyper-compression considered in our minimal dark matter scenario; and 
(e) optimization of the selection cuts with full detector simulation, such as the energy and angular distribution of the photon.

\subsection{Comparison of Channels}
\begin{figure}[h!]
\centering
\includegraphics[width = 0.8\textwidth]{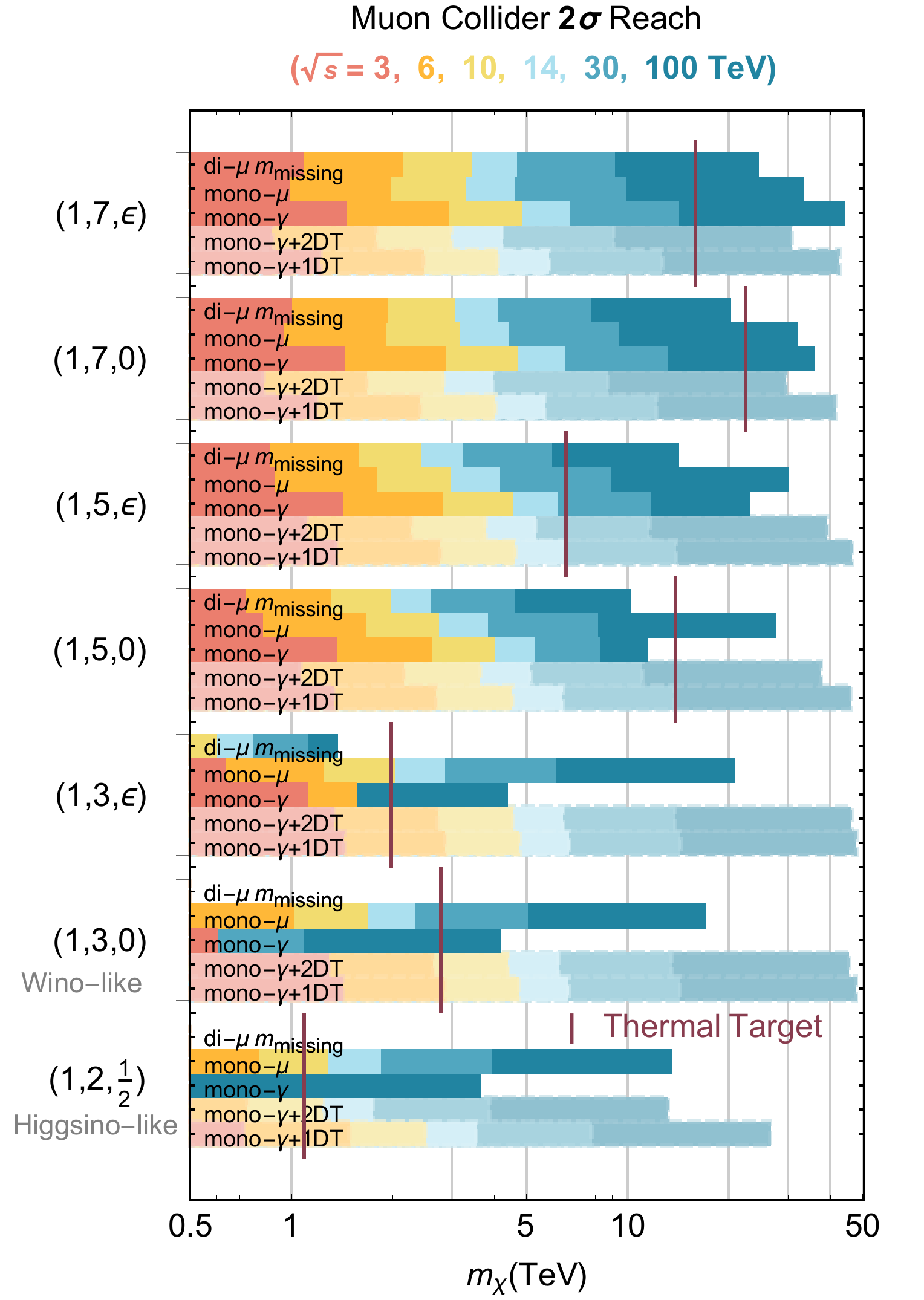}
\caption{Comparison of different channels discussed in this paper. The faint bars represent our estimation of the mono-photon plus one- or two-disappearing track searches. The burgundy vertical bars represent the thermal target for a given EW-multiplet model.
}
\label{fig:channels}
\end{figure}

In this section, we summarize the reach from various search channels and compare their relative strengths. We present the reach from different channels for different EW multiplets in
\autoref{fig:channels}, with various muon collider running scenarios listed in \autoref{eq:para} 
as indicated by the color codes. Our observations are as follows:
\begin{itemize}
\item The mono-muon channel, a unique signal for muon collider, shows a lot of potential as a high signal-background ratio search.
The signal production processes mainly come from the contributions $\gamma Z \to \chi\chi$ and $ZW, \gamma W \to \chi \chi$, more favorable for $m_\chi \ll \sqrt{s}/2$. This channel is especially promising for lower-dimensional EW multiplets, with reach stronger than that of the mono-photon channel for $n \leq 3$. For an EW doublet (Higgsino), the search in this channel can cover the thermal target at a 10 TeV muon collider. The rates of higher-dimensional representations are higher, due to the combinatorics of possible final states and larger couplings from the larger EW isospin. Hence, the reaches for these larger representations are better. However, mono-muon could not quite cover the (much higher) thermal target for these EW multiplets. As the signal decreases with the dark matter mass as $1/m_\chi^4$, the reach in this channel falls short of the kinematic limit of $\sqrt{s}/2$. As the energy increases, the reaches in mass increase approximately linearly with energy if the total integrated luminosity also scales as ${{\mathcal L} \propto E^2}$.
\item The traditional mono-photon channel at lepton colliders (e.g., LEP), on the other hand, is suitable for higher-dimensional EW multiplets. This is due to the coupling enhancement for high EW $n$-plets ($\sim n^2$), as well as the high multiplicity of the final state ($\sim n$).  This channel's reach is stronger than that of the mono-muon for EW multiplets with $n\geq 5$. The reach still does not quite reach the kinematical limit of $\sim \sqrt{s}/2$ but it gets close in the case of the $n=7$. The main challenge for this channel is the large irreducible mono-photon background, leading to a small signal-to-background ratio at the level of $S/B < 10^{-2}$. 
\item The di-muon plus missing-mass signal captures the processes with $ZZ$, $Z\gamma$, and $\gamma\gamma$ initial states, but event selection suffers from the severe penalty of inability to tag two muons in the forward-backward regions, as we conservatively assumed a muon angular acceptance of $|\eta_\mu|<2.5$. Nevertheless, for higher-dimensional EW multiplets, such as $n=7$, this channel provides competitive coverage compared with the mono-photon channel and is more robust against systematics. Advanced detector design, covering more forward regime, such as  $2.5<|\eta_\mu|<4.0$ (perhaps even up to $|\eta_\mu|$ of 8 \cite{delphesTalk}), 
can significantly boost the signal acceptance and help separating background from different origins.
\item Disappearing track will play an indispensable role in the search for EW multiplets. However, making accurate projections for this channel's reach is hampered by the lack of knowledge of the detailed detector design and the beam-induced background. Our estimate, based on requiring tens of signal events, allows us to demonstrate its potential for discovery of various EW multiplets. The mono-photon channel with one disappearing track will have the largest signal rate and can extend the reach significantly for all odd-dimensional cases. It does not entirely reach the kinematical threshold, since some boost is still needed to allow the charged $\pm 1$ particle to have enough hits in the tracker before decaying, particularly for the cases $n\geq 5$.\footnote{In this case, the delayed signal using timing information provides complementary information that can potentially enhance the reach~\cite{Liu:2018wte}. However, BIB also has a significant out-of-time contribution and hence requires detailed studies.}
The triplet receives the most boost in sensitivity from the disappearing track signal, bringing the reach very close to the kinematical threshold. It can help with the doublet case, with a reach stronger than the mono-muon channel. Requiring disappearing-track pairs will reduce the reach. However, it is a cleaner signal and could be more important if the single disappearing track signature does not provide enough background suppression.
\end{itemize}

\section{Summary and Outlook}
\label{sec:Sum}

\begin{figure}[h!]
\centering
\includegraphics[width = 0.8\textwidth]{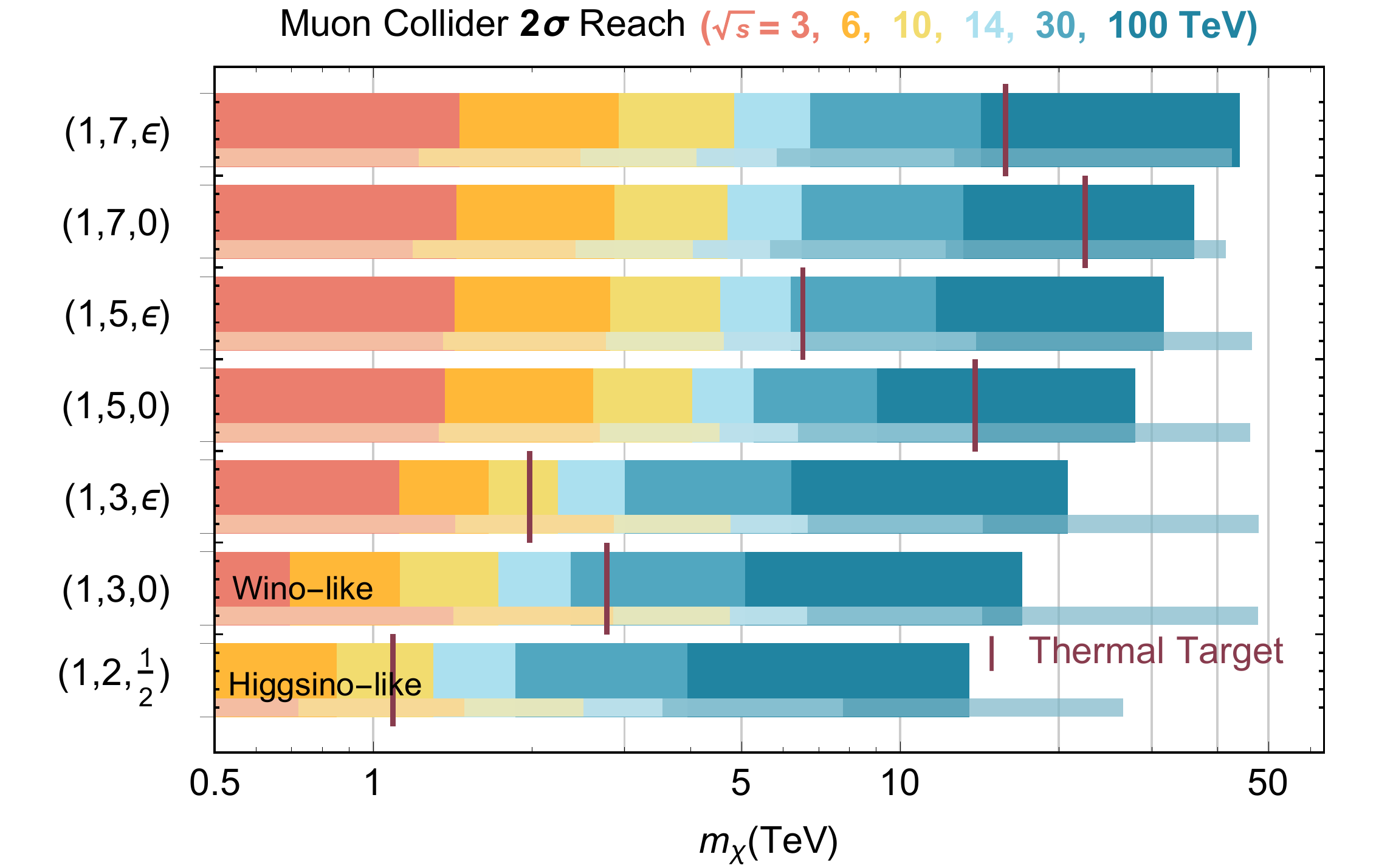}
\includegraphics[width = 0.8\textwidth]{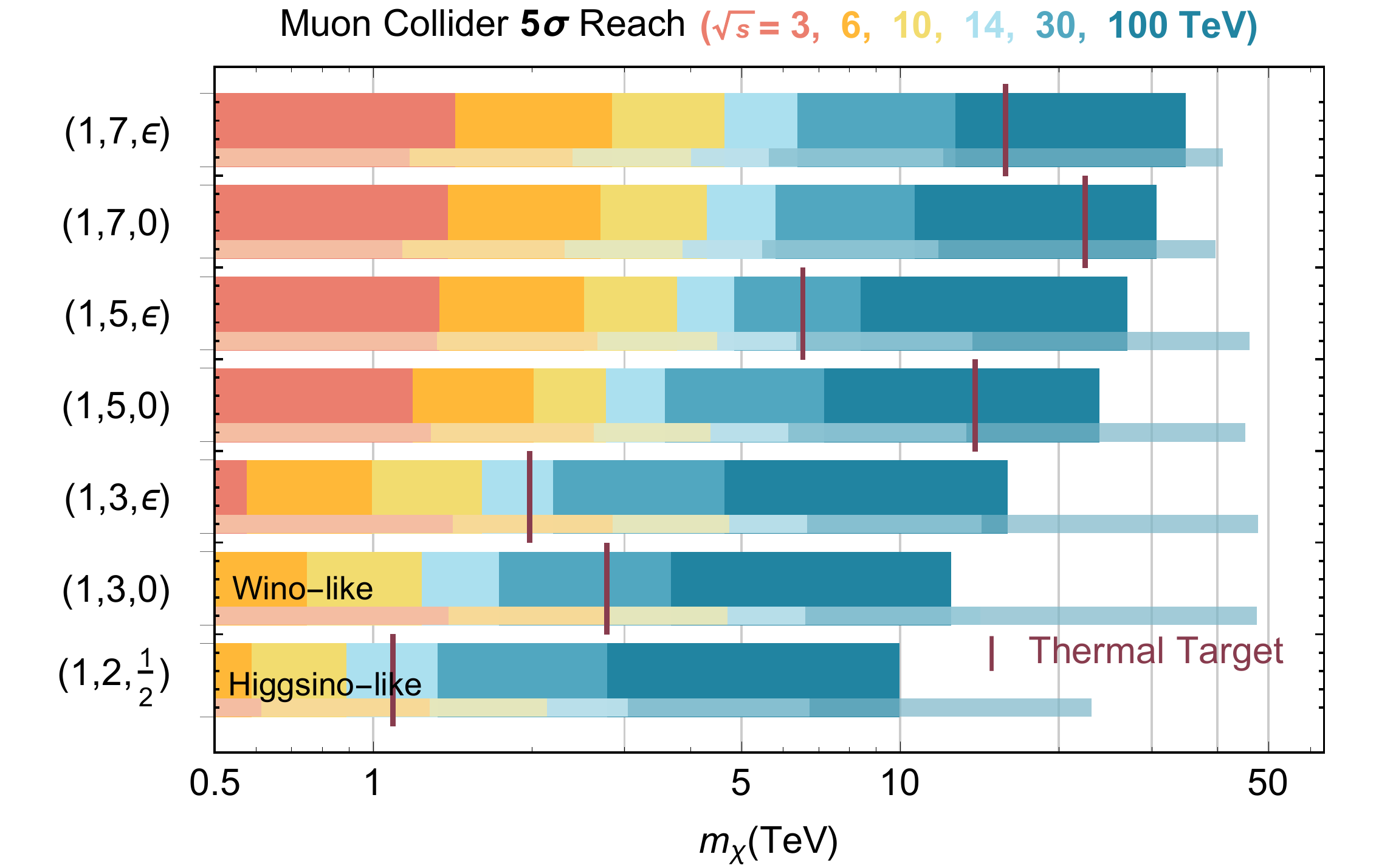}
\caption{Summary of the exclusion (upper panel)  and  discovery (lower panel) reaches of various muon collider running scenarios. The thick bars represent the combined reach from missing mass searches through mono-photon, mono-muon, and VBF di-muon channels. The thin and faint bars represent our estimates of the mono-photon plus one disappearing track search. The burgundy vertical bars represent the thermal target for a given EW-multiplet model.
}
\label{fig:summary}
\end{figure}

The WIMP paradigm remains one of the most attractive scenarios for particle dark matter, owing to its predictive power. Consistency of the thermal DM relic abundance also hints at a  
potential connection to the EW scale. 
Therefore, the search for WIMP DM at high-energy colliders is of fundamental significance. It also provides pivotal complementary information to the direct detection in the underground labs and the indirect detection from astro-particle observations. 

A muon collider running at high energies will open new physics thresholds and thus offer great potential in reaching an unprecedented mass scale, necessary for testing the paradigm of thermal WIMP DM conclusively. In our analyses, we focused on the search at a high-energy muon collider. 
We adopted the benchmark collider energies and the corresponding integrated luminosities as listed in \autoref{eq:para}.

EW multiplets, of which dark matter particle is the lightest member, furnish arguably the simplest candidates of dark matter. They also serve as one of the most challenging WIMP DM scenarios given its minimal signature of missing momentum and high thermal target mass in the 1 TeV$-$20 TeV range. 
We begin with the simple, yet conservative and challenging, signals,  in which the states in the EW multiplets are pairly produced in association with another energetic SM particle. 
We identified three leading search channels: mono-muon, mono-photon, and VBF di-muon plus missing energy-momenta. In contrast to the hadron colliders, the kinematics at a muon collider allows us to construct a missing mass variable for this class of signals with sharp features. 
We then explored an additional signature of disappearing tracks from a long-lived charged particle in the EW multiplet, which yields a short track before it  decays in the detector. Without a full knowledge of the optimal detector design and the beam-induced background, we estimated sensitivity by making the analogous requirements to existing searches and requiring at least $20-50$ signal events, after imposing the mono-photon selection cuts (\autoref{eq:angle} and \autoref{eq:ecut}) and disappearing track selection cuts (\autoref{eq:dtcut}). 
Such an analysis illustrates the potential of this signal while setting a target for the detector performance. 

Our main results are summarized in \autoref{fig:summary}. The reaches for $2\sigma$ exclusion (upper panel) and the $5\sigma$ discovery (lower panel) are shown,  and various muon collider running scenarios are indicated by the color codes. The thick (darker) bars represent the mass reach (horizontal axis) by combining the channels of inclusive missing-mass signals. The thin (fainter) bars are our estimates of the mono-photon plus one disappearing track search. 
For comparison, we have also included the target masses (burgundy vertical bars) for which the dark matter thermal relic abundance is saturated by the EW multiplets DM  under consideration.  When combining the inclusive (missing mass) channels, the overall reach is less than the kinematical limit $m_\chi \sim \sqrt{s}/2$, especially for EW multiplets with $n \leq 3$ due to the low signal-to-background ratio. It is possible to cover (with $2 \sigma$) the thermal targets of the doublet and Dirac triplet with a 10 TeV muon collider. For the Majorana triplet, a 30 TeV option would suffice. 
The thermal targets of Dirac(Majorana) 5-plet would be covered by 30 (100) 
TeV muon colliders. The 100 TeV option will also cover the thermal target for the 7-plet. It is important to emphasize that, in order to cover the thermal target, the necessary center of mass energy and luminosity in many cases, as shown in \autoref{fig:app_2} in the appendix, can be much lower than the benchmark values we showed in \autoref{eq:para}. A Majorana triplet can be covered by a 20 TeV muon collider (still assuming integrated luminosity scales with $s$). A Majorana 5-plet can be covered by a 50 TeV muon collider, while a 70 TeV muon collider is enough to cover the case of the 7-plet. 
A 75 TeV muon collider is sufficient to reach $5 \sigma$ discovery potential for all EW multiplets with their thermal mass targets considered in this paper. 
At the same time, the disappearing track signal has excellent potential. Based on our study, it could bring the reach very close to the kinematical threshold $m_\chi \sim \sqrt{s}/2$. We note here, a 6 TeV muon collider with disappearing track search can cover the thermal target of the doublet case, motivating further detailed studies in this direction.

Both direct and indirect detections may also be able to detect the WIMPs considered in this paper. They should be regarded as complementary probes focusing on different kinematical regimes (in the case of direct detection) and depending on potentially large astrophysical uncertainties (in the case of indirect detection). Hence, even though it can be fruitful to combine the reaches of these very different probes, the collider searches are important in its own right. Only when two discoveries of the WIMP DM reach agreement, can we reveal the nature of the particle dark matter.

In summary, we have demonstrated that muon colliders running at high energies have great potential in searching for the EW multiplets and can make a decisive statement about their viability as WIMP dark matter candidates. This should serve as a main physics driver for the high energy muon colliders. Our results give detailed projection based on more inclusive signals. Our studies on the disappearing tracks, as a first look, identify cases and regions where potential gains in reach can be significant. 
More refined studies are needed for detecting soft charged particles and disappearing tracks when realistic detector design and performance optimization become available. The study can be further extended to general explorations of general electroweak states, where the additional reconstructable signatures from heavy particle decays to enhance the sensitivity would be highly desirable. 

\begin{acknowledgments}
The work of TH was supported in part by the U.S.~Department of Energy under grant No.~DE-FG02- 95ER40896 and in part by the PITT PACC. ZL was supported in part by the NSF grants PHY-1620074, PHY-1914480 and PHY-1914731, and by the Maryland Center for Fundamental Physics (MCFP). LTW is supported by the
DOE grant de-sc0013642. XW was supported by the National Science Foundation under Grant No.~PHY-1915147.
\end{acknowledgments}

\appendix

\section{Details of the sensitivity results}
In this appendix, we present our detailed results of the 2$\sigma$ and 5$\sigma$ reaches for the six different high-energy muon collider running scenarios in \autoref{eq:para}, for fermionic WIMP DM candidates from a broad variety of EW multiplets and several proposed search channels. For the missing mass search, we also include a combined sensitivity of the three search channels (in the column labeled ``combined''). For the disappearing track (DT) search, we show the results with two different background assumptions, namely zero background and 100 background events. 
The results shown in the main text and the summary figures corresponds to 100 background events after cuts, and here we also show the DT reaches with zero background for reference.
The number in the parentheses in the DT searches indicates the number of signal events required. 
For the $n=4,~6$ even-plet with $Y=1/2$, due to the dependence on additional mass splitting terms to ensure the lightest statement being electrically neutral, the lifetime critically depending on the non-minimal model parameter choices. For this reason, we do not show the DT projections for these even-plet models.

\begin{figure}[tb]
\centering
\begin{subfigure}[t]{0.48\textwidth}\centering
\includegraphics[width=\textwidth]{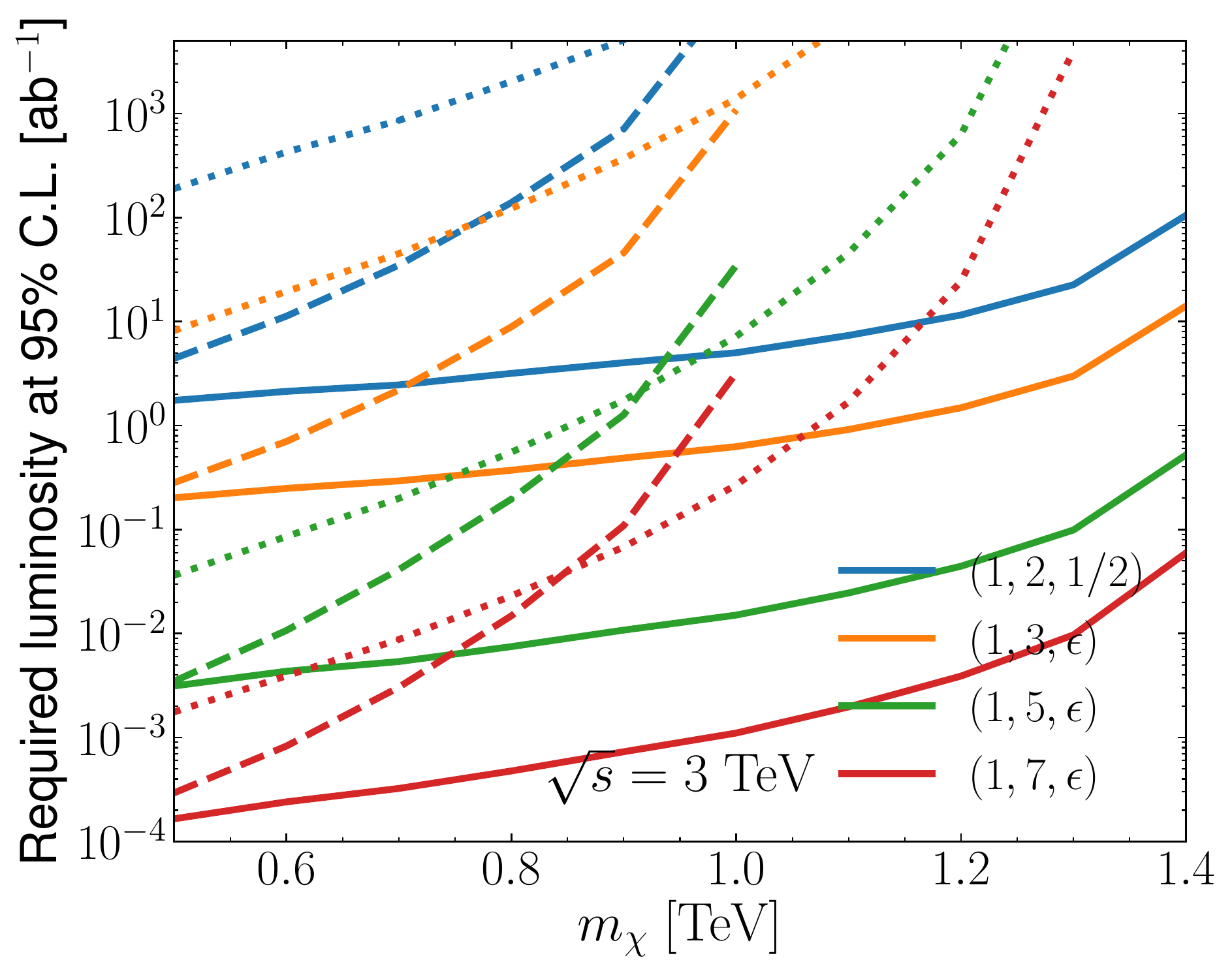}
\caption{}
\end{subfigure}
\begin{subfigure}[t]{0.48\textwidth}\centering
\includegraphics[width=\textwidth]{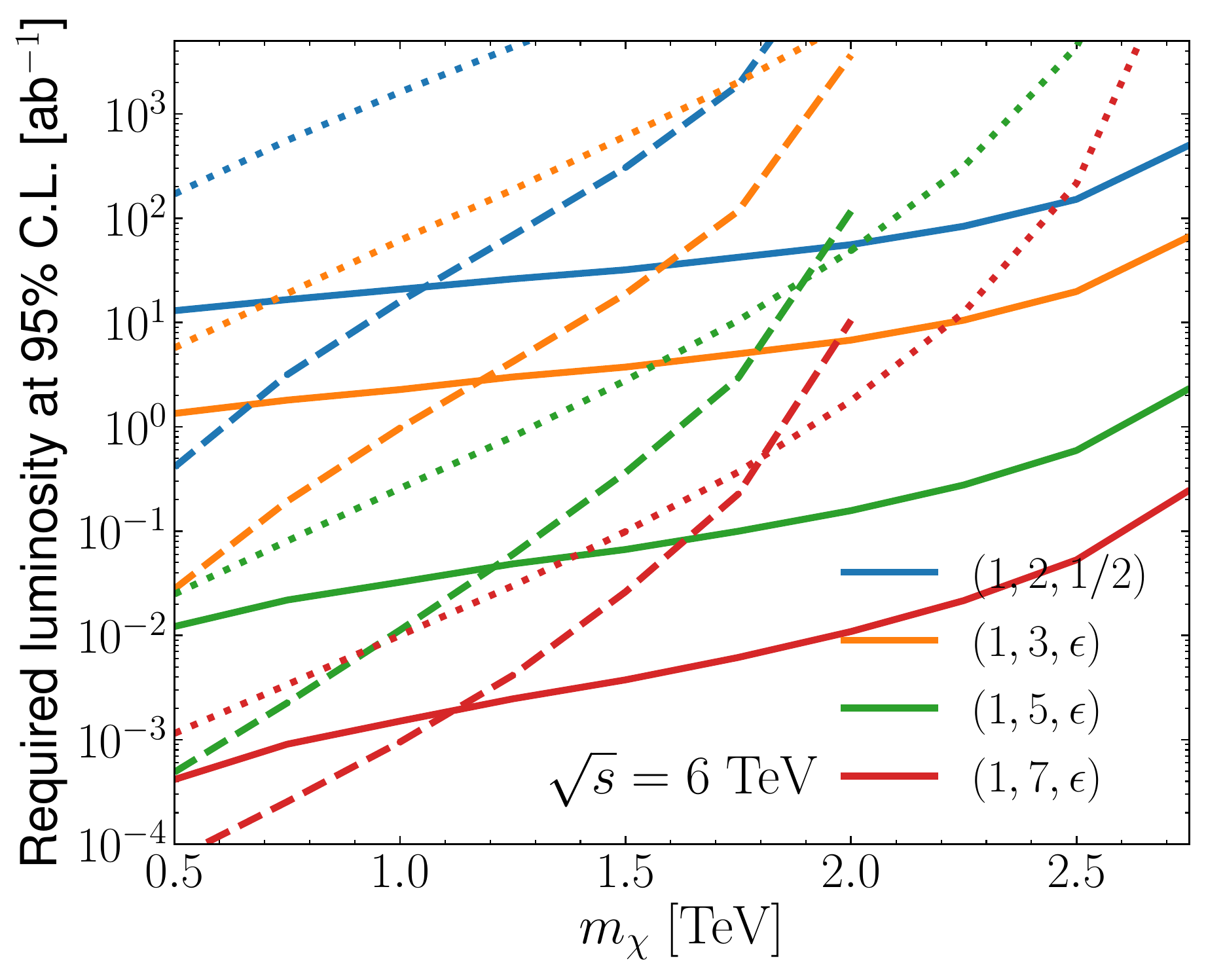}
\caption{}
\end{subfigure} \\ 
\begin{subfigure}[t]{0.48\textwidth}\centering
\includegraphics[width=\textwidth]{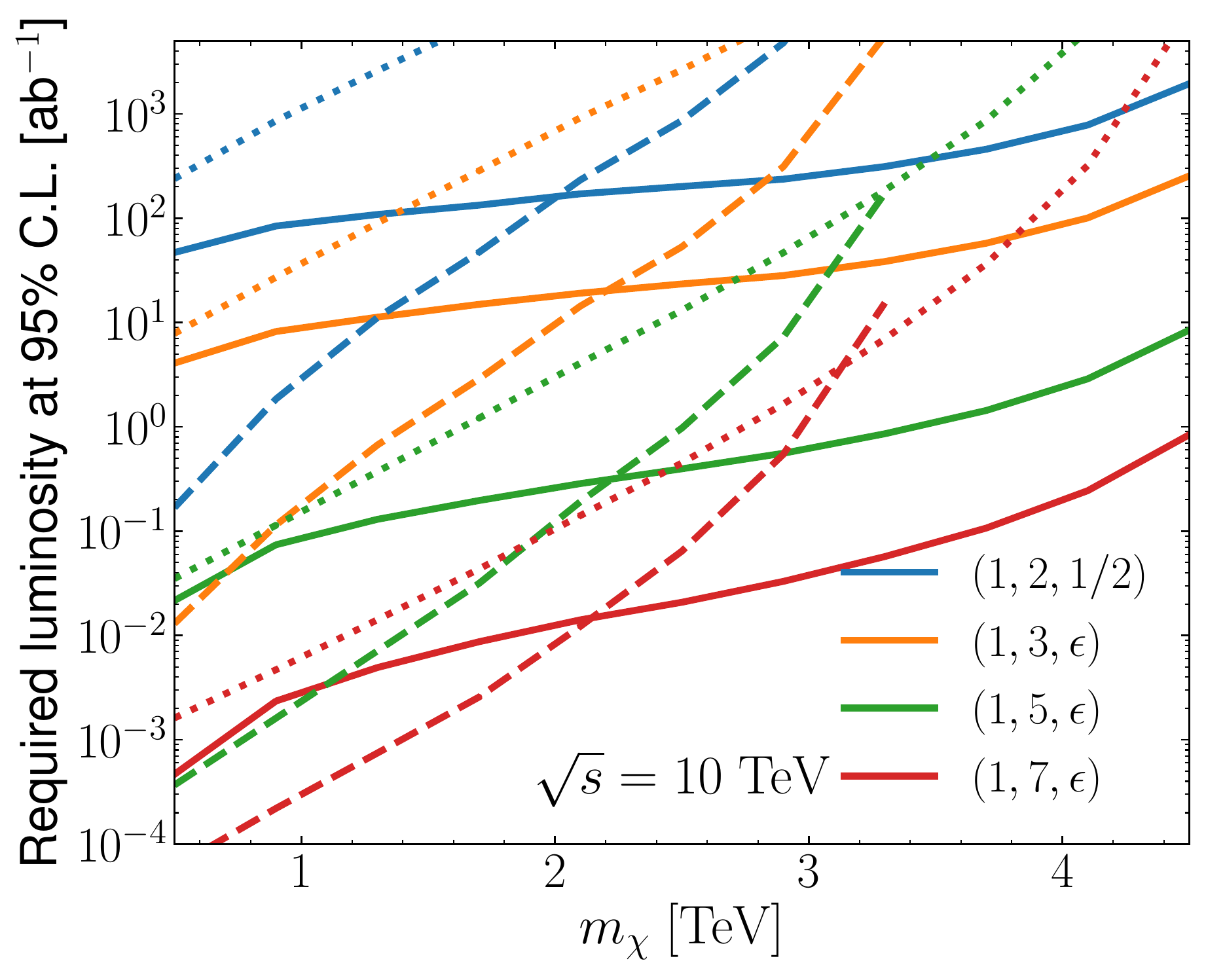}
\caption{}
\end{subfigure} 
\begin{subfigure}[t]{0.48\textwidth}\centering
\includegraphics[width=\textwidth]{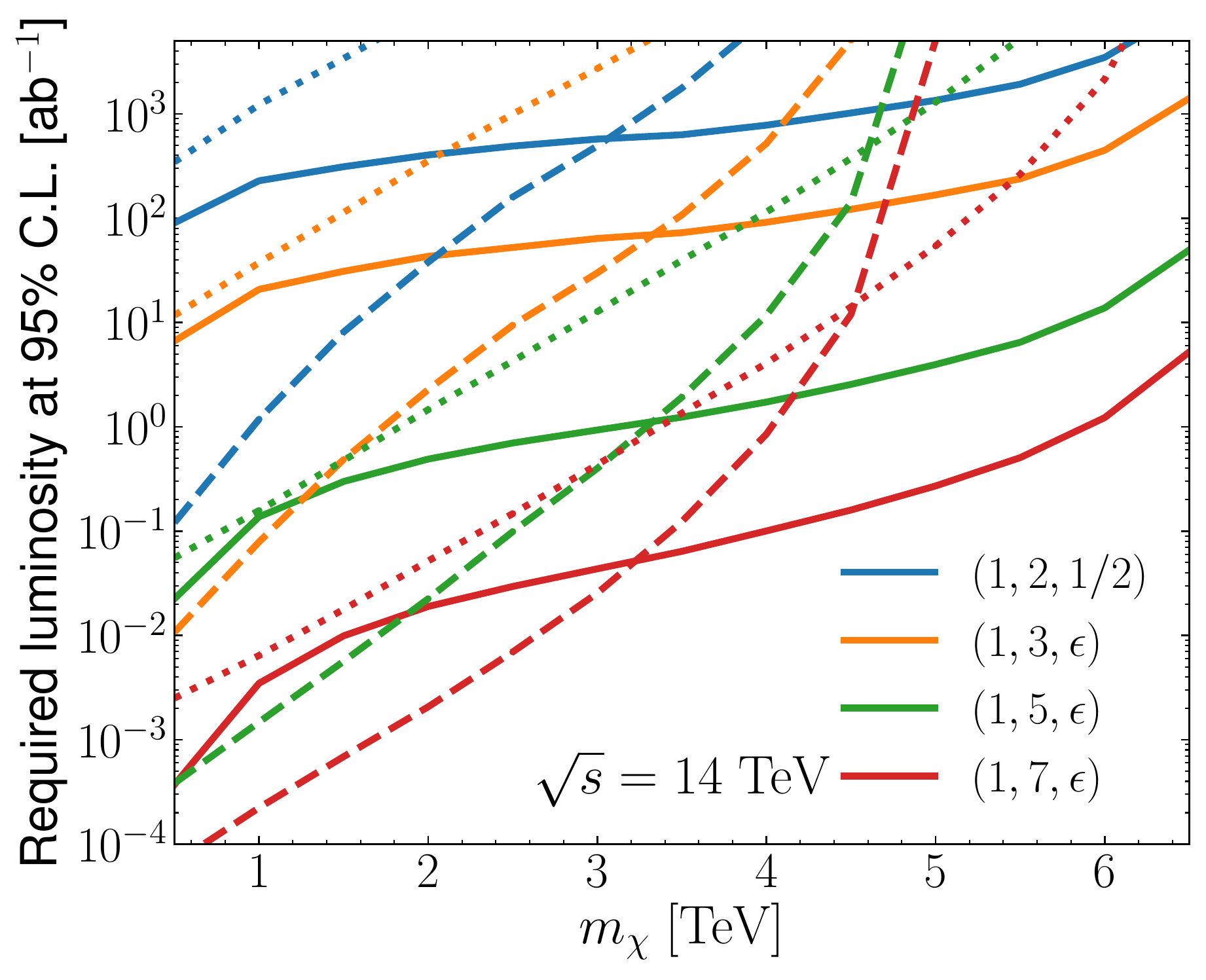}
\caption{}
\end{subfigure} \\ 
\begin{subfigure}[t]{0.48\textwidth}\centering
\includegraphics[width=\textwidth]{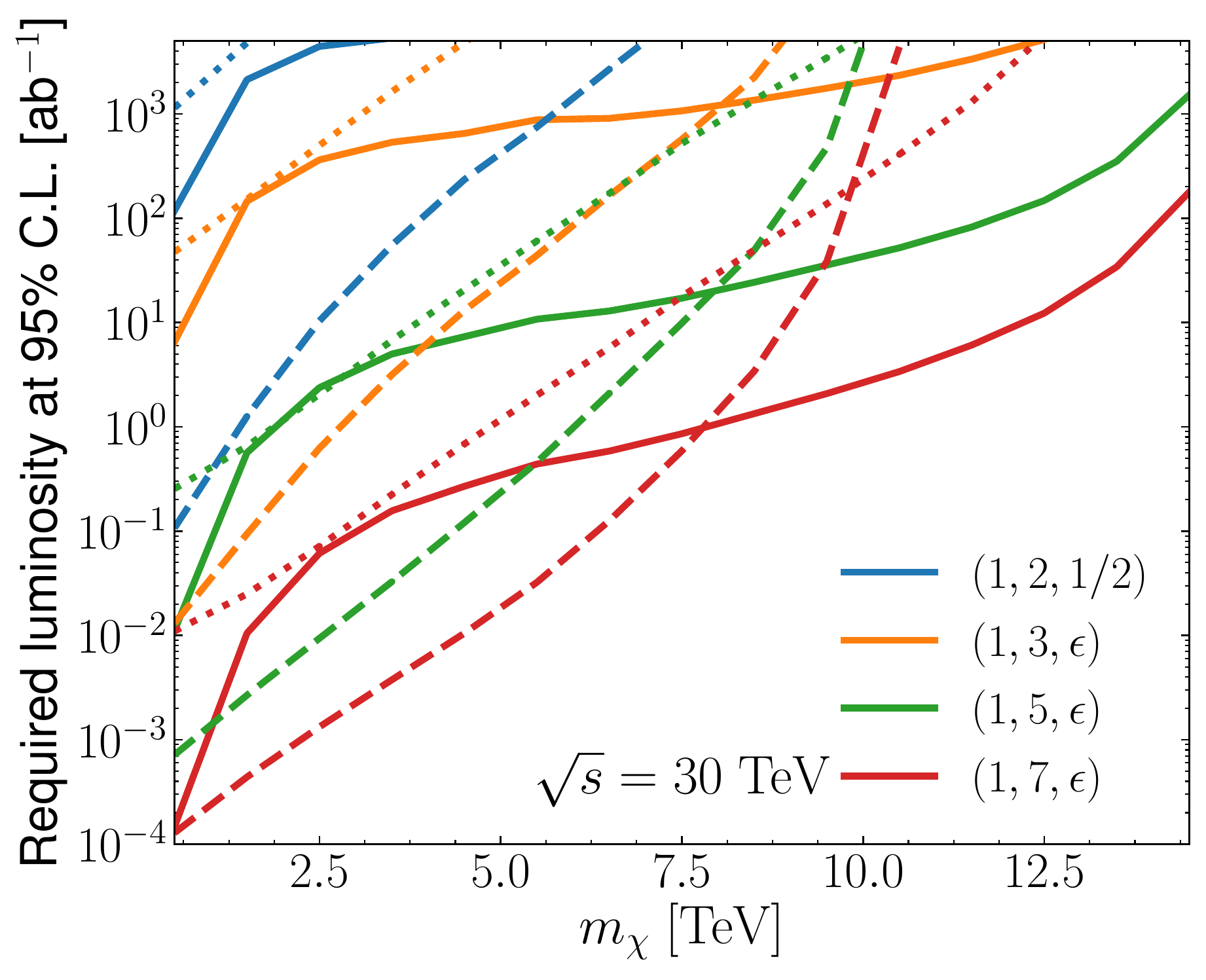}
\caption{}
\end{subfigure}
\begin{subfigure}[t]{0.48\textwidth}\centering
\includegraphics[width=\textwidth]{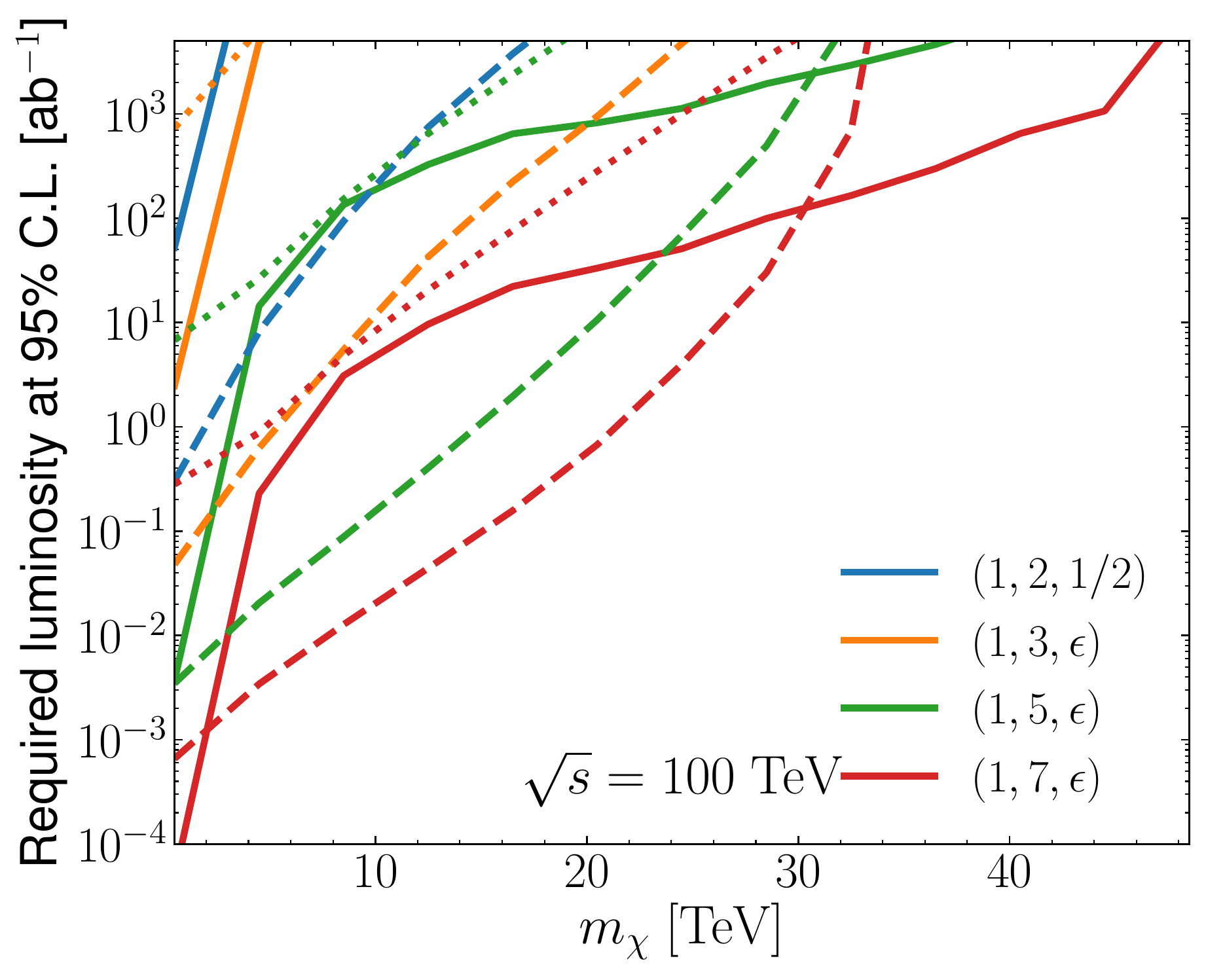}
\caption{}
\end{subfigure}
\caption{Integrated luminosities needed for mono-photon (solid), mono-muon (dashed), and di-muon (dotted) channels, to reach $2\sigma$ statistical significance at $\sqrt s=3$, 6, 10, 14, 30, and 100 TeV.}
\label{fig:app_1}
\end{figure}

\begin{figure}[tb]
\centering
\begin{subfigure}[t]{0.48\textwidth}\centering
\includegraphics[width=\textwidth]{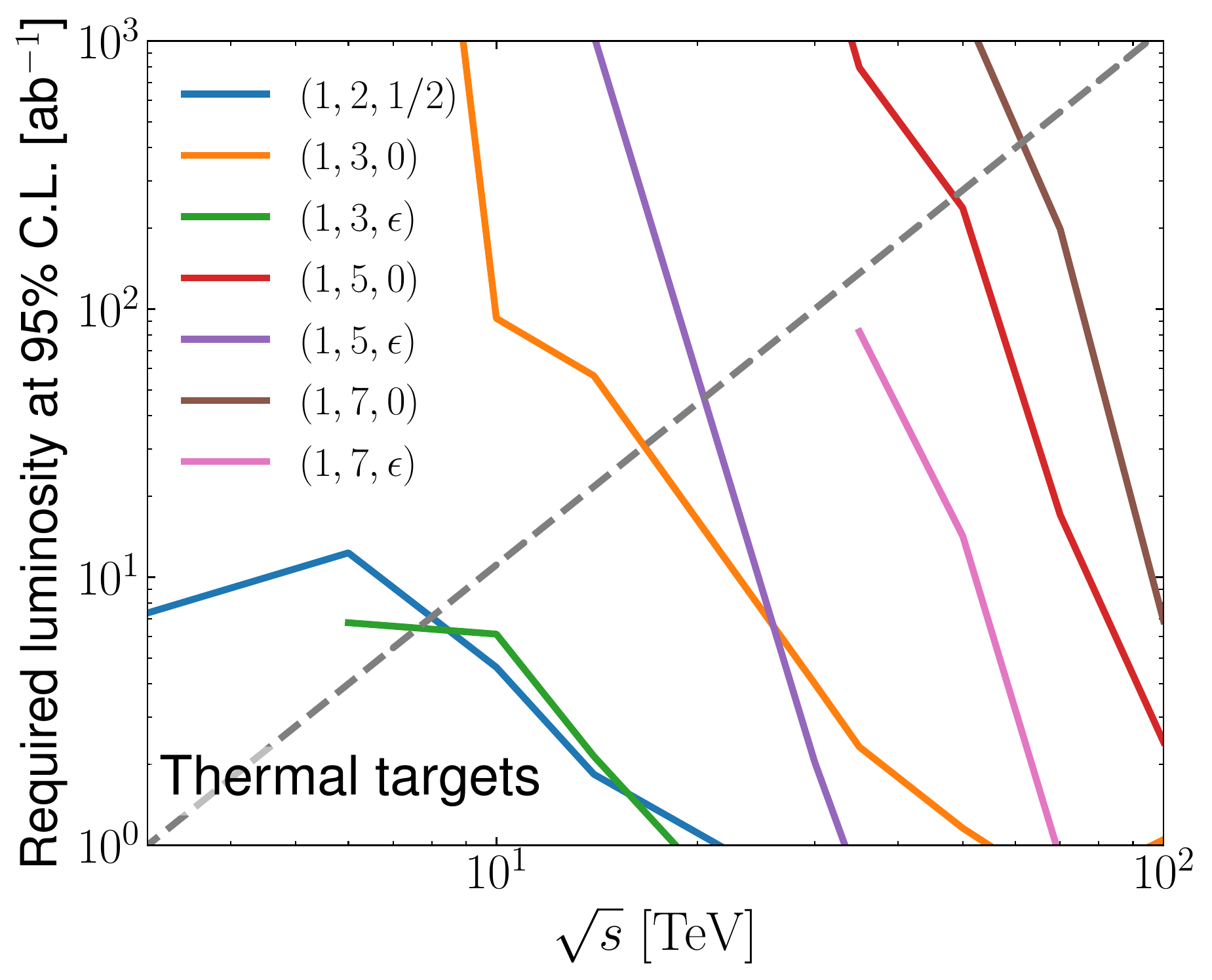}
\caption{}
\end{subfigure} 
\begin{subfigure}[t]{0.48\textwidth}\centering
\includegraphics[width=\textwidth]{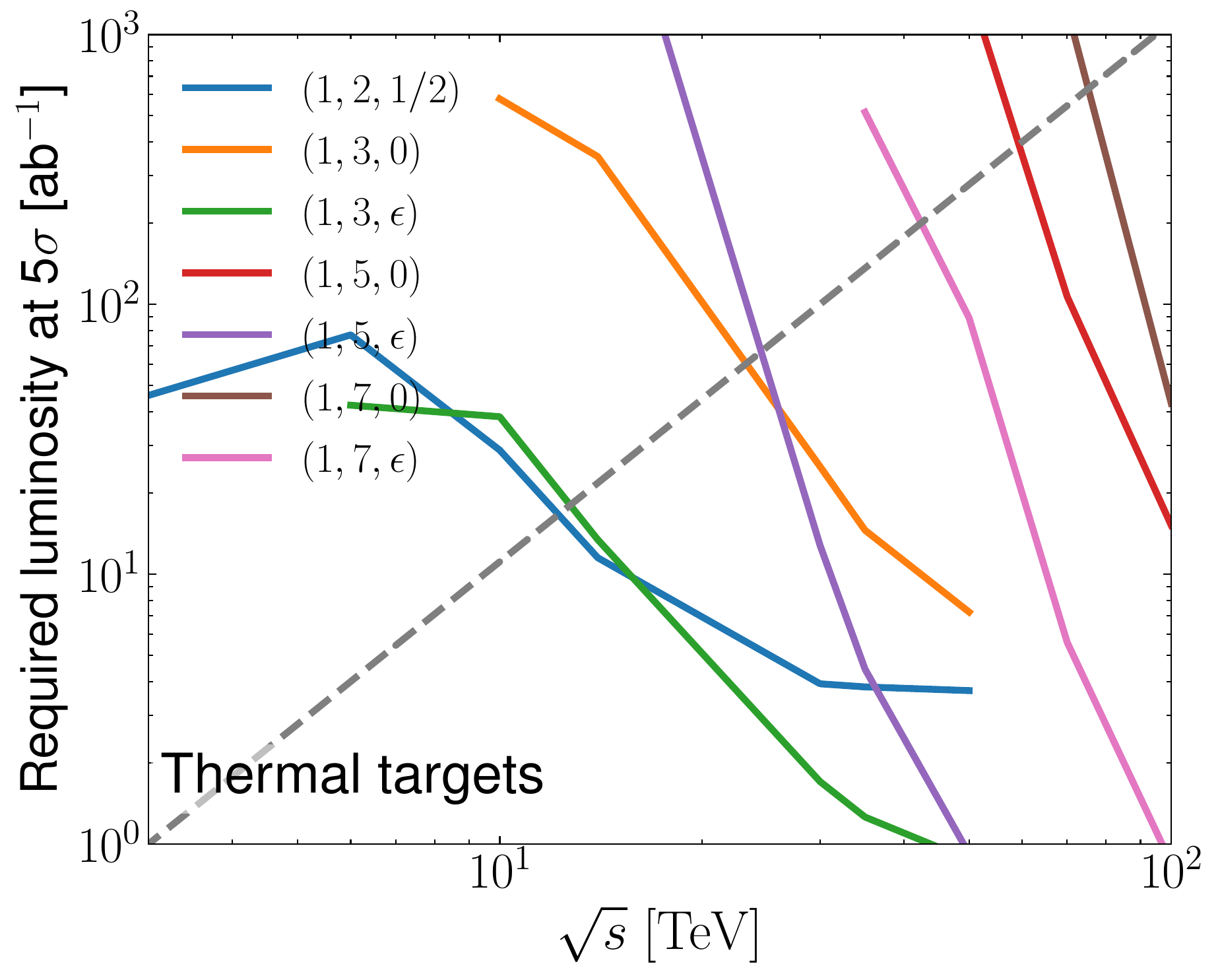}
\caption{}
\end{subfigure}
\caption{Integrated luminosities needed for the combined missing mass search, to reach the thermal targets with (a) $2\sigma$ and (b) $5\sigma$ statistical significance. The diagonal dashed line indicates the benchmark luminosity v.s. center of mass energy relation used in this study.}
\label{fig:app_2}
\end{figure}

We show, in \autoref{fig:app_1}, the luminosities needed for mono-photon (solid), mono-muon (dashed), and di-muon (dotted) channels, to reach $2\sigma$ statistical significance at $\sqrt s=3$, 6, 10, 14, 30, and 100 TeV. For fixed thermal masses given in \autoref{tab:WIMP}, \autoref{fig:app_2} shows the luminosities needed for the combined missing mass search to reach $2\sigma$ and $5\sigma$ statistical significance.

\pagestyle{empty} 

\begin{table}[htbp]
\footnotesize
  \centering
    \begin{tabular}{c|c|c|c|c|c|c|c|c|c}
\multicolumn{10}{c}{2$\sigma$ reaches (TeV)} \\
\hline 
\multicolumn{1}{c|}{Collider} & Model & \multicolumn{4}{c|}{Missing Mass Searches} & \multicolumn{4}{c}{mono-$\gamma$ + Disappearing Track} \\
\cline{3-10}    
\multicolumn{1}{p{4.555em}|}{Parameter} & (color,n,Y)  & mono-$\gamma$ & mono-$\mu$ & di muon & Combined & 1DT(3) & 1DT(20) & 2DT(3) & 2DT(20) \\
       \hline
    \multicolumn{1}{c|}{\multirow{4}[8]{*}{3 TeV}} & (1,2,1/2) & --- & --- & --- & --- & 0.9 & 0.7 & --- & --- \\
\cline{2-10}      & (1,3,0) & 0.6 & --- & --- & 0.7 & 1.4 & 1.4 & 1.4 & 1.3 \\
\cline{2-10}      & (1,3,$\epsilon$) & 1.1 & 0.6 & --- & 1.1 & 1.4 & 1.4 & 1.4 & 1.3 \\
\cline{2-10}      & (1,4,1/2) & 1.4 & 0.8 & 0.7 & 1.4 & --- & --- & --- & --- \\
\cline{2-10}    \multicolumn{1}{c|}{\multirow{5}[10]{*}{1 $\abi$}} & (1,5,0) & 1.4 & 0.8 & 0.7 & 1.4 & 1.4 & 1.3 & 1.2 & 1.1 \\
\cline{2-10}      & (1,5,$\epsilon$) & 1.4 & 0.9 & 0.9 & 1.4 & 1.4 & 1.4 & 1.2 & 1.1 \\
\cline{2-10}      & (1,6,1/2) & 1.4 & 1.0 & 1.0 & 1.4 & --- & --- & --- & --- \\
\cline{2-10}      & (1,7,0) & 1.4 & 0.9 & 1.0 & 1.4 & 1.3 & 1.2 & 0.9 & 0.8 \\
\cline{2-10}      & (1,7,$\epsilon$) & 1.5 & 1.0 & 1.1 & 1.5 & 1.3 & 1.2 & 1.0 & 0.9 \\
    \hline
    \multicolumn{1}{c|}{\multirow{4}[8]{*}{6 TeV}} & (1,2,1/2) & --- & 0.8 & --- & 1.9 & 1.5 & 1.0 & 0.7 & --- \\
\cline{2-10}      & (1,3,0) & --- & 1.0 & --- & 1.1 & 2.9 & 2.9 & 2.8 & 2.6 \\
\cline{2-10}      & (1,3,$\epsilon$) & 1.6 & 1.2 & --- & 1.7 & 2.9 & 2.9 & 2.8 & 2.7 \\
\cline{2-10}      & (1,4,1/2) & 2.6 & 1.6 & 1.2 & 2.6 & --- & --- & --- & --- \\
\cline{2-10}    \multicolumn{1}{c|}{\multirow{5}[10]{*}{4 $\abi$}} & (1,5,0) & 2.6 & 1.7 & 1.3 & 2.6 & 2.8 & 2.7 & 2.4 & 2.2 \\
\cline{2-10}      & (1,5,$\epsilon$) & 2.8 & 1.8 & 1.6 & 2.8 & 2.8 & 2.8 & 2.5 & 2.3 \\
\cline{2-10}      & (1,6,1/2) & 2.9 & 1.9 & 1.9 & 2.9 & --- & --- & --- & --- \\
\cline{2-10}      & (1,7,0) & 2.9 & 1.9 & 1.9 & 2.9 & 2.6 & 2.4 & 1.9 & 1.7 \\
\cline{2-10}      & (1,7,$\epsilon$) & 2.9 & 2.0 & 2.1 & 2.9 & 2.6 & 2.5 & 2.0 & 1.8 \\
    \hline
    \multicolumn{1}{c|}{\multirow{4}[8]{*}{10 TeV}} & (1,2,1/2) & --- & 1.3 & --- & 1.3 & 3.1 & 2.5 & 1.7 & 1.2 \\
\cline{2-10}      & (1,3,0) & --- & 1.7 & --- & 1.7 & 4.8 & 4.8 & 4.7 & 4.4 \\
\cline{2-10}      & (1,3,$\epsilon$) & 1.2 & 2.0 & 0.6 & 2.2 & 4.8 & 4.8 & 4.7 & 4.6 \\
\cline{2-10}      & (1,4,1/2) & 4.1 & 2.7 & 1.8 & 4.1 & --- & --- & --- & --- \\
\cline{2-10}    \multicolumn{1}{c|}{\multirow{5}[10]{*}{10 $\abi$}} & (1,5,0) & 4.0 & 2.7 & 2.0 & 4.0 & 4.7 & 4.5 & 4.0 & 3.6 \\
\cline{2-10}      & (1,5,$\epsilon$) & 4.5 & 3.0 & 2.4 & 4.5 & 4.7 & 4.6 & 4.1 & 3.8 \\
\cline{2-10}      & (1,6,1/2) & 4.8 & 3.2 & 3.1 & 4.8 & --- & --- & --- & --- \\
\cline{2-10}      & (1,7,0) & 4.7 & 3.2 & 3.1 & 4.7 & 4.3 & 4.0 & 3.2 & 2.8 \\
\cline{2-10}      & (1,7,$\epsilon$) & 4.8 & 3.3 & 3.4 & 4.8 & 4.4 & 4.1 & 3.4 & 3.0 \\
    \hline
    \multicolumn{1}{c|}{\multirow{4}[8]{*}{14 TeV}} & (1,2,1/2) & --- & 1.8 & --- & 1.9 & 4.4 & 3.9 & 2.4 & 1.7 \\
\cline{2-10}      & (1,3,0) & --- & 2.3 & --- & 2.4 & 6.7 & 6.7 & 6.6 & 6.2 \\
\cline{2-10}      & (1,3,$\epsilon$) & 1.0 & 2.9 & 0.8 & 3.0 & 6.7 & 6.7 & 6.6 & 6.4 \\
\cline{2-10}      & (1,4,1/2) & 5.2 & 3.8 & 2.3 & 5.3 & --- & --- & --- & --- \\
\cline{2-10}    \multicolumn{1}{c|}{\multirow{5}[10]{*}{20 $\abi$}} & (1,5,0) & 5.3 & 3.8 & 2.6 & 5.3 & 6.6 & 6.4 & 5.7 & 5.1 \\
\cline{2-10}      & (1,5,$\epsilon$) & 6.2 & 4.2 & 3.2 & 6.2 & 6.6 & 6.5 & 5.8 & 5.3 \\
\cline{2-10}      & (1,6,1/2) & 6.6 & 4.4 & 4.1 & 6.6 & --- & --- & --- & --- \\
\cline{2-10}      & (1,7,0) & 6.5 & 4.4 & 4.1 & 6.5 & 6.1 & 5.7 & 4.5 & 4.0 \\
\cline{2-10}      & (1,7,$\epsilon$) & 6.7 & 4.6 & 4.7 & 6.7 & 6.1 & 5.8 & 4.7 & 4.2 \\
    \hline
    \multicolumn{1}{c|}{\multirow{4}[8]{*}{30 TeV}} & (1,2,1/2) & --- & 3.9 & --- & 3.9 & 9.6 & 7.8 & 5.2 & 3.9 \\
\cline{2-10}      & (1,3,0) & 1.1 & 5.0 & --- & 5.1 & 14 & 14 & 14 & 14 \\
\cline{2-10}      & (1,3,$\epsilon$) & 1.4 & 6.1 & 1.1 & 6.2 & 14 & 14 & 14 & 14 \\
\cline{2-10}      & (1,4,1/2) & 7.9 & 8.0 & 4.1 & 8.8 & --- & --- & --- & --- \\
\cline{2-10}    \multicolumn{1}{c|}{\multirow{5}[10]{*}{90 $\abi$}} & (1,5,0) & 8.3 & 8.1 & 4.6 & 9.0 & 14 & 14 & 12 & 11 \\
\cline{2-10}      & (1,5,$\epsilon$) & 12 & 8.9 & 5.9 & 12 & 14 & 14 & 13 & 12 \\
\cline{2-10}      & (1,6,1/2) & 14 & 9.4 & 7.8 & 14 & --- & --- & --- & --- \\
\cline{2-10}      & (1,7,0) & 13 & 9.4 & 7.8 & 13 & 13 & 12 & 9.9 & 8.7 \\
\cline{2-10}      & (1,7,$\epsilon$) & 14 & 9.9 & 9.1 & 14 & 13 & 13 & 10 & 9.1 \\
    \hline
    \multicolumn{1}{c|}{\multirow{4}[8]{*}{100 TeV}} & (1,2,1/2) & 3.7 & 13 & --- & 13 & 33 & 27 & 18 & 13 \\
\cline{2-10}      & (1,3,0) & 4.2 & 17 & --- & 17 & 48 & 48 & 47 & 46 \\
\cline{2-10}      & (1,3,$\epsilon$) & 4.4 & 21 & 1.4 & 21 & 48 & 48 & 47 & 46 \\
\cline{2-10}      & (1,4,1/2) & 9.3 & 27 & 8.2 & 28 & --- & --- & --- & --- \\
\cline{2-10}    \multicolumn{1}{c|}{\multirow{5}[10]{*}{$10^3 \abi$}} & (1,5,0) & 11 & 28 & 10 & 28 & 47 & 46 & 41 & 38 \\
\cline{2-10}      & (1,5,$\epsilon$) & 23 & 30 & 14 & 32 & 47 & 47 & 43 & 39 \\
\cline{2-10}      & (1,6,1/2) & 37 & 32 & 20 & 37 & --- & --- & --- & --- \\
\cline{2-10}      & (1,7,0) & 36 & 32 & 20 & 36 & 43 & 41 & 33 & 30 \\
\cline{2-10}      & (1,7,$\epsilon$) & 44 & 33 & 25 & 44 & 45 & 43 & 34 & 31 \\
    \hline
    \end{tabular}%
  \label{tab:2sigmatable}%
\end{table}%

\begin{table}[htbp]
\footnotesize
  \centering
    \begin{tabular}{c|c|c|c|c|c|c|c|c|c}
\multicolumn{10}{c}{5$\sigma$ reaches (TeV)} \\
\hline 
\multicolumn{1}{c|}{Collider} & Model & \multicolumn{4}{c|}{Missing Mass Searches} & \multicolumn{4}{c}{mono-$\gamma$ + Disappearing Track} \\
\cline{3-10}    
\multicolumn{1}{p{4.555em}|}{Parameter} & (color,n,Y)  & mono-$\gamma$ & mono-$\mu$ & di muon & Combined & 1DT(10) & 1DT(50) & 2DT(10) & 2DT(50) \\
    \multicolumn{1}{c|}{\multirow{4}[8]{*}{3 TeV}} & (1,2,1/2) & --- & --- & --- & --- & 0.8 & 0.7 & --- & --- \\
\cline{2-10}      & (1,3,0) & --- & --- & --- & --- & 1.4 & 1.4 & 1.3 & 1.2 \\
\cline{2-10}      & (1,3,$\epsilon$) & --- & --- & --- & 0.6 & 1.4 & 1.4 & 1.4 & 1.3 \\
\cline{2-10}      & (1,4,1/2) & 1.2 & 0.7 & --- & 1.2 & --- & --- & --- & --- \\
\cline{2-10}    \multicolumn{1}{c|}{\multirow{5}[10]{*}{1 $\abi$}} & (1,5,0) & 1.2 & 0.7 & 0.5 & 1.2 & 1.4 & 1.3 & 1.1 & 1.0 \\
\cline{2-10}      & (1,5,$\epsilon$) & 1.3 & 0.8 & 0.7 & 1.3 & 1.4 & 1.3 & 1.2 & 1.0 \\
\cline{2-10}      & (1,6,1/2) & 1.4 & 0.9 & 0.9 & 1.4 & --- & --- & --- & --- \\
\cline{2-10}      & (1,7,0) & 1.4 & 0.9 & 0.9 & 1.4 & 1.2 & 1.1 & 0.9 & 0.8 \\
\cline{2-10}      & (1,7,$\epsilon$) & 1.4 & 0.9 & 1.0 & 1.4 & 1.3 & 1.2 & 0.9 & 0.8 \\
    \hline
    \multicolumn{1}{c|}{\multirow{4}[8]{*}{6 TeV}} & (1,2,1/2) & --- & 0.6 & --- & 0.6 & 1.6 & 1.3 & 0.8 & 0.6 \\
\cline{2-10}      & (1,3,0) & --- & 0.7 & --- & 0.7 & 2.9 & 2.8 & 2.7 & 2.5 \\
\cline{2-10}      & (1,3,$\epsilon$) & --- & 1.0 & --- & 1.0 & 2.9 & 2.8 & 2.8 & 2.6 \\
\cline{2-10}      & (1,4,1/2) & 2.0 & 1.4 & 0.8 & 2.0 & --- & --- & --- & --- \\
\cline{2-10}    \multicolumn{1}{c|}{\multirow{5}[10]{*}{4 $\abi$}} & (1,5,0) & 2.0 & 1.4 & 0.9 & 2.0 & 2.8 & 2.6 & 2.3 & 2.0 \\
\cline{2-10}      & (1,5,$\epsilon$) & 2.5 & 1.6 & 1.2 & 2.5 & 2.8 & 2.7 & 2.3 & 2.1 \\
\cline{2-10}      & (1,6,1/2) & 2.8 & 1.7 & 1.6 & 2.8 & --- & --- & --- & --- \\
\cline{2-10}      & (1,7,0) & 2.7 & 1.7 & 1.6 & 2.7 & 2.5 & 2.3 & 1.8 & 1.6 \\
\cline{2-10}      & (1,7,$\epsilon$) & 2.8 & 1.9 & 1.9 & 2.8 & 2.5 & 2.4 & 1.9 & 1.7 \\
    \hline
    \multicolumn{1}{c|}{\multirow{4}[8]{*}{10 TeV}} & (1,2,1/2) & --- & 0.9 & --- & 0.9 & 2.8 & 2.1 & 1.4 & 1.0 \\
\cline{2-10}      & (1,3,0) & --- & 1.2 & --- & 1.2 & 4.8 & 4.7 & 4.6 & 4.2 \\
\cline{2-10}      & (1,3,$\epsilon$) & --- & 1.6 & --- & 1.6 & 4.8 & 4.7 & 4.6 & 4.4 \\
\cline{2-10}      & (1,4,1/2) & 2.3 & 2.3 & 1.1 & 2.7 & --- & --- & --- & --- \\
\cline{2-10}    \multicolumn{1}{c|}{\multirow{5}[10]{*}{10 $\abi$}} & (1,5,0) & 2.5 & 2.3 & 1.3 & 2.8 & 4.6 & 4.4 & 3.8 & 3.4 \\
\cline{2-10}      & (1,5,$\epsilon$) & 3.8 & 2.6 & 1.8 & 3.8 & 4.7 & 4.5 & 3.9 & 3.6 \\
\cline{2-10}      & (1,6,1/2) & 4.4 & 2.9 & 2.5 & 4.4 & --- & --- & --- & --- \\
\cline{2-10}      & (1,7,0) & 4.3 & 2.9 & 2.5 & 4.3 & 4.1 & 3.9 & 3.0 & 2.6 \\
\cline{2-10}      & (1,7,$\epsilon$) & 4.6 & 3.1 & 2.9 & 4.6 & 4.3 & 4.0 & 3.1 & 2.8 \\
    \hline
    \multicolumn{1}{c|}{\multirow{4}[8]{*}{14 TeV}} & (1,2,1/2) & --- & 1.3 & --- & 1.3 & 3.9 & 3.0 & 2.0 & 1.4 \\
\cline{2-10}      & (1,3,0) & --- & 1.7 & --- & 1.7 & 6.7 & 6.6 & 6.4 & 5.9 \\
\cline{2-10}      & (1,3,$\epsilon$) & --- & 2.2 & --- & 2.2 & 6.7 & 6.7 & 6.5 & 6.1 \\
\cline{2-10}      & (1,4,1/2) & 2.4 & 3.2 & 1.5 & 3.5 & --- & --- & --- & --- \\
\cline{2-10}    \multicolumn{1}{c|}{\multirow{5}[10]{*}{20 $\abi$}} & (1,5,0) & 2.7 & 3.3 & 1.8 & 3.6 & 6.5 & 6.1 & 5.3 & 4.8 \\
\cline{2-10}      & (1,5,$\epsilon$) & 4.8 & 3.7 & 2.4 & 4.8 & 6.5 & 6.4 & 5.6 & 5.1 \\
\cline{2-10}      & (1,6,1/2) & 6.0 & 4.1 & 3.3 & 6.0 & --- & --- & --- & --- \\
\cline{2-10}      & (1,7,0) & 5.8 & 4.0 & 3.3 & 5.8 & 5.8 & 5.5 & 4.2 & 3.7 \\
\cline{2-10}      & (1,7,$\epsilon$) & 6.4 & 4.3 & 3.9 & 6.4 & 6.0 & 5.6 & 4.4 & 3.9 \\
    \hline
    \multicolumn{1}{c|}{\multirow{4}[8]{*}{30 TeV}} & (1,2,1/2) & --- & 2.8 & --- & 2.8 & 8.5 & 6.7 & 4.3 & 3.2 \\
\cline{2-10}      & (1,3,0) & --- & 3.7 & --- & 3.7 & 14 & 14 & 14 & 13 \\
\cline{2-10}      & (1,3,$\epsilon$) & 0.9 & 4.6 & --- & 4.6 & 14 & 14 & 14 & 13 \\
\cline{2-10}      & (1,4,1/2) & 2.5 & 6.8 & 2.4 & 7.0 & --- & --- & --- & --- \\
\cline{2-10}    \multicolumn{1}{c|}{\multirow{5}[10]{*}{90 $\abi$}} & (1,5,0) & 3.1 & 7.0 & 3.1 & 7.2 & 14 & 13 & 12 & 10 \\
\cline{2-10}      & (1,5,$\epsilon$) & 6.9 & 7.8 & 4.2 & 8.4 & 14 & 14 & 12 & 11 \\
\cline{2-10}      & (1,6,1/2) & 11 & 8.7 & 6.1 & 11 & --- & --- & --- & --- \\
\cline{2-10}      & (1,7,0) & 11 & 8.6 & 6.1 & 11 & 13 & 12 & 9.1 & 8.1 \\
\cline{2-10}      & (1,7,$\epsilon$) & 13 & 9.2 & 7.4 & 13 & 13 & 12 & 10 & 8.6 \\
    \hline
    \multicolumn{1}{c|}{\multirow{4}[8]{*}{100 TeV}} & (1,2,1/2) & 2.2 & 10 & --- & 10 & 29 & 23 & 15 & 10 \\
\cline{2-10}      & (1,3,0) & 3.6 & 12 & --- & 12 & 48 & 48 & 46 & 43 \\
\cline{2-10}      & (1,3,$\epsilon$) & 4.1 & 16 & --- & 16 & 48 & 48 & 47 & 45 \\
\cline{2-10}      & (1,4,1/2) & 5.8 & 23 & 4.0 & 23 & --- & --- & --- & --- \\
\cline{2-10}    \multicolumn{1}{c|}{\multirow{5}[10]{*}{$10^3 \abi$}} & (1,5,0) & 6.9 & 24 & 5.8 & 24 & 47 & 45 & 39 & 35 \\
\cline{2-10}      & (1,5,$\epsilon$) & 9.4 & 27 & 8.7 & 27 & 47 & 46 & 40 & 37 \\
\cline{2-10}      & (1,6,1/2) & 23 & 30 & 15 & 31 & --- & --- & --- & --- \\
\cline{2-10}      & (1,7,0) & 22 & 29 & 15 & 31 & 43 & 40 & 31 & 27 \\
\cline{2-10}      & (1,7,$\epsilon$) & 34 & 31 & 19 & 35 & 43 & 41 & 32 & 29 \\
    \hline

    \end{tabular}%
  \label{tab:5sigmatable}%
\end{table}%

\clearpage

\pagestyle{headings} 
%

\clearpage

\bibliography{refs_MuCDM}

\bibliographystyle{JHEP}

\end{document}